\documentclass[12pt]{report}

\usepackage{setspace}
\usepackage{feynmp}
\usepackage{amsmath}
\usepackage{amsfonts}
\usepackage{amssymb}
\usepackage{dsfont}
\usepackage{dcolumn}
\usepackage{bm}
\usepackage{psfrag}
\usepackage{graphics,graphicx,epsfig}
\usepackage{latexsym}

\def\beq{\begin{equation}}
\def\be{\begin{equation}}
\def\beqn{\begin{eqnarray}}
\def\ee{\end{equation}}
\def\eeq{\end{equation}}
\def\eeqn{\end{eqnarray}}

\def \cha{\widetilde{\chi}^{\pm}_1}

\def \na{\widetilde{\chi}^{0}_1}
\def \nb{\widetilde{\chi}^{0}_2}
\def \nc{\widetilde{\chi}^{0}_3}

\def \g{\widetilde{g}}

\def \ta{\widetilde{t}_1}

\def \sta{\widetilde{\tau}_1}
\def \stb{\widetilde{\tau}_2}

\def \slr{\widetilde{l}_R}

\def \snl{\widetilde{\nu}_{\tau}}
\def \snm{\widetilde{\nu}_{\mu}}

\def \hc{H^{\pm}}

\newcommand{\sellpm}{\widetilde{\ell}^{\pm}}
\newcommand{\sell}{\widetilde{\ell}}
\newcommand{\el}{\widetilde{e}_L}
\newcommand{\ml}{\widetilde{\mu}_L}

\def \non{nonuniversalities~}

\def\s{Stueckelberg~}

\def\s1{$s_{\alpha}$}
\def\s2{$s_{\gamma}$}
\def\s3{$s_{\delta}$}
\def\c1{$c_{\alpha}$}
\def\c2{$c_{\gamma}$}
\def\c3{$c_{\delta}$}

\newcommand{\mathsym}[1]{{}}

\begin{document}

\singlespacing

\begin{center}

\null\vfill
{\large PROBING SUPERGRAVITY UNIFIED THEORIES 
    \\AT THE LARGE HADRON COLLIDER}\\

\vfill\vfill\vfill
A dissertation presented \\
\vfill
 by \\ 
\vfill
 Zuowei Liu \\
\vfill\vfill\vfill
to \\
The Department of Physics \\
\vfill\vfill\vfill
In partial fulfilment of the requirements for the degree of \\
Doctor of Philosophy \\
\vfill
in the field of \\
\vfill
Physics\\
\vfill\vfill\vfill
Northeastern University\\
Boston, Massachusetts\\
August, 2008
\vfill\null
\end{center}

\newpage
\null\vfill
\begin{center}
\copyright Zuowei Liu, 2008\\
ALL RIGHTS RESERVED
\end{center}
\clearpage

\doublespacing

\addcontentsline{toc}{chapter}{Abstract}

\singlespacing

\newpage
\begin{center}

\null\vfill
{\large PROBING SUPERGRAVITY UNIFIED THEORIES 
    \\AT THE LARGE HADRON COLLIDER}\\

\vfill
 by \\ 
\vfill
 Zuowei Liu \\
\vfill\vfill
{\large ABSTRACT OF DISSERTATION}\\
\vfill\vfill
Submitted in partial fulfillment of the requirement \\
for the degree of Doctor of Philosophy in Physics\\
in the Graduate School of Arts and Sciences of \\
Northeastern University, August, 2008\\
\vfill\vfill\vfill\vfill  

\end{center}

\newpage
\doublespacing

\begin{flushleft}
\textbf{\Huge Abstract}
\end{flushleft}
\vspace{1cm}

The discovery of supersymmetry is one of the major goals of the
current experiments at the Tevatron and in proposed experiments 
at the Large Hadron Collider (LHC). 
However when sparticles are produced the signatures of  their production 
will to a significant degree depend on their hierarchical mass patterns.
Here we investigate hierarchical mass patterns for the four lightest
sparticles within one of the leading candidate theories - the SUGRA model.
Specifically we analyze the hierarchies for the four lightest sparticles
for the mSUGRA as well as for a general class of supergravity unified models
including nonuniversalities in the soft breaking sector.
It is shown that out of nearly $10^4$ possibilities of sparticle mass hierarchies,
only a small number  survives the rigorous constraints of radiative electroweak symmetry
breaking, relic density and other experimental constraints.
The signature space of these mass patterns at the LHC is investigated using
a large set of final states including   multi-leptonic states,
hadronically decaying $\tau$s,  tagged $b$ jets and other hadronic jets.
In all, we analyze more than 40 such lepton plus jet and missing energy
signatures along with several kinematical signatures such as missing
transverse momentum,  effective mass, and invariant mass  distributions of
final state observables. It is shown that a composite analysis can
produce significant discrimination among sparticle mass patterns
allowing for a possible identification of the source of soft
breaking. While the analysis given is for supergravity models, the techniques
used in the analysis are applicable to wide class of models including string
and brane models.	
\addcontentsline{toc}{chapter}{Acknowledgements}

\newpage
\thispagestyle{plain}
\begin{flushleft}
\textbf{\Huge Acknowledgements}
\end{flushleft}
\vspace{1cm}

\noindent I am deeply indebted to my advisor, Professor Pran Nath, for 
bringing me to the field of theoretical physics, and also 
for the enormous amount of time and efforts that he has taken 
to help me progress in my understanding of physics. 
My thesis would not have been possible without his constant guidance. 
His commitment to research and dedication to physics, 
and even his unique style of language and his sense of humor 
have all greatly influenced me. 
Being his student is a great fortune of my life. \\

\noindent I thank the other members of my committee, 
Professor George Alverson and Professor Tomasz Taylor for their 
help and patience during my Ph.D.\ study. 
I am also thankful to other members of the high energy group 
for many fruitful interactions throughout the years, 
especially Professor Brent Nelson and Professor Darien Wood. \\

\noindent I wish to thank Daniel Feldman, my collaborator, with whom 
the research contained in this thesis was completed 
and whose extraordinary energy and enthusiasm 
for physics has set a great example for me. I also thank him 
for his constant encouragement and help over the years.  \\

\noindent I thank the Physics department for giving me 
this wonderful opportunity to study here, and for their financial 
support during the process of my graduate study. 
I would also like to thank the Office of the Provost and Dr.\ Luis M.\ Falcon 
for awarding me the Dissertation Writing Fellowship during my thesis writing. \\

\noindent I am very grateful to my fellow graduate students: 
Ismet Altunkaynak, Tanmoy Das, 
Gabriel Facini, Peng He, 
Yongjian Huang, Jing Lou, 
Eugen Panaitescu, Thayaparan Paramanathan, 
Romain Scheck, Fei Wang, 
Zhen Wu, Weiqiao Zeng. 
With their friendship, I have spent a good time at Northeastern University. \\

\noindent Last and most, I thank my parents who have always been there for me. 
None of these would be made possible without the everlasting love and support of 
my parents who have devoted their lives to making me an educated person.

\tableofcontents

\chapter{Introduction}
\label{ch:intro}

\begin{figure*}[htb]
\centering
\includegraphics[width=12.0cm,height=6.0cm]{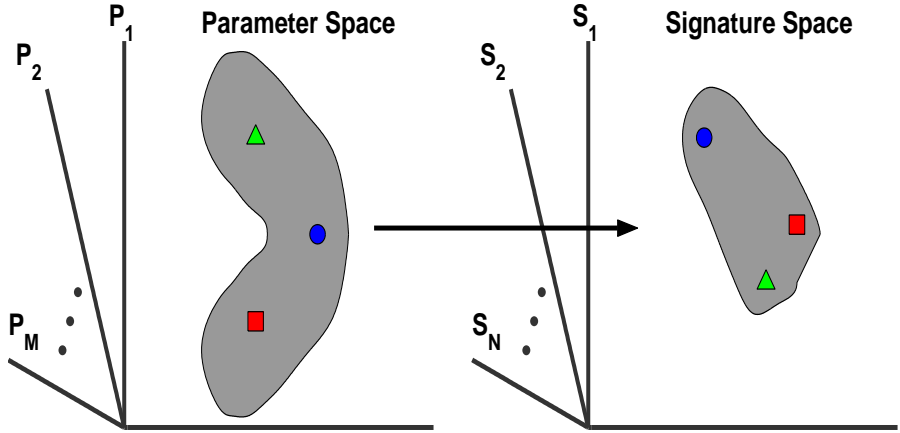}
\caption{The map between the parameter space and the signature space.}
\label{fig:map}
\end{figure*}

\noindent While the Standard Model of particle interactions is highly successful, 
important gaps remain in extending the model to a more complete unification, 
including the electroweak and the strong, and eventually the gravitational interactions. 
Over the  past decades, supersymmetry (SUSY) has turned out to be one of the leading candidates  
for physics beyond the Standard Model. In this thesis, we investigate the signatures  
at the CERN Large Hadron Collider (LHC) for some of the supersymmetric theories. 
This analysis can be thought as a map from the parameter space of the underlying theories 
onto the signature space of the LHC, as indicated in Fig.~(\ref{fig:map}). 
The parameter space is spanned by the input parameters of the theoretical models, 
and the number of the independent input parameter can be as many as 
$\sim 110$  in the Minimal Supersymmetric Standard Model (MSSM). 
It is not realistic to investigate models with such a large parameter space  
for their LHC signatures. However, it is possible to do so in some well motivated models 
where the dimensionality of the parameter space reduces significantly. 
Specifically we will focus on models where the dimension $M$ of 
the parameter space is small, as often $M=4$. 

The strategy of this analysis is as follows. 
We start with a point in the M-dimensional parameter space, 
investigate the theoretical predictions of this model point, 
simulate its behavior within the Large Hadron Collider, 
and finally extract many possible LHC signatures which is 
represented by a point in the N-dimensional signature space as shown in Fig.~(\ref{fig:map}). 
The dimensionality of the signature space can be expanded by adding 
more LHC signatures. However, many of the signature channels are 
strongly correlated so the independent set of signatures is typically much 
smaller than the number of signatures investigated.

The analysis in this thesis focuses on supergravity models, including the 
minimal supergravity grand unification models (mSUGRA) and SUGRA models 
with \non (NUSUGRA). 
The mSUGRA model depends on four soft breaking parameters, and 
the SUGRA with \non that we investigate here contain two more 
which characterize the \non in different sectors. 
We perform an exhaustive scan with Monte Carlo simulations 
in the mSUGRA soft parameter space (4-D parameter space) and in  
NUSUGRA parameter space (6-D parameter space). 
The model points that  
pass the various cosmological and collider constraints are classified by their mass 
hierarchical patterns. These hierarchical patterns are defined 
by the four lightest sparticles. We will see that the hierarchical mass patterns to a great 
degree influence the LHC signatures. 
Most of the analysis given here will attempt to correlate the hierarchical mass patterns 
with specific signatures at the LHC in the mSUGRA model. Extension of the analysis to 
NUSUGRA will also be discussed.

The thesis is organized as following. In chapter (\ref{ch:sm}) we give 
a brief introduction to the Standard Model (SM), and in chapter (\ref{ch:mssm}) 
we discuss Supersymmetry (SUSY), 
and the Minimal Supersymmetric Standard Model (MSSM). In chapter (\ref{ch:sugra}), 
we give an introduction to supergravity unified models. 
We list the various experimental constraints that are imposed 
on the supergravity models in chapter (\ref{ch:constraint}). 

An analysis of hierarchical mass patterns within the mSUGRA and 
NUSUGRA models is given in chapter (\ref{ch:msp}). 
Here the correlations between the sparticle patterns and 
the nature of the soft breaking are also analyzed. 
In chapter (\ref{ch:lhc}), we give a detailed description of our simulations 
of the CERN Large Hadron Collider which includes the various steps of our simulation, the 
detector cuts, and the various signatures we investigate. 
In chapter (\ref{ch:sig}), we analyze the signatures arising from the sparticle patterns, and also 
use the signatures to discriminate the patterns. Several important kinematical signatures are analyzed in 
chapter (\ref{ch:kin}), including the missing transverse momenta, effective mass, and the 
dileptonic invariant mass distributions. In order to utilize as many signatures as possible, 
a global analysis using the so called fuzzy vector technique is given in chapter (\ref{ch:fuzzy}). 
An analysis regarding the signature degeneracies from different models and the ability of 
resolving the parameter space using the LHC data is carried out in chapter (\ref{ch:resolution}). 

In addition to the signature analysis for the LHC, 
we also investigate other experimental signatures 
within the context of the sparticle pattern discrimination. These include the 
analysis of Higgs production at Tevatron in chapter (\ref{ch:higgs}), 
the $B_s\to\mu^+\mu^-$ constraints in chapter (\ref{ch:bsmm}), 
and the direct detection of dark matter in chapter (\ref{ch:directdark}).

\chapter{The Standard Model}
\label{ch:sm}

The Standard Model (SM) \cite{Glashow:1961tr,Gross:1973id} 
of particle physics is a theory that describes three of the four known interactions, 
which are the electrodynamics, the weak interactions and the strong interactions. 
The SM is based on the gauge group $SU(3)_C\times SU(2)_L \times U(1)_Y$ where 
C stands for color, L stands for left chiral, and Y stands for hypercharge. 
There are three generations of quarks and leptons in the SM which are represented by  
left-handed doublets and right handed singlets in the $SU(2)_L$ gauge group, 
\begin{equation}
q_i=\left(\begin{matrix}u_L\cr d_L\end{matrix}\right)_i;\hspace{1cm}
\ell_i=\left(\begin{matrix}\nu_L\cr e_L\end{matrix}\right)_i;\hspace{1cm}
u_{Ri};\hspace{1cm}
d_{Ri};\hspace{1cm}
e_{Ri}
\end{equation}
where $\Psi_{L,R}=P_{L,R}\Psi$, $P_L=(1-\gamma_5)/2$, 
$P_R=(1+\gamma_5)/2$, 
and $i=1,2,3$ is the generation index. 
In order to give mass to quarks and leptons, as well as to vector bosons, 
an additional scalar $SU(2)_L$ doublet is introduced in the theory
\begin{equation}
\phi=\left(\begin{matrix}H^{(+)}\cr H^{(0)}\end{matrix}\right).
\label{eq:sm-higgs}
\end{equation}
The three fundamental interactions (the electroweak and the strong interactions) are 
mediated by eight $SU(3)$ color gluons, $G_\mu^a$; three $SU(2)_L$ gauge bosons, 
$A_\mu^i$; and one $U(1)$ hypercharge gauge field, $B_\mu$. 
All the above gauge bosons are realized in the adjoint representations of their 
corresponding gauge groups, and the strength of the interactions are described by 
their coupling constants $g_3$, $g_2$ and $g'$.

The dynamics of the Standard Model consists of the following three parts:
\begin{enumerate}
\item Gauge interactions,
\item Yukawa interactions,
\item Higgs potential.
\end{enumerate}  
The gauge interactions arise via the gauge covariant derivative 
\begin{equation}
D_\mu = \partial_\mu-i\left[ g_3\sum_{a=1}^8 G_\mu^aT_C^a+
g_2\sum_{i=1}^3A_\mu^iT_L^i+g'B_\mu\frac{Y}{2}\right]
\end{equation}
where $T_C^a$=($\lambda^a/2$; 0) for (quarks; leptons, Higgs), where 
the $\lambda^a$ are the eight Gell-Mann matrices; 
$T_L^i$=($\sigma^i/2$; 0) for $SU(2)_L$ (doublets; singlets), where 
$\sigma^i$ are the three Pauli matrices; and $Y$ is the hypercharge 
defined by $Q=T_L^3+Y/2$.

Since the explicit fermion mass terms violate the gauge symmetries, 
adding mass terms in the SM is done through the Yukawa interactions 
and the mechanism of spontaneous symmetry breaking. 
Thus we consider the Higgs potential given by 
\begin{equation}
V(\phi)=-\mu^2 \phi^{\dagger}\phi+\lambda(\phi^{\dagger}\phi)^2
\end{equation}
where $\mu^2$ and $\lambda$ are positive. The tachyonic mass term 
for the Higgs field gives rise to a spontaneous symmetry breaking which results 
in a non-vanishing vacuum expectation value (VEV) of the Higgs field
\begin{equation}
\left<\phi\right>_0=\left(\begin{matrix}0\cr v/\sqrt{2}\end{matrix}\right)
\end{equation}
where $v=\mu/\sqrt{\lambda}$. The neutral Higgs gets redefined with respect 
to this VEV, $H^0=(v+h)/\sqrt{2}$ and the new Higgs boson possesses a positive 
tree level (mass)$^2$ of $M_h^2=2\mu^2>0$. 

The Yukawa interactions which preserve the gauge symmetries are given by 
\begin{equation}
V_{Y}=\lambda^{(e)}_{ij}\bar{\ell}_i\phi e_{Rj}
+\lambda^{(u)}_{ij}\bar{q}_i\bar{\phi} u_{Rj}
+\lambda^{(d)}_{ij}\bar{q}_i\phi d_{Rj}
+h.c.
\label{eq:sm-yukawa}
\end{equation}
where $\bar{\phi}=i\sigma^2\phi^*$ is the complex conjugate of the Higgs doublet 
given in Eq.~(\ref{eq:sm-higgs}), and $\lambda^{(e,u,d)}_{ij}$ are 
the Yukawa coupling constants where $i,j=1,2,3$ are the generation indices. 
The Yukawa interactions are responsible for giving rise to the fermion masses. 
The flavor basis of the fermion species are not necessary the same as the mass eigen states of 
the particles, and the unitary transformations between the flavor eigen states and the mass eigen 
states is accomplished via the Cabibbo-Kobayashi-Maskawa (CKM) matrix. 

The spontaneous symmetry breaking  
also rotates the four $SU(2)_L\times U(1)_Y$ gauge bosons to their 
mass eigen states by means of the gauge interaction term of Higgs fields, 
$\{A^1,A^2\}\to\{W^1,W^2\}$ and $\{A^3,B\}\to\{A,Z\}$ 
where the photon field $A$ remains massless. Introducing the weak mixing 
angle, $\theta_W$ defined by $\tan\theta_W=g'/g$, 
we can express the photon and $Z$ boson fields as follows  
\begin{eqnarray}
A=\cos\theta_W B+\sin\theta_W A^3,\\
Z=-\sin\theta_W B+\cos\theta_W A^3.
\end{eqnarray}
The $W^{\pm}$ and $Z$ boson masses at tree level are given by
\begin{equation}
M_W=\frac{g_2v}{2}=\frac{g_2}{2\sqrt{2\lambda}}M_h,~~
M_Z=\frac{M_W}{\cos\theta_W},
\end{equation}
showing that the Higgs mass sets the electroweak mass scale.

Standard Model is a highly successful model. 
It predicted the existence of the $W$ and $Z$ bosons before these particles were observed. 
Since then, the SM has passed a variety of experimental tests with great accuracy. 
However, from the theoretical side, many aspects of the Standard Model are 
not satisfactory. There is no real unification 
among the electroweak and the strong interactions as the SM gauge group is 
a product of three different gauge groups.

However, the most serious problem within the Standard Model seems to be the 
``gauge hierarchy'' problem, or ``naturalness'' problem which concerns the renormalization 
of the Higgs mass. Within the SM, the one loop corrections to $M_h$ are quadratic 
in the momentum cutoff. If the cutoff momentum were to placed at the GUT scale 
$M_G\sim 10^{16}$ GeV, the tree level Higgs mass is forced to the same scale, 
in order to reproduce the electroweak scale. This then results in the ``fine tuning'' 
problem because of the huge difference between the GUT scale and the electroweak 
scale. Another shortcoming of the Standard Model 
is that gravity is not explained in its framework.

\chapter{Supersymmetry \& the Minimal Supersymmetric Standard Model}
\label{ch:mssm}

Supersymmetry (SUSY) \cite{Wess:1973kz} is a symmetry between bosons and fermions, 
i.e. it requires that the number of bose and fermi helicity states in a multiplet be equal. 
A supersymmetry transformation turns a bosonic state into a fermionic state, and vice versa. 
The generator $Q$ and its hermitian conjugate $Q^{\dagger}$ 
of a supersymmetry transformation obey the so called 
``Graded Lie Algebra'' which includes the following commutation and 
anti-commutation relations 
\begin{eqnarray}
\{Q_\alpha,Q_{\dot{\alpha}}^{\dagger} \}&=&
2\sigma_{\alpha\dot{\alpha}}^{\mu}P_{\mu} ,\\
\{Q_\alpha,Q_\beta \}=
\{Q_{\dot{\alpha}}^{\dagger},Q_{\dot{\beta}}^{\dagger} \}&=&0 ,\\
\lbrack Q_\alpha,P_{\mu}\rbrack =
\lbrack Q_{\dot{\alpha}}^{\dagger},P_{\mu}\rbrack&=&0 ,\\
\lbrack P_{\mu},P_{\nu} \rbrack &=&0 ,
\end{eqnarray}
where $\sigma^{\mu}=(-1,-\vec{\sigma})$ with $\sigma^{i}$ being the Pauli matrices, 
and the undotted (dotted) indices, $\alpha=1,2$ ($\dot\alpha=1,2$) are introduced when 
a four-component Dirac spinor is decomposed into two two-component Weyl spinors.

The simplest SUSY multiplets are the massless chiral multiplet and the vector multiplet. 
The left chiral multiplet consists of one left handed Weyl spinor (spin 1/2) and its superpartner,  
one complex scalar field $\phi$ (spin 0, left-handed).
The Weyl spinors can be used to represent fermionic matter fields such as quarks and leptons 
and the scalar fields are the superpartners, i.e. ``squarks'' and ``sleptons''. 
The vector multiplet in the Wess-Zumino gauge consists of one vector field (spin 1) and 
one Majorana spinor (spin 1/2). The vector fields represent the gauge bosons, and 
the additional spinors are the superpartners of the gauge bosons called ``gauginos''.

\begin{table}[htb]
\begin{center}
\begin{tabular}{|c|c|c|c|}
\hline
quark & $q_i=\left(\begin{matrix}u_L\cr d_L\end{matrix}\right)_i$ &
 squark & $\tilde{q}_i=\left(\begin{matrix}\tilde{u}_L\cr \tilde{d}_L\end{matrix}\right)_i$ \\
 & $u_{Ri}$ & & $\tilde{u}_{Ri}$ \\
 & $d_{Ri}$ & & $\tilde{d}_{Ri}$ \\
 \hline
lepton & $\ell_i=\left(\begin{matrix}\nu_L\cr e_L\end{matrix}\right)_i$ &
 slepton & $\tilde{\ell}_i=\left(\begin{matrix}\tilde{\nu}_L\cr \tilde{e}_L\end{matrix}\right)_i$ \\
 & $e_{Ri}$ & & $\tilde{e}_{Ri}$ \\ 
 \hline
Higgsino & $\tilde{H}_1=\left(\begin{matrix}\tilde{H}_1^0\cr \tilde{H}_1^-\end{matrix}\right)$ &
Higgs & $H_1=\left(\begin{matrix}H_1^0\cr H_1^-\end{matrix}\right)$ \\
 & $\tilde{H}_2=\left(\begin{matrix}\tilde{H}_2^+\cr \tilde{H}_2^0\end{matrix}\right)$ &
 & $H_2=\left(\begin{matrix}H_2^+\cr H_2^0\end{matrix}\right)$ \\
 \hline
\end{tabular}
\caption{Chiral supermultiplets in the Minimal Supersymmetric Standard Model.}
\label{tab:chiral}
\end{center}
\end{table}

\begin{table}[htb]
\begin{center}
\begin{tabular}{|c|c|c|c|}
\hline gluon & $G_{\mu}^a, (a=1,...,8)$ & gluino & $\tilde{G}^a, (a=1,...,8)$ \\
\hline $SU(2)$ gauge boson & $A_{\mu}^i, (i=1,2,3)$ & $SU(2)$ gaugino & $\tilde{A}^i,(i=1,2,3)$ \\
\hline $U(1)$ gauge boson & $B_{\mu}^Y$ & $U(1)$ gaugino & $\tilde{B}^Y$ \\
\hline \end{tabular}
\caption{Vector supermultiplets in the Minimal Supersymmetric Standard Model.}
\label{tab:vector}
\end{center}
\end{table}

The Minimal Supersymmetric Standard Model (MSSM) is the simplest 
supersymmetric extension of the Standard Model. One promotes each of the 
Standard Model particles to either a chiral or a vector multiplet, which makes  
the particle content roughly twice as big as in SM as shown in 
Table (\ref{tab:chiral}, \ref{tab:vector}). In the Higgs sector one has 
two Higgs doublets, one of which ($H_2$) gives mass to the up quarks and 
the other ($H_1$) gives mass to the down quarks and the charged leptons.

In total there are 32 supersymmetric particles. 
These include 4 Higgs boson states,  of which three  $(h,H,A)$ are neutral,
the first two being CP even and the third CP odd, and one charged Higgs $H^{\pm}$. 
In the gaugino-Higgsino sector there are two charged mass eigenstates (charginos) 
$\tilde\chi^{\pm}_{i=1,2}$, four charge neutral states (neutralinos) 
$\tilde\chi^0_{i=1,2,3,4}$, and the gluino $\tilde g$.
In the sfermion sector, before diagonalization, there are
9  scalar leptons (sleptons) which are superpartners of the leptons 
with left and right chirality and are denoted as:
$\{\tilde e_{L,R},\tilde \mu_{L,R}, \tilde\tau_{L,R} ,
  \tilde\nu_{e_L},\tilde \nu_{{\mu}_L},\tilde \nu_{{\tau}_L} \}$.
Finally there are 12 squarks which are the superpartners of the
quarks and are represented by:
$\{\tilde u_{L,R},\tilde c_{L,R}, \tilde t_{L,R},
  \tilde d_{L,R},\tilde s_{L,R}, \tilde b_{L,R} \}$.
Mass diagonal slepton and squark states will in general be mixtures of
$L$, $R$ states.

In MSSM the superpotential with R-parity conservation is given by
\beq
W = \hat{U}^C Y_u \hat{Q} \hat{H}_u + \hat{D}^C Y_d \hat{Q} \hat{H}_d + \hat{E}^C Y_e \hat{L} \hat{H}_d + \mu \hat{H}_u \hat{H}_d
\eeq
where $Y_{u,d,e}$ are matrices in family space. 
One can add a large number of soft terms to the Lagrangian. 
Examples of R-parity conserving terms are 
\begin{equation}
{\cal L}_{soft}={\cal L}_{soft}^{(2)}+{\cal L}_{soft}^{(3)}.
\end{equation}
The soft SUSY-breaking 
Lagrangian contains scalar couplings
\begin{equation}
{\cal L}_{soft}^{(3)} = \tilde{u}^C h_u \tilde{Q} H_u + \tilde{d}^C h_d \tilde{Q} H_d 
+ \tilde{e}^C h_e \tilde{L} H_d + B H_u H_d + h.c. 
\end{equation}
where $h_{u,d,e}$ are $3 \times 3$ matrices. There are also scalar masses
\begin{eqnarray}
{\cal L}_{soft}^{(2)} = m_{H_u}^2 H_u^{\dagger} H_u + m_{H_d}^2 H_d^{\dagger} H_d 
+ \tilde{Q}^{\dagger} M_{\tilde Q}^2 \tilde{Q} 
+ \tilde{L}^{\dagger} M_{\tilde L}^2 \tilde{L} \nonumber\\
+ \tilde{u}^{C \dagger} m_{\tilde u}^2 \ \tilde{u}^C + 
\tilde{d}^{C \dagger} m_{\tilde d}^2 \ \tilde{d}^C + 
\tilde{e}^{C \dagger} m_{\tilde e}^2 \ \tilde{e}^C
\end{eqnarray}
where $M_{\tilde Q}^2$, $M_{\tilde L}^2$, $m_{\tilde u}^2$, 
$m_{\tilde d}^2$, and $m_{\tilde e}^2$ are $3 \times 3$ matrices 
in family space. More generally, as already noted, there can be as many as $\sim 110$ parameters 
in the soft terms. 
However, as we will see, the sets of parameters decrease significantly 
in Supergravity Unified Models.

\chapter{Supergravity Unified Models}
\label{ch:sugra}

Supersymmetry provides a solution to the gauge hierarchy problem, which makes it 
an attractive candidate for the new physics beyond the Standard Model. 
The main hurdle in the development of realistic supersymmetric models in the 
early days was the difficulty of breaking supersymmetry in a phenomenologically 
viable manner. In the framework of supergravity, this problem is solved by 
the inclusion of the gravity into the analysis, which promotes supersymmetry 
from a global symmetry to a local symmetry \cite{Nath:1975nj, Freedman:1976xh, msugra}.

To construct viable supergravity models  one must couple 
 $N=1$ supergravity with vector gauge fields and with matter. This construction is often
referred to as ``applied supergravity'' \cite{msugra,applied,Cremmer:1982en} 
and herein one couples $N=1$ supergravity 
with $N=1$ Yang-Mills  fields in the adjoint representation of 
the gauge group G (where G could be $SU(3)_C\times SU(2)_L\times U(1)_Y$,
or $SU(5), SO(10), E(6)$ etc), and with $N=1$ matter which contains quarks and leptons and 
Higgs fields which belong to anomaly free combinations of representations of the gauge
group. The most general effective Lagrangian  thus constructed depends on three functions
which are the superpotential $W (\phi_i)$, the Kahler potential $K (\phi_i,\phi_i^{\dagger})$, 
and the gauge kinetic function $f_{\alpha\beta} (\phi_i, \phi_i^{\dagger})$ 
where $\alpha$, $\beta$ are adjoint representation gauge indices, $\phi_i$ are 
the spin zero components of the left handed chiral multiplet consisting of $(\phi_i, \chi_i)$.
We note  that  $W, K$ and $f_{\alpha\beta}$ are hermitian.
In fact $W$ and  $K$ enter in the effective theory only in the following fixed combination
\begin{equation}
{\cal G} = \kappa^2 K +\ell n [\kappa^6 W W^{\dagger}] 
\label{4.1}
\end{equation} 
where 
\begin{equation}
\kappa  =  1/M_{\rm Pl}
\label{4.2}
\end{equation} 
and $M_{\rm Pl}$ is the Planck mass defined in terms of the Newton's constant $G_N$ by 
\begin{equation}
M_{\rm Pl} =(8 \pi G_N)^{-1/2}  =2.4\times 10^{18} {\rm GeV}.
\label{4.3}
\end{equation} 
The above implies that the
supergravity Lagrangian effectively depends  only on two functions which are 
$f_{\alpha\beta}$~and~${\cal G}$.  
From the above one can easily check that the effective theory is invariant under the 
Kahler transformation as given below
\begin{equation}
K \rightarrow K - f (\phi_i) - f^{\dagger} (\phi_i),~W \rightarrow e^{\kappa^2 f} W.
\label{4.4}
\end{equation} 
It is useful to introduce the  so called Kahler metric as follows
\begin{equation}
K_j^i = K_{,j}^{,i} \equiv \frac{\partial^2 K}{\partial \phi_i \phi_j^{\dagger}}~=  
\kappa^{-2} {\cal G}_{,j}^{,i} .
\label{4.5}
\end{equation}
The case $K= \sum_i \phi_i \phi_i^{\dagger}$  gives $K_j^i =
\delta_j^i$
which is referred to as the  flat Kahler metric.

One of the most important results that  emerges from the  applied supergravity  analysis is 
that the effective potential of $N=1$ theory takes the form \cite{msugra,Cremmer:1982en}
\begin{equation}
V  =  \kappa^{-4} e^{-{\cal G}}~\left[({\cal G}^{-1})_j^i {\cal G}_{,i} {\cal G}^{,j} - 3\right]  
+  \frac{g^2}{2}~\left[Re (f^{-1})_{\alpha\beta}\right]~D_{\alpha}D_{\beta} .
\label{4.6}
\end{equation}
In the above $g$ is the gauge coupling constant, and 
$(f^{-1})_{\alpha\beta}$ and $({\cal G}^{-1})_j^i$
 are the matrix inverses of $f_{\alpha\beta}$
 and  ${\cal G}_{,j}^{,i}$ while $D_{\alpha}$ is given by
\begin{equation}
D_{\alpha} = \kappa^{-2} {\cal G}^{,i} (T^{\alpha})_{ij} z_j 
\label{4.7}
\end{equation} 
where $T^{\alpha}$ is the group generator.  
An alternative  form which is often useful is to  write  the scalar potential explicitly in term of  
${\tilde W}$~and~$K$, and one then has 
\begin{equation}
V = e^{\kappa K}~\left[(K^{-1})_j^i \left(\frac{\partial {\tilde W}}{\partial z_i}~+
\kappa^2 K_{,i} {\tilde W}\right) \left(\frac{\partial {\tilde W}}{\partial z_j}~+
\kappa^2 K_{,j} {\tilde W}\right)^{\dagger} - 3 \kappa^2 | {\tilde W} |^2\right]~+ V_D
\label{4.8}
\end{equation}

\noindent
where $V_D$ is as given as before.  In the applied supergravity construction 
the  kinetic energy of scalar fields is given by
\begin{equation}
- K_{,j}^{,i} (D^{\mu} \phi_i)~(D_{\mu} \phi_j)^{\dagger} 
\label{4.9}
\end{equation} 
where $D_{\mu}$ is the gauge covariant derivative. 

  Before proceeding further we comment on the $\mu$ term that arises in the Higgs bilinear 
  term in the form $\mu H_1 H_2$ in the superpotential. For phenomenological reasons $\mu$
  must be of electroweak size, and thus one might speculate on the origin of this term in
  the superpotential. In fact it is not difficult to see how such a term can arise. The simplest  way
  to envision the  generation of such a  term is via the Kahler potential. Thus one can write the 
  Kahler potential  for the MSSM case so that 
  \begin{equation}
K= K_0(\phi_i\phi_i^{\dagger}) + c_0 H_1 H_2
\label{4.10}
\end{equation}
where $\phi_i$ are the MSSM scalar and $c_0$ is dimensionless. 
Note that $\mu_0 H_1H_2$ is the most general bilinear term which
one write without introducing dimensioned parameters in the Kahler potential. Specifically operators
of dimension greater than 2 will be suppressed by $1/M_{Pl}$ and thus their  contributions will be small.
Next one can make a Kahler transformation and move the $c_0 H_1H_2$  term from the Kahler potential
to the superpotential \cite{Giudice:1988yz}.  Indeed a term of this type arises naturally in 
string constructions \cite{Antoniadis:1994hg}.

After the transformation the superpotential has the form
  \begin{equation}
W e^{\kappa^2f}  = W + c_0 \kappa^2 W H_1 H_2 + \cdot\cdot 
\label{4.11}
\end{equation}
As is discussed below spontaneous breaking of supersymmetry by gravity mediation gives  
a non-vanishing VEV for $W$ and  we define 
  \begin{equation}
\mu_0 = c_0\kappa^2 \langle W\rangle   
\label{4.12}
\end{equation}
where as will be seen below the quantity $\kappa^2 \langle W\rangle$ is of electroweak size.  

 We now turn to the issue of breaking of supersymmetry. One of the reasons that globally supersymmetric
 models do not have phenomenologically acceptable breaking of supersymmetry is that here one has 
 a positive definite potential which after spontaneous breaking of supersymmetry leads to 
 a large non-vanishing vacuum energy. This problem is corrected in supergravity
 models. Here as seen in   Eq.~(\ref{4.8}) one finds that the potential contains a term with a negative sign
 and thus after spontaneous breaking the vacuum energy can be fine tuned to zero. 
 
The central assumption of breaking of supersymmetry in supergravity models is that  
supersymmetry is broken in a hidden sector and the breaking is transmitted 
by the gravitational interactions to the visible sector. A specific illustration of this comes about as follows:
One writes the superpotential in the form \cite{msugra,barbi}
 \begin{equation}
 W= W_{\rm vis} + W_{\rm hid}
 \end{equation}
where $W_{\rm vis}$ contains fields of the visible sector which are the MSSM fields including quarks and 
leptons and Higgs fields, and $W_{\rm hid}$ contains fields in the hidden sector where supersymmetry breaks.
The breaking gives a non-vanishing VEV  so that $\langle W\rangle =\langle W_{\rm hid}\rangle$. The size of 
$ \langle W_{\rm hid}\rangle$ is estimated to be $m^2 M_{\rm Pl}$ where $m$ is an intermediate scale so 
that $m\sim 10^{11}$ GeV. The breaking is transmitted to the visible sector by gravitational interactions 
producing soft breaking terms. The gravitino develops a mass  which is 
$M_{3/2} = \kappa^2 e^{{\cal K}/2} |W| $. Similarly 
the scale of soft breaking that enters the scalar sector is given by
 \begin{equation}
m_0^2 \sim (\kappa^2  \langle W_{\rm hid}\rangle)^2.
\end{equation} 
Thus with $m\sim 10^{11}$ one finds $m_0 \sim 10^3$ GeV, i.e., of electroweak size.
Other soft  terms can be generated in a similar way. 
A remarkable aspect of the analysis is that in supergravity grand unified models the soft breaking
is independent of the grand unified unification scale $M_G$ which cancels out in the low energy
theory \cite{msugra,hlw,Nath:1983aw}

The gauge kinetic energy terms that generate masses for the gauginos 
are exhibited below
\begin{eqnarray}
{\cal L_{\rm gauge}}=&& -\frac{1}{4}\Re\!\left[f_{\alpha\beta}F_{\mu\nu}^{\alpha}
F^{\beta\mu\nu}\right] + \frac{1}{4} i\Im\!\left[
f_{\alpha\beta}F_{\mu\nu}^{\alpha}\tilde{F}^{\beta\mu\nu}\right]
+ \frac{1}{2}\Re\!\left[ f_{\alpha\beta}\left(
- \frac{1}{2}\bar{\lambda}^{\alpha}D\!\!\!\!/\lambda^{\beta}\right) \right]
\nonumber \\
& & - \frac{1}{8} i \Im\!\left[ f_{\alpha\beta} e^{-1}D_{\mu}
(e \bar{\lambda}^{\alpha}\gamma^{\mu}\gamma_5\lambda^{\beta})\right]
+ \frac{1}{4}\bar{e}^{G/2}G^a(G^{-1})^b_a(\partial f^*_{\alpha\beta}
/\partial z^{*b}\lambda^{\alpha}\lambda^{\beta}) + {\rm h.c.}\nonumber 
\end{eqnarray}
In general the gauge kinetic energy function 
$f_{\alpha\beta}$  has a non-trivial field dependence
involving fields which transform as a singlet or a non-singlet
irreducible representation of the underlying gauge group. 
After the spontaneous breaking of supersymmetry the above lead
 to gaugino masses. If one assumes that the fields transform as
 singlets of the underlying gauge groups, then the gaugino masses
 at the GUT scale will be universal and generate a term of the form 
 $m_{1/2} \bar \lambda_{\alpha} \lambda_{\alpha}$. Splitting of the gauge
 masses can be obtained by the assumption that 
 $f_{\alpha\beta}$ have fields which transform  as non-singlet
irreducible representation of the underlying gauge group.


The phenomenology of supergravity (SUGRA)  models has been discussed since
the inception of these models  (for reviews see \cite{nilles,nath20,Arnowitt:1993qp,Martin:1997ns}
and there exists now  a considerable amount of  literature
regarding the implications of SUGRA
(for early works see \cite{earlypheno}, for more recent works see 
\cite{modern,Trotta,Allanach0607} and 
\cite{Feldman:2007zn,Feldman:2007fq,Feldman:2008hs,Feldman:2008en,Feldman:2008jy}, 
for works with \non  see \cite{NU}, 
and for works with hierarchical breaking and with
$U(1)$ gauge extensions
see \cite{Kors:2004hz,Feldman:2006wd,Barger:2004bz}).
While many analyses of the mSUGRA parameter space have been
limited to the  case of vanishing trilinear couplings,  several recent works 
\cite{Ellis:2004tc,Stark:2005mp,Djouadi:2006be,Ellis:2006ix,Trotta,Allanach0607,uc,Bringmann:2007nk,Bhattacharyya:2008zi}
 have appeared relaxing this assumption, and new portions of
the parameter space have been found consistent with all known
experimental constraints on the model.


 As mentioned already in the previous chapter, there are 32 different 
 supersymmetric particles, or sparticles. 
 If all the 32 sparticle masses are treated as essentially all independent, aside
from sum rules (for a pedagogical analysis on sum rules in the
context of unification and RG analysis see \cite{Martin:1993ft}) on
the Higgs, sfermions, chargino and neutralino masses, then without
imposition of any phenomenological constraints, the number of
hierarchical patterns for the sparticles could be as many as
$O(10^{25})$ or larger. This represents a mini landscape in a loose way  reminiscent
of  the string landscape (which, however, is much larger with as
many as $O(10^{1000})$ possibilities) \cite{landscape}.
(Here we refer to the landscape of mass hierarchies and not to the
landscape of vacua as is  the case when one talks of a string
landscape. For the string case the landscape consists of a countably
discrete set, while for the case considered here, since the
parameters can vary continuously, the landscape of vacua is indeed
much larger.  However, our focus will be the landscape of mass
hierarchies.) Now, the number of
possibilities can be reduced by very significant amounts in
supergravity models with the imposition of  the constraints of
radiative electroweak symmetry breaking (REWSB) which we 
discuss below.

\section{Radiative Electroweak Symmetry Breaking}

In the minimal supergravity unification (mSUGRA) 
the potential at the GUT scale which gives rise to the soft breaking of 
supersymmetry is given by 
\begin{equation}
V=\sum_a|\frac{\partial W}{\partial \phi_a}|^2
+m_0^2\sum_i\phi_i^{\dagger}\phi_i
+A_0W^{(3)}+B_0W^{(2)}
+m_{1/2}\sum_{\alpha=3,2,1}\bar{\lambda}_{\alpha}\lambda_{\alpha}
\end{equation}
where $\phi_i$ are the scalar fields, $\lambda_{\alpha}(\alpha=3,2,1)$ are 
the gauginos corresponding to $SU(3)_C$, $SU(2)_L$ and $U(1)_Y$ gauge groups, 
$A_0$ is the coefficient of the trilinear coupling, and $B_0$ is the coefficient of 
the bilinear coupling. For MSSM, $W^{(2)}$ takes the form 
\begin{equation}
W^{(2)}=\mu_0\epsilon_{ij}H_1^iH_2^j
\end{equation}
where $i,j=1,2$, and $W^{(3)}$ takes the form 
\begin{equation}
W^{(3)}=\lambda_{ij}^{(u)} q_i H_2 u_j^C
+\lambda_{ij}^{(d)} q_i H_1 d_j^C
+\lambda_{ij}^{(e)} \ell_i H_1 e_j^C
\end{equation}
where $\lambda_{ij}^{(u,d,e)}$ are the Yukawa couplings 
(analogous to those in Eq.~(\ref{eq:sm-yukawa})) and 
$H_1$ and $H_2$ are the two Higgs doublets. 


The soft breaking for the minimal supergravity model is thus given by the parameters  
\begin{equation}
m_0,~~m_{1/2},~~A_0,~~B_0.
\end{equation}
While the mSUGRA model is initialized at the GUT scale, the experiments are carried out 
at the electroweak scale. One needs the Renormalization Group Equations (RGE) to 
connect these two domains \cite{vaughn}. 
For discussion of the electroweak symmetry breaking, 
one needs the effective Higgs potential which is $V=V_0+\Delta V_1$ where 
$V_0$ is the tree part and $\Delta V_1$ is the one loop corrections \cite{Coleman:1973jx,Arnowitt:1992qp}
\begin{equation}
V_0=m_1^2|H_1|^2+m_2^2|H_2|^2-m_3^2(H_1H_2+h.c.)
+\frac{1}{8}(g_2^2+g_Y^2)(|H_1|^2-|H_2|^2)^2, 
\end{equation}
and
\begin{equation}
\Delta V_1 = \frac{1}{64\pi^2}\sum_a(-1)^{2s_a}n_aM_a^4
\ln \left[ \frac{M_a^2}{e^{3/2}Q^2} \right].
\end{equation}\label{eq:higgs-loop}
Here, $m_i(t)$, $g_2(t)$, $g_Y(t)$ are all ``running'' parameters at scale $Q$ 
where $t=\ln(M_G^2/Q^2)$. Thus \cite{Arnowitt:1993qp}
\begin{equation}
m_i^2(t) = m_{H_i}^2(t) + \mu^2(t), ~~ i = 1,2;
\end{equation}
\begin{equation}
m_3^2(t) = -B(t)\mu(t); 
\end{equation}
with the boundary conditions at the GUT scale $Q=M_G(t=0)$:
\begin{equation}
m_i^2(0) = m_0^2 + \mu_0^2, ~~ i = 1,2;
\end{equation}
\begin{equation}
m_3^2(0) = -B_0\mu_0.
\end{equation}
The coupling constants are unified at the GUT scale too  
\begin{equation}
\alpha_2(0) = (5/3)\alpha_Y(0) = \alpha_G.
\end{equation}
In Eq.~(\ref{eq:higgs-loop}), $M_a\equiv M_a(v_1,v_2)$ is the tree level mass 
of particle $a$ as functions of the Higgs VEVs, $v_i=\langle H_i \rangle$, and 
$s_a$ and $n_a$ are the spin and number of helicity states of particle $a$.


At the GUT scale, all scalar particles are of positive mass$^2$ values $m_0^2$. 
When one integrates down in the energy scale with RGEs, some mass$^2$ values can turn 
negative which will break down the $SU(2)\times U(1)$ symmetry as we have seen 
in the chapter (\ref{ch:sm}). The $SU(2)\times U(1)$ symmetry breaking requires 
two conditions: (1) the determinant of the mass$^2$ matrix be negative so that 
there exists one negative eigenvalue; (2) the potential be bounded from below. 
We apply these two conditions to the tree level Higgs potential and get 
\begin{equation}
{\cal D}=m_1^2m_2^2-m_3^4<0, 
\end{equation}
\begin{equation}
{\cal L}=m_1^2+m_2^2-2|m_3^2|>0. 
\end{equation}
The above two conditions cannot be satisfied simultaneously at the GUT scale since 
all scalar particles have the same mass. However,  when one evolves down from the 
GUT scale to the electroweak scale, the heavy top quark contributions to  
$m_{H_2}^2$ naturally pushes it to turn negative  
so that electroweak symmetry breaking can be achieved. 
The fact that the top quark must be heavy was one of the 
predictions of the supergravity theories \cite{AlvarezGaume:1983gj}
(see \cite{Ibanez:2007pf} for a recent review of radiative  breaking). 
Because the electroweak symmetry breaking is driven 
by the quantum loop corrections, this mechanism is therefore known as radiative electroweak 
symmetry breaking (REWSB). 

Turning now to the minimization of the Higgs potential, one has the following relations 
\begin{equation}
\mu_1^2-m_3^2\tan\beta+\frac{1}{2}M_Z^2\cos(2\beta)=0, 
\end{equation}
\begin{equation}
\mu_2^2-m_3^2\cot\beta-\frac{1}{2}M_Z^2\cos(2\beta)=0, 
\end{equation}
where $\tan\beta\equiv v_2/v_1$,  $\mu_i^2=m_i^2+\Sigma_i(i=1,2)$ and 
$\Sigma_i$ is the loop corrections arising from the loop Higgs potential 
$\Delta V_1$. Taking $\mu_i^2$ and $m_3^2$ as the input parameters, 
one can solve for $\tan\beta$ and $M_Z$ 
\begin{equation}
\sin(2\beta) = \frac{2m_3^2}{\mu_1^2+\mu_2^2};
\label{eq:bmu1}
\end{equation}
\begin{equation}
\frac{1}{2}M_Z^2 = \frac{\mu_1^2-\mu_2^2\tan^2\beta}{\tan^2\beta-1}.
\label{eq:bmu2}
\end{equation}
One can also treat the $\tan\beta$ and $M_Z$ as the input parameters, and eliminate 
two of the GUT scale parameters, say $B_0$ and $\mu_0$, 
using Eqs.~(\ref{eq:bmu1}, \ref{eq:bmu2}). 
Since only $\mu^2$ enters the Eqs.~(\ref{eq:bmu1}, \ref{eq:bmu2}), the sign of 
$\mu$ is undetermined. 
Thus one has that the low energy physics depends on the following parameters 
\begin{equation}
m_0,~~m_{1/2},~~A_0,~~\tan\beta,~~sign(\mu)
\end{equation}
where $m_0$ is the universal scalar mass, $m_{1/2}$ is the universal gaugino mass,
$A_0$ is the universal trilinear coupling, $\tan\beta$ is the ratio of the two
Higgs VEVs in the MSSM, and $\mu$
is the Higgs mixing parameter that enters via the term $\mu H_1 H_2$ in the superpotential.


\section{Hyperbolic Branch (HB) of Radiative Symmetry Breaking }

The symmetry breaking condition Eq.~(\ref{eq:bmu2}) 
can be rewritten as \cite{hb/fp,Lahanas:2003bh} 
\begin{equation}
\Phi = \frac{1}{4}+\frac{\mu^2}{M_Z^2}, 
\end{equation}
where the new parameter 
\begin{equation}
\Phi^{-1}\equiv 4 \frac{\lambda^2-\mu^2}{\lambda^2+\mu^2}
\end{equation}
is introduced for the purpose of the measure of naturalness. 
Using the radiative electroweak symmetry breaking constraint and 
ignoring the $b$-quark couplings, we may express the parameter $\Phi$ as 
\begin{equation}
\Phi=-\frac{1}{4}+\left(\frac{m_0}{M_Z}\right)^2C_1
+\left(\frac{A_0}{M_Z}\right)^2C_2
+\left(\frac{m_{1/2}}{M_Z}\right)^2C_3
+\left(\frac{m_{1/2}A_0}{M_Z^2}\right)C_4
+\frac{\Delta\mu_{loop}^2}{M_Z^2},
\end{equation}\label{eq:naturalness}
where
\begin{eqnarray}
C_1&=&\frac{1}{\tan^2\beta-1}\left(1-\frac{3D_0-1}{2}\tan\beta\right) ,\\
C_2&=&\frac{\tan^2\beta}{\tan^2\beta-1}k ,\\
C_3&=&\frac{1}{\tan^2\beta-1}\left(g-e\tan^2\beta \right) ,\\
C_4&=&-\frac{\tan^2\beta}{\tan^2\beta-1}f ,\\
\Delta\mu_{loop}^2&=&\frac{\Sigma_1-\Sigma_2\tan^2\beta}{\tan^2\beta-1}.
\end{eqnarray}
Here $D_0=1-(m_t/m_f)^2$ with $m_f\sim 200 \sin\beta$ GeV, and $e$, $f$, $g$, $k$ are 
defined in \cite{Ibanez:1984vq}. To investigate the limits of $m_0$ and $m_{1/2}$ consistent 
with the symmetry breaking for some given value of $\Phi$, we rewrite the Eq.~(\ref{eq:naturalness}) as 
\begin{equation}
C_1m_0^2+C_3m_{1/2}^{'2}+C_2^{'}A_0^2+\Delta\mu_{loop}^2=M_Z^2\left(\Phi_0+\frac{1}{4}\right),
\end{equation}\label{eq:mmm}
where 
\begin{equation}
m'_{1/2}=m_{1/2}+\frac{A_0C_4}{2C_3},~~C'_2=C_2-\frac{C_4^2}{4C_3}.
\end{equation}

For small to moderate values of $\tan\beta$, when the loop corrections are small and 
$C_1$, $C'_2$, and $C_3$ are all positive from the renormalization group analysis, 
Eq.~(\ref{eq:mmm}) can be rewritten as 
\begin{equation}
\frac{m_{1/2}^{'2}}{a^2}+\frac{m_0^2}{b^2}+\frac{A_0^2}{c^2}\simeq 1.
\end{equation}
Here one finds that the radiative symmetry breaking demands that the allowed set of 
soft parameters lie on the surface of an Ellipsoid for a fixed value of $\mu$.  
However, for the case with large values of $\tan\beta$, the loop corrections to $\mu$ become significant. 
In this case, the size of the loop corrections depends sharply on the scale $Q_0$ where 
the minimization of the effective potential is carried out. If we choose the scale at which 
the loop corrections are minimized, or even vanish, the loop corrections can be omitted for 
the analysis again except that the sign of $C_1$ can be flipped for some region of the parameter 
space. Typically the scale is not distant from the average of the smallest and 
largest sparticle masses. Under these conditions, the minimization condition takes the form 
\begin{equation}
\frac{m_{1/2}^{'2}}{\alpha^2(Q_0)}-\frac{m_0^2}{\beta^2(Q_0)}\simeq \pm1
\end{equation}
where 
\begin{eqnarray}
\alpha^2&=&|\frac{(\Phi_0+1/4)M_Z^2-C'_2A_0^2}{C_3}|,\\
\beta^2&=&|\frac{(\Phi_0+1/4)M_Z^2-C'_2A_0^2}{C_1}|.
\end{eqnarray}
The set of parameters which satisfy the above relation lie on the surface of a Hyperboloid, 
and thus this branch is known as the Hyperbolic Branch (HB). The interesting thing about the HB region 
is that $m_0$ and $m_{1/2}$ can become quite large with $m_0$ lying in the multi TeV 
region even with small fine tuning \cite{hb/fp}.

We note in passing that while the phenomena with soft breaking discussed above are within the
framework of supergravity models many aspects of these  results translate to soft breaking within the
framework of heterotic  string models and for models based on intersecting D branes 
(see, e.g.,  \cite{Nath:2002nb,Chattopadhyay:2004dt,Kors:2003wf,Lust:2004dn,Kane:2004hm}).

\section{Sparticle Masses}

Now let us list the mass matrices for the sparticles in the MSSM. Due to the effect of 
electroweak symmetry breaking, the higgsinos and the electroweak gauginos are mixed to 
form the mass eigenstates called neutralinos and charginos. For the neutralinos, the mass 
matrix takes the form 
\begin{equation}
M_{{\tilde \chi}^0}=\left(\begin{matrix}
		M_1 & 0 & -M_Z s_W  c_\beta    & M_Z    s_W  s_\beta \cr
                 0  & M_2 & M_Z c_W   c_\beta & -M_Z c_W  s_\beta \cr
		 -M_Z s_W  c_\beta &  M_Z c_W c_\beta & 0 & -\mu \cr
		 M_Z s_W s_\beta  & -M_Z  c_W s_\beta & -\mu & 0
		 \end{matrix}\right) 
\end{equation}
where $\theta_W$ is the weak angle, $s_W=\sin \theta_W$, $c_W=\cos \theta_W$, 
$s_\beta= \sin\beta$, and $c_\beta = \cos \beta$. 
For the charginos, the mass matrix takes the form 
\begin{equation}
M_{{\tilde \chi}^{\pm}}=\left(\begin{matrix}
M_2 & \sqrt{2}M_W s_\beta \cr
\sqrt{2}M_W c_\beta & \mu \cr
		 \end{matrix}\right) 
\end{equation}

The up squark (mass)$^2$  matrix at the electroweak scale is given by
\begin{equation}
M_{\tilde{u}}^2=\left(\begin{matrix}M_{\tilde{Q}}^2+m{_u}^2 + M_{Z}^2(\frac{1}{2}-Q_u
s^2_W)\cos2\beta & m_u(A_{u}^{*} - \mu \cot\beta) \cr
   	          	m_u(A_{u} - \mu^{*} \cot\beta) & 
m_{\tilde{u}}^2+m{_u}^2 + M_{Z}^2 Q_u s^2_W \cos2\beta
		 \end{matrix}\right) 
\end{equation}
where $Q_u=\frac{2}{3}$.  And the down squark (mass)$^2$  matrix 
at the electroweak scale is given by
\begin{equation}
M_{\tilde{d}}^2=\left(\begin{matrix}M_{\tilde{Q}}^2+m{_d}^2-M_{Z}^2(\frac{1}{2}+Q_d
s^2_W)\cos2\beta & m_d(A_{d}^{*} - \mu \tan\beta) \cr
                        m_d(A_{d} - \mu^{*} \tan\beta) & m_{\tilde{d}
}^2+m{_d}^2+M_{Z}^2 Q_d s^2_W \cos2\beta
                \end{matrix} \right) 
\end{equation}
where $Q_d=-\frac{1}{3}$. In deducing the above, the relations $h_{u,d,e}=Y_{u,d,e} A_{u,d,e}$ are used 
for the trilinear terms. Finally, the slepton (mass)$^2$ matrix is given by 
\begin{equation}
M_{\tilde {\it l}}^2=\left(\begin{matrix} 
M_{\tilde L}^2+m_{e}^2 -M_Z^2(\frac{1}{2}
- s^2_W)\cos 2\beta & m_{e}(A_{e}^* - \mu\tan\beta)\cr
m_{e}(A_{e} - \mu^*\tan\beta) &  m_{\tilde e}^2 + m_{e}^2 -M_Z^2 s^2_W
	\cos 2\beta  
	 \end{matrix}            \right)
  \end{equation}

\section{Dark Matter in Supergravity}

In most of the allowed parameter space of the mSUGRA model 
consistent with radiative breaking of the EW symmetry, the lightest 
neutralino is the LSP and hence a candidate for cold dark matter. 
We briefly discuss the computation of the relic density of neutralino 
$\Omega_{\chi} \equiv \rho_{\chi}/\rho_c$ where 
$\rho$ is the mass density of relic neutralinos in the universe and 
$\rho_c$ is the critical mass density needed to close the universe, i.e. 
\begin{equation}
\rho_c=\frac{3H_0^2}{8\pi G_N}.
\end{equation}
Here $H_0$ is the Hubble parameter at current time and $G_N$ is the 
newtonian constant. Numerically, one has 
\begin{equation}
\rho_c=1.9h_0^2\times 10^{-29} {\rm gm/cm}^3
\end{equation}
where $h_0$ is now the Hubble parameter in unit of 100 km/secMpc. 
The current value of $h_0$ is $h_0=0.7\pm 0.013$. In the analysis 
of $\Omega_{\chi}h_0^2$ we need to solve the Boltzman equation 
for $n$, the number density of neutralinos in the early universe, which 
is given by \cite{Lee:1977qs} 
\begin{equation}
\frac{dn}{dt}=-3Hn-\langle \sigma v \rangle (n^2-n_0^2). 
\end{equation} 
In the above, $n_0$ is the value of $n$ at thermal equilibrium, 
$\langle \sigma v \rangle$ is the thermal average of the neutralino 
annihilation cross section $\sigma(\na\na\to X)$ and $v$ is the relative 
$\na$ velocity, and $H$ is the Hubble parameter at time $t$. 
In the computation of the thermal average one can assume that 
the neutralinos are non-relativistic and thus one can approximate 
$\langle \sigma v \rangle$ by the relation 
\begin{equation}
\langle \sigma v \rangle =
\frac{\int_0^\infty dvv^2 ( \sigma v )e^{-v^2/4x}}
{\int_0^\infty dvv^2 e^{-v^2/4x}}.
\end{equation}
Here $x$ is defined by $x=kT/m_{\chi}$ where $T$ is the temperature 
and $k$ is Boltzman constant. A solution to the Boltzman equation gives \cite{Arnowitt:1993qp} 
\begin{equation}
\Omega_{\chi}h_0^2 = 2.5 \times 10^{-11} 
\left(\frac{T_{\chi}}{T_{\gamma}}\right)^3
\left(\frac{T_{\gamma}}{2.75}\right)^3
\frac{N_f^{1/2}}{J(x_f)}
\end{equation}
where $T_{\gamma}$ is the current microwave background temperature, 
$x_f$ is the ``freeze out'' temperature corresponding to the temperature where 
the annihilation rate becomes smaller than the expansion rate, so that $\na$ 
decouples from the background. $x_f$ is typically small with a value 
$x_f\sim 0.04$. $N_f$ is the number of degrees of freedom at freeze out and 
typically $N_f\simeq 289.5/8$ \cite{Olive:1980wz}. 
The factor $(T_{\gamma}/T_{\chi})^3$ 
is estimated to be $(T_{\gamma}/T_{\chi})^3\simeq 18.5$ 
(for more recent evaluation, see \cite{Baer:1995nc}). 
Finally, $J(x_f)$ is the integral defined by 
\begin{equation}
J(x_f) = \int_0^{x_f} dx \langle \sigma v \rangle x ~~({\rm GeV}^{-2}).
\end{equation}

Away from the poles one can carry out a power series expansion for 
$\sigma v$ so that $\sigma v = a + bv^2/6 + \cdot\cdot$. In this case 
the thermal average is straightforward. However, the above approximation 
is invalid near poles 
\cite{Griest:1990kh, Gondolo:1990dk, Arnowitt:1993mg, Nath:1992ty} 
and the integration becomes 
tricky since one has a double integration over a pole. However, it is possible 
to overcome this problem by an interchange in the order of 
integration as discussed in \cite{Arnowitt:1993mg, Nath:1992ty}. 

A realistic computation of the relic density in supersymmetric models is however, more complicated as 
co-annihilations contribute to the relic density \cite{Griest:1990kh,co1,co2,co3}
 
For instance, one can have 
co-annihilations involving staus $\tilde \tau$ so that 
\begin{eqnarray}
\tilde \tau \chi &\to& \tau Z, \tau h, \tau\gamma \\
\tilde \tau \tilde \tau^{*}  &\to& f_i \bar{f_i}, W^+W^-, ZZ, \gamma Z, \gamma\gamma \\
\tilde \tau \tilde \tau  &\to& \tau \tau \\
\tilde \tau \tilde \ell_i(i\ne \tau)  &\to& \tau \ell_i 
\end{eqnarray}

For the case of co-annihilation, one must consider the total density 
$n=\sum_i n_i$ where $i$ runs over all the sparticles that enter in the 
co-annihilations, where $n$ now obeys the equation
\begin{equation}
\frac{dn}{dt} = -3 H n - \langle \sigma_{\rm eff} v_{\rm rel} \rangle(n^2-n_{\rm eq}^2)
\end{equation}
where
\begin{equation}
\sigma_{\rm eff} = \sum_{ij}\sigma_{ij} \gamma_i \gamma_j.
\end{equation}
Here $\sigma_{ij}$ is the cross section of annihilation of particles $i$ and $j$, 
and $\gamma_i=n^i_{\rm eq}/n_{\rm eq}$ where $n^i_{\rm eq}$ refers 
to the number density of sparticle $i$ at thermal equilibrium. The relic density 
for these processes requires numerical integration programs and we use 
micrOMEGAs \cite{MICRO} in our analysis. We note in passing that relic density is affected
by Yukawa unification (see, e.g. \cite{Chattopadhyay:2001va} and the references therein) but we do not take such effects into  account here. 

\section{CP violation}
MSSM contains many sources of CP violation which arise from the soft breaking sector of the theory
(for a review of CP violation in SUGRA, strings and branes see \cite{Ibrahim:2007fb}).
The number of  phases in SUGRA models is reduced drastically. Specifically in mSUGRA one has 
just two CP phases which can be chosen to be the phase of the trilinear coupling A and the phase
of $\mu$. With non-universalities one can bring in new phases. A similar  situation holds in string 
and D brane models.  A stringent constraint on CP phases arises from the electric dipole moment (EDM) of 
the electron and of the neutron (see, e.g., \cite{Ellis:2008zy})
 which would naively imply that SUSY phases are all very small. 
However, this need not be the case \cite{Nath:1991dn} because of the cancellation 
mechanism \cite{cancellation}.
With the cancellation mechanism, one finds that the phases can be large and at the same time one can 
satisfy the EDM constraints.  Such phases can affect low energy phenomena such as the 
the gaugino and sfermion masses and Higgs masses \cite{cphiggs}.  CP phases can affect LHC signatures.
However, in this study we do not take the effect of CP phases into account. 

\section{Proton stability}
Another constraint on unified models of particle interactions arises from proton 
decay constraints (for a recent review see \cite{pdecay}).
In grand unified models and also in string and brane models, one has several sources of 
proton decay.  First of all, both in supersymmetric as well as in non-supersymmetric grand unified 
models one has baryon and lepton number violating dimension six operators due to the exchange
of lepto-quarks. The most prominent decay mode from these is $p\to e^+\pi^0$ and the current 
limit on this decay is $\tau (p\to e^+\pi^0) > 1.6\times 10^{33}$ yr. 
In supersymmetric theories the most dominant decay mode is $p\to \bar \nu K^+$ and the current 
limit on it is $\tau(p\to \bar \nu K^+) > 2.3\times 10^{33}$ yr.  
This latter  limit, i.e., on the mode $\bar \nu K^+$,
puts very stringent limits on grand unified models and on string models.
At the same time these modes are subject to a much greater
degree  of model dependence because of the unknown nature of physics at high scales.
In contrast the soft parameters are known to be independent of the high scale, specifically of 
the GUT scale \cite{msugra}. For this reason we will not consider any specific high scale model
but rather focus on weak scale supersymmetry which is determined by the soft  parameters
independent of the GUT scale \cite{msugra}.

\chapter{Experimental Constraints on Unified Models}
\label{ch:constraint}

Below we summarize the relevant constraints from collider and from astrophysical data
that are applied throughout the analysis unless stated otherwise.

\begin{enumerate}
\item WMAP 3 year data:
The lightest R-Parity odd supersymmetric particle (LSP) is assumed charge neutral.
The constraint on the relic abundance of
dark matter under the assumption that the relic abundance of neutralinos is the dominant component
places the bound: $0.0855<\Omega_{\na} h^2<0.1189~~(2\sigma)$ \cite{Spergel:2006hy}.
\item
As is well known sparticle loop exchanges make a contribution to the FCNC process
$b\rightarrow s\gamma$ which
is of the same order as  the Standard Model contributions (for an update of SUSY contributions
see  \cite{susybsgamma}).  The
 experimental limits on  $b\to s\gamma$ impose severe constraints
 on the SUSY parameter space and we use here the  constraints from
the Heavy Flavor Averaging Group (HFAG) \cite{hfag} along with the
BABAR, Belle and CLEO
   experimental results: ${\mathcal Br}(b\rightarrow s\gamma) =(355\pm 24^{+9}_{-10}\pm 3) \times 10^{-6}$.
A new estimate of ${\mathcal Br}(\bar B \to X_s \gamma)$ at
$O(\alpha^2_s)$ gives \cite{Misiak:2006zs} ${\mathcal
Br}(b\rightarrow s\gamma) =(3.15\pm 0.23) \times 10^{-4}$  which
moves the previous SM mean value of $3.6\times 10^{-4}$ a bit lower. In order
to accommodate this recent analysis on the SM mean, as well as the previous analysis, we have taken a wider
$3.5\sigma$ error corridor around the HFAG
value in our numerical analysis. The total ${\mathcal Br}(\bar B \to X_s \gamma)$ including
the sum of SM and SUSY contributions  are constrained by  this
corridor. With a 2$\sigma$ corridor, while some of the allowed points in our analysis will be eliminated,
the main results of our pattern analysis remain unchanged.
\item
The process $B_s\to \mu^+\mu^-$ can become significant for large
$\tan\beta$ since
the decay  has a leading $\tan^6\beta$ \cite{tb}
dependence and thus large $\tan\beta$ could be constrained by the experimental
limit ${\mathcal Br}( B_s \to \mu^{+}\mu^{-})$ $< 1.5 \times10^{-7}$ (90\% CL), $ 2.0 \times
10^{-7}$  (95\% CL) \cite{Abulencia:2005pw}.  This limit has
just recently been updated  \cite{Abazov:2007iy}
and gives ${\mathcal Br}( B_s \to \mu^{+}\mu^{-}) < 1.2 \times10^{-7}$  (95\% CL).
Preliminary analyses \cite{bsmumu07} have reported
the possibility of even more stringent constraints by a factor of 10.
We take a more conservative
approach in this analysis and allow model points subject to the bound
${\mathcal Br}( B_s \to \mu^{+}\mu^{-})  < 9\times 10^{-6}$ (for a review see \cite{Anikeev:2001rk}).

\item Additionally, we also impose a
lower limit  on the lightest  CP even  Higgs boson mass. For the Standard
Model like Higgs boson this limit is
$\approx$ 114.4 {~\rm GeV} \cite{smhiggs}, while a limit of 108.2 {\rm GeV} at 95\% CL
is set on the production of an invisibly decaying Standard Model like Higgs by
OPAL \cite{OPAL2007}. For the MSSM we take the constraint to be $m_h> 100 ~{\rm GeV}$.
A relaxation of the light Higgs mass constraint by 8 - 10 GeV
affects mainly the analysis of SUGRA models where the stop mass can be light.
However, light stops are possible even with the strictest imposition of the LEP bounds on the SM Higgs Boson.
We take the other sparticle mass
constraints to be $m_{\cha}>104.5 ~{\rm GeV}$ \cite{lepcharg} for the lighter
chargino,  $m_{\ta}>101.5 ~{\rm GeV}$  for the lighter stop,  and $m_{\sta}>98.8 ~{\rm  GeV}$ for the
lighter stau.
\end{enumerate}
In addition to the above one may also consider the constraints from the anomalous magnetic moment
of the muon. It  is known that the supersymmetric electroweak corrections to $g_{\mu}-2$ can be as large or
larger than the Standard Model electroweak corrections \cite{yuan}.
The implications of recent  experimental data  has
been discussed in several works (see, e.g.\cite{g2T}). As in \cite{Djouadi:2006be}, here we use a rather
conservative bound $-11.4\times 10^{-10}<g_{\mu}-2<9.4\times 10^{-9}$.

\chapter{The SUGRA Sparticle Patterns}
\label{ch:msp}

In this chapter, we discuss the possible sparticle mass hierarchical 
patterns arising from SUGRA models utilizing the Monte Carlo simulations. 
We also discuss the correlation between the sparticle patterns and 
the nature of the soft parameter space. 
Finally we give a collection of benchmarks for each of 
the sparticle patterns analyzed here.

\section{The Sparticle  Landscape \label{A}}

The analysis proceeds by specifying the model input parameters at
the GUT scale, $M_{G}\sim 2\times 10^{16}$ GeV, 
(for our analysis, no flavor mixing is considered at the GUT scale)
and using the renormalization group equations (RGEs)
to predict the sparticle masses and mixing angles at the electroweak scale. 
The RGE code used to obtain the mass spectrum is SuSpect 2.34 \cite{SUSPECT}, 
which is the default RGE calculator in MicrOMEGAs version 2.0.7 \cite{MICRO}.  
We have also investigated other RGE programs 
including ISASUGRA/ISAJET \cite{ISAJET}, SPheno
\cite{SPHENO} and SOFTSUSY \cite{SOFTSUSY}. 
We have cross checked our analysis using different codes 
and find no significant disagreement in most regions of the parameter space. 
The largest sensitivity appears to arise for the case of large $\tan\beta$ and
the analysis is also quite sensitive to the running bottom mass and to the
top pole mass (we take $m^{\overline{\rm MS}}_b(m_b)= 4.23~ {\rm GeV} $ 
and $m_t(\rm pole) = 170.9 ~{\rm GeV}$  in this analysis).
\begin{table}[htbp]
    \begin{center}
\begin{tabular}{||l||l||c||c||}
\hline\hline
mSP&     Mass Pattern & $\mu >0$ & $\mu<0$
\\\hline\hline
mSP1    &   $\na$   $<$ $\cha$  $<$ $\nb$   $<$ $\nc$   & Y  & Y    \cr
mSP2    &   $\na$   $<$ $\cha$  $<$ $\nb$   $<$ $A/H$  & Y  & Y  \cr
mSP3    &   $\na$   $<$ $\cha$  $<$ $\nb$ $<$ $\sta$    & Y  & Y  \cr
mSP4    &   $\na$   $<$ $\cha$ $<$ $\nb$   $<$ $\g$     & Y  & Y  \cr
\hline
mSP5    &   $\na$ $<$ $\sta$  $<$ $\slr$  $<$ $\snl$      & Y  & Y  \cr
mSP6 &   $\na$   $<$ $\sta$  $<$ $\cha$  $<$ $\nb$      & Y  & Y  \cr
mSP7    &   $\na$   $<$ $\sta$  $<$ $\slr$  $<$ $\cha$  & Y  & Y  \cr
mSP8    &   $\na$ $<$ $\sta$  $<$ $A\sim H$             & Y  & Y  \cr
mSP9    &   $\na$   $<$ $\sta$  $<$ $\slr$ $<$ $A/H$    & Y  & Y  \cr
mSP10   &   $\na$   $<$ $\sta$ $<$ $\ta$ $<$ $\slr$     & Y & \cr
 \hline
mSP11   &   $\na$ $<$ $\ta$ $<$ $\cha$  $<$ $\nb$       & Y  & Y  \cr
mSP12 &   $\na$ $<$ $\ta$   $<$ $\sta$ $<$ $\cha$   & Y  & Y  \cr
mSP13   & $\na$   $<$ $\ta$ $<$ $\sta$  $<$ $\slr$      & Y  & Y  \cr
\hline
mSP14   &   $\na$   $<$  $A\sim H$ $<$ $\hc$        & Y  & \cr
mSP15   &   $\na$   $<$ $ A\sim H$ $<$ $\cha$   & Y  & \cr
mSP16   &   $\na$   $<$ $A\sim H$ $<$$\sta$         & Y  & \cr
\hline
mSP17   &   $\na$   $<$ $\sta$ $<$ $\nb$ $<$ $\cha$     & & Y \cr
mSP18   &  $\na$   $<$ $\sta$  $<$ $\slr$  $<$ $\ta$    & & Y \cr
mSP19   &  $\na$ $<$ $\sta$ $<$ $\ta$   $<$ $\cha$  & & Y \cr
 \hline
mSP20  & $\na$ $<$ $\ta$   $<$ $\nb$   $<$ $\cha$   & & Y \cr
mSP21   & $\na$   $<$ $\ta$   $<$ $\sta$  $<$ $\nb$     & & Y \cr
\hline
mSP22   & $\na$   $<$ $\nb$   $<$ $\cha$  $<$ $\g$  & & Y \cr
\hline\hline
 \end{tabular}
\caption{ Hierarchical mass patterns for  the four lightest
sparticles  in mSUGRA when $\mu <0$ and $\mu>0$.
The patterns can be classified according to the next to the lightest
sparticle. For the mSUGRA analysis the next to the lightest sparticle
is found to be either a chargino, a stau, a stop, a CP
even/odd Higgs, or the next lightest neutralino $\nb$. The
notation $A/H $ stands for either $A$ or $H$.
In mSP14-mSP16 it is
possible that the Higgses become
lighter than the LSP. Y stands
for appearance of the pattern for the sub case.}
\label{msptable}
\end{center}
 \end{table}
Such sensitivities and their implications for the analysis of relic density calculations  are
well known in the literature \cite{sensitivity} and a detailed comparison for
various codes  can be found in  Refs.~(\cite{Baer:2005pv}, \cite{Belanger:2005jk}, 
\cite{Allanach:2004rh}, \cite{Allanach:2003jw}).


\subsection{The mSUGRA landscape  for the 4 lightest sparticles}
\label{A1}

As discussed in chapter (\ref{ch:sugra}), 
one mSUGRA model is a point in a 4 dimensional parameter space spanned by 
the soft parameters $m_0$, $m_{1/2}$, $A_0$,  $\tan\beta$, and the sign of $\mu$.
Typically scans of the parameter space
are done by taking a vanishing
trilinear coupling, and/or by looking at fixed values of
$\tan\beta$ while varying ($m_0$, $m_{1/2}$). In this work we carry
out a random scan in the 4-D  input parameter space for both signs  of $\mu$
with Monte Carlo
simulations using flat priors under the following ranges of the
input parameters
\beqn
0 < m_0 < 4 {~\rm TeV}, ~~~ 0 < m_{1/2} < 2 {~\rm TeV}~~~
|A_0/m_0| < 10,~~~ 1 < \tan\beta < 60. \label{softranges}
\eeqn
Since SUGRA models with $\mu>0$ are favored by the  experimental
constraints  much of the analysis presented here  focuses  on this case.
Specifically for the  $\mu>0$ mSUGRA case, we perform
a scan of the parameter space with a total of $2 \times 10^6$
trial parameter points. We delineate the patterns that emerge for the first four
lightest sparticles.
Here we find that at least  sixteen  hierarchical mass  patterns emerge
which are labeled as mSPs (minimal SUGRA Pattern).
These mSPs can be generally classified according to the type of particle which is next
heavier than the LSP, and we find four classes of patterns in mSUGRA:  
the chargino patterns (CP), the stau patterns (SUP),
the stop patterns (SOP),  and the Higgs patterns (HP), as exhibited below
\begin{enumerate}
\item Chargino patterns (CP) : mSP1, mSP2, mSP3, mSP4; 
\item Stau patterns (SUP) : mSP5, mSP6, mSP7, mSP8, mSP9, mSP10; 
\item Stop patterns (SOP) : mSP11, mSP12, mSP13; 
\item Higgs patterns  (HP) : mSP14, mSP15, mSP16.
\end{enumerate}
The hierarchical mass patterns  mSP1-mSP16 are defined in Table (\ref{msptable}).
We note  that the pattern mSP7 appears in the analyses of 
\cite{ArnowittTexas,Gounaris:2007gx,Buchmueller:2007zk}.
\begin{figure}[t]
\begin{center}
\includegraphics[width=9.5 cm,height=8cm]{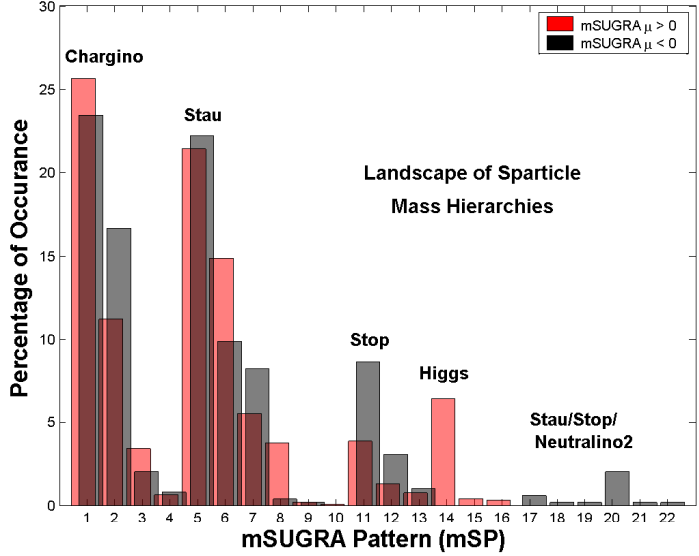}
\caption{Distribution of the surviving hierarchical mass patterns in the landscape
for the mSUGRA model with $\mu >0$ (light) and $\mu <0$ (dark),
under various constraints as discussed in the text.
 } \label{m-land}
\end{center}
\end{figure}
\begin{table}[h]
\begin{center}
\begin{tabular}{|c|c |  }
\hline \hline Snowmass  &  mSP  \\\hline
\hline SPS1a, SPS1b, SPS5 & mSP7 \\
\hline SPS2 & mSP1 \\
\hline SPS3 & mSP5  \\
\hline SPS4, SPS6& mSP3  \\
\hline \hline\end{tabular}
\begin{tabular}{|c|c |  }\hline \hline Post-WMAP3 &  mSP\\\hline
\hline $A',B',C',D',G',H',J', M' $ & mSP5\\
\hline $I', L'$ &  mSP7\\
\hline $E'$ & mSP1 \\
\hline $K'$  & mSP6 \\
\hline \hline\end{tabular}
\begin{tabular}{|c|c |  }\hline \hline CMS LM/HM &  mSP\\\hline
\hline LM1, LM6, HM1 & mSP5\\
\hline LM2, LM5, HM2 &  mSP7\\
\hline LM3, LM7, LM8, LM9, LM10, HM4 & mSP1 \\
\hline LM4, HM3  & mSP3 \\
\hline \hline\end{tabular}
\caption{Mapping between the mSPs and
the Snowmass, Post-WMAP3, and CMS benchmark points. The points $B'
=$~LM1, $I' =$~LM2, $C' =$~LM6. HM1 in SuSpect  has $m_{\na} >
m_{\sta}$, but this is not the case for ISAJET, SPheno, and
SOFTSUSY. 
Among the CMS benchmarks, only LM1, LM2, LM6, and HM1, HM2 are capable of giving the
correct relic density. Thus the mapping above applies only to the
mass pattern, while all of our mSP and NUSP  benchmark points
satisfy the relic density constraints from MicrOMEGAs with SuSpect.
The CMS test points do a better job of representing mSP1 which is
the dominant pattern found in our analysis. There are no HP test
points or SOP test points in any of the previous works.
}
 \label{pattern}\end{center}
\end{table}
We also performed a similar scan for the  mSUGRA  with $\mu<0$ case using
the Monte Carlo simulation with flat priors and the same parameter ranges
as specified in Eq.~(\ref{softranges}). Most of the mSP patterns that
appear in the $\mu>0$ case also appear in the $\mu<0$ case (see
Table (\ref{msptable})). However, in addition one finds new
patterns shown below
\begin{enumerate}
\item Stau patterns  (SUP) : mSP17, mSP18, mSP19; 
\item Stop patterns  (SOP) : mSP20, mSP21; 
\item Neutralino patterns  (NP) : mSP22.
\end{enumerate}
We note that  the analysis of Ref.~\cite{SPM} has a sparticle spectrum which
corresponds to mSP11 and contains light stops.
Light stops  have also been discussed  recently in \cite{Gladyshev:2007ec,uc}.

While the earlier works which advocated benchmark points
and slopes made good progress in systematizing the search for supersymmetry,
we find that they do not cover the more broad set of possible
mass hierarchies we discuss here. 
That is, many of the mSP patterns do not appear in the earlier works that
advocated benchmark points for SUSY searches. For example, the
Snowmass mSUGRA points (labeled SPS) \cite{Allanach:2002nj} and the
Post-WMAP benchmark points of \cite{Battaglia:2003ab}, make up only
a small fraction of the possible mass hierarchies listed in Table (\ref{msptable}).
The CMS benchmarks classified as  Low Mass (LM) and
High Mass (HM) \cite{Ball:2007zza} (for a recent review see \cite{Hubisz:2008gg, Spiropulu:2008ug})
does a good job covering the mSP1 pattern which appears as the most dominant
pattern in our analysis, but there are no Higgs patterns or stop patterns discussed
in the CMS benchmarks as well as in SPS or in Post-WMAP benchmarks.
We exhibit the mapping of mSPs with other benchmarks points in a tabular form in Table (\ref{pattern}).

In Fig.~(\ref{m-land}) we give the relative distribution of these
hierarchies found in our Monte Carlo scan. 
Because the scan is done randomly within the soft parameter space,
the distribution of sparticle patterns in Fig.~(\ref{m-land}) represents the probability 
of finding these patterns in the parameter space. 
The most common patterns found here are CPs and SUPs,  especially mSP1 and mSP5.
However there exists a significant region of the parameter space where
SOPs and HPs can be realized. The  percentages of occurrence
of the various patterns in the mSUGRA landscape  for  both $\mu$ positive and
$\mu$ negative are exhibited in Fig.~(\ref{m-land}).
The analysis of Fig.~(\ref{m-land}) shows that
the  chargino patterns (CP) are the most  dominant patterns, followed by
the stau patterns (SUP),  the stop patterns (SOP), and the Higgs patterns (HP).
In contrast, most  emphasis in the literature,  specifically in the context of relic
density analysis,  has focused on the stau patterns, with much less attention
on other patterns. Specifically the Higgs patterns have hardly been investigated
or discussed.  The exceptions to this, in the context of the Higgs patterns, are the  more recent works of
Refs. \cite{Feldman:2007zn,Feldman:2007fq}, and  similar mass ranges for the Higgs bosons have
been studied in  \cite{WP}  (see also \cite{Ellis:2007ss}).

\subsection{The landscape of the 4 lightest  sparticles  in NUSUGRA  \label{A2} }

Next we discuss the landscape of the 4 lightest sparticles for the case of nonuniversal
supergravity models. Here we consider \non in the Higgs sector
(NUH),  in the third  generation sector (NU3), and  in the
gaugino sector (NUG). Such \non appear quite naturally in supergravity models
with a non-minimal K\"{a}hler potential, and  in string and D-Brane models.
The parametrization of the \non is given by
\begin{equation}
\begin{array}{lcl}
{\rm NUH}&:&M_{H_u}=  m_0(1+\delta_{H_u}),~~ M_{H_d} = m_0(1+\delta_{H_d}),\cr
{\rm NU3}&:&M_{q3} =m_0(1+\delta_{q3}), ~~M_{u3,d3}=m_0(1+\delta_{tbR}),\cr
{\rm NUG}&:&M_{1}=m_{1/2}, ~~~M_{2,3}=m_{1/2}(1+\delta_{M_{2,3}}).
\end{array}\label{nonuni}
\end{equation}
In the above $\delta_{H_u}$ and $\delta_{H_d}$ define the \non for
the up and down Higgs  mass parameters, $M_{q3}$ is the left-handed squark
mass for the 3rd generation, and $M_{u3}$ ($M_{d3}$) are  the
right-handed u-squark (d-squark) masses for the 3rd generation.
The \non  in the gaugino
sector are parameterized here by $\delta_{M_2}$ and $\delta_{M_3}$.
We have carried out a Monte Carlo scan with flat priors using $10^6$ model
points in each of the three types of NUSUGRA models,  taking the same input parameter
ranges
as specified in Eq.~(\ref{softranges})  and $-0.9\leqslant\delta\leqslant1$.
Almost all of the mSP patterns seen for the mSUGRA cases were found in
supergravity models with nonuniversal soft breaking,  as the mSUGRA  model is contained
within the nonuniversal supergravity models.  In addition
we find many new patterns labeled NUSPs (nonuniversal SUGRA pattern), and they are
exhibited in  Table (\ref{nusptable}).
 As in the mSUGRA case one finds several pattern classes,
CPs, SUPs, SOPs, and HPs as exhibited below. In addition,
 we  find several  Gluino patterns (GP) where  the gluino is the NLSP.
\begin{enumerate}
\item Chargino patterns (CP) : NUSP1, NUSP2, NUSP3, NUSP4; 
\item Stau patterns (SUP) : NUSP5, NUSP6, NUSP7, NUSP8, NUSP9; 
\item Stop patterns (SOP) : NUSP10, NUSP11; 
\item Higgs patterns (HP) : NUSP12; 
\item Gluino patterns (GP) : NUSP13, NUSP14, NUSP15.
\end{enumerate}
It is  interesting to note that for the  4 sparticle landscape we find saturation in the number of
mass hierarchies that are present.   For example, for the case $\mu>0$ in
mSUGRA , increasing the soft parameter scan from $1\times 10^6$ parameter
model points to $2\times 10^6$ model points does not increase the
number of 4 sparticle patterns. In this context it becomes relevant to
examine as to what degree the relic density and other experimental
constraints play a role in constraining the parameter space and  thus
\begin{table}[htbp]
    \begin{center}
\begin{tabular}{||l||l||c|c||}
\hline\hline NUSP &   Mass Pattern  &     NU3   &   NUG\\\hline\hline
NUSP1   &   $\na$   $<$ $\cha$  $<$ $\nb$   $<$ $\ta$       & Y &   Y \cr
NUSP2   &   $\na$   $<$ $\cha$  $<$ $A\sim H$               &Y     &    \cr
NUSP3   &   $\na$   $<$ $\cha$  $<$ $\sta$  $<$ $\nb$       &    & Y \cr
NUSP4   &   $\na$   $<$ $\cha$  $<$ $\sta$  $<$ $\slr$          & &   Y \cr
\hline
NUSP5   &   $\na$   $<$ $\sta$  $<$ $\snl$  $<$ $\stb$          & Y  &    \cr
NUSP6  &   $\na$   $<$ $\sta$ $<$ $\snl$  $<$ $\cha$        & Y  &\cr
NUSP7   &   $\na$ $<$ $\sta$  $<$ $\ta$   $<$ $A/H$         & & Y \cr
NUSP8   &   $\na$   $<$ $\sta$  $<$ $\slr$  $<$ $\snm$      & &   Y \cr
NUSP9   &   $\na$   $<$ $\sta$  $<$ $\cha$  $<$ $\slr$          & &   Y \cr
\hline
NUSP10  &   $\na$   $<$ $\ta$   $<$ $\g$    $<$ $\cha$          & &   Y \cr
NUSP11  &   $\na$   $<$ $\ta$   $<$ $A\sim H$                       & & Y \cr
\hline
NUSP12  &   $\na$   $<$ $A\sim H$   $<$ $\g$                    && Y \cr
\hline
NUSP13  &   $\na$   $<$ $\g$    $<$ $\cha$ $<$ $\nb$        & &   Y \cr
NUSP14  &   $\na$   $<$ $\g$    $<$ $\ta$ $<$ $\cha$            & &   Y \cr
NUSP15  &   $\na$   $<$ $\g$    $<$ $A\sim H$               &&   Y \cr
  \hline\hline
\end{tabular}
\caption{New  4 sparticle  mass patterns  that arise in NUSUGRA
over  and above the mSP patterns of  Table (\ref{msptable}). 
 These  are labeled nonuniversal SUGRA patterns (NUSP) and  at least
 15 new patterns are seen to emerge which are denoted by NUSP1-NUSP15.}
 \label{nusptable}
\end{center}
 \end{table}
reducing the number of patterns. This is exhibited in
Table (\ref{tab:spnum}) where we demonstrate how the relic density
and the other experimental constraints  decrease the number of
admissible model points in the allowed parameter space  for the mSUGRA
models with both $\mu>0$ and $\mu<0$, and also for the cases with \non
 in the Higgs sector, \non in the third generation sector,
and with \non in the gaugino sector. In each case we start with
$10^6$ model points at the GUT scale, and find that the electroweak symmetry
breaking constraints reduce the number of viable models to about
$1/4$ of what we started with. We find that the allowed number of models
translates into SUGRA mass patterns which are typically less than 100.  The
admissible set of parameter points reduces drastically when the
relic density constraints are imposed and are then found to typically reduce
the number of models
by a factor of about 200 or more, with a reduction in the number of
allowed patterns by a factor of 2 or more. Inclusion of all other
experimental constraints further reduces the  number of admissible points by
a factor between 30\% and 50\%,
with a corresponding
reduction in the number of patterns by up to 40\%.
The above
analysis shows that there is an enormous reduction in the number of
admissible models and the corresponding number of  hierarchical mass patterns  after the
constraints of radiative breaking of the electroweak symmetry, relic
density constraints, and other experimental constraints are imposed.
\begin{table}[htbp]
    \begin{center}
 \scriptsize{
\begin{tabular}{|c|c|c|c|c|c|c|c|}
\hline
Model  &   Trial      &   Output  &   No. of    &   Relic Density          &  No. of  &   All          & No. of  \\
Type    &   Models  &   Models   &  Patterns  &   Constraints    & Patterns & Constraints & Patterns\\
\hline mSUGRA($\mu>0$) &   10$^6$   & 265,875 &   55  &   1,360   &   22  & 902 & 16  \\
\hline mSUGRA($\mu<0$) & 10$^6$   &   226,991 &   63  & 1,000   & 31  &   487 & 18  \\
\hline NUH($\mu>0$)   &   10$^6$   & 222,023 &   59 &   1,024   & 24  &   724 &   15  \\
\hline NU3($\mu>0$) & 10$^6$  &   229,928 &   73  &   970 &   28  &   650 &   20  \\
\hline NUG($\mu>0$) &   10$^6$   &   273,846 &   103 &   1,788   & 36 & 1,294   &   28  \\  \hline
\end{tabular}
}
\caption{An analysis of  mass patterns  for the four lightest sparticles. 
Exhibited  here are  the model type, the number of trial input points
for each model, the number surviving the radiative  electroweak
symmetry breaking scheme as given by SuSpect (column 3), the
number surviving when the relic density  constraints  are applied with MicrOMEGAs
(column 5),  the number surviving with inclusion of all experimental  collider
constraints (column 7), along with the corresponding number of  hierarchical mass
patterns in each case  (column 8).
} \label{tab:spnum}
    \end{center}
 \end{table}

\subsection{Hierarchical patterns for the full sparticle spectrum \label{A4}}

\begin{figure*}[htbp]
\centering
\hspace*{-.1in}\includegraphics[width=6.5cm,height=5cm]{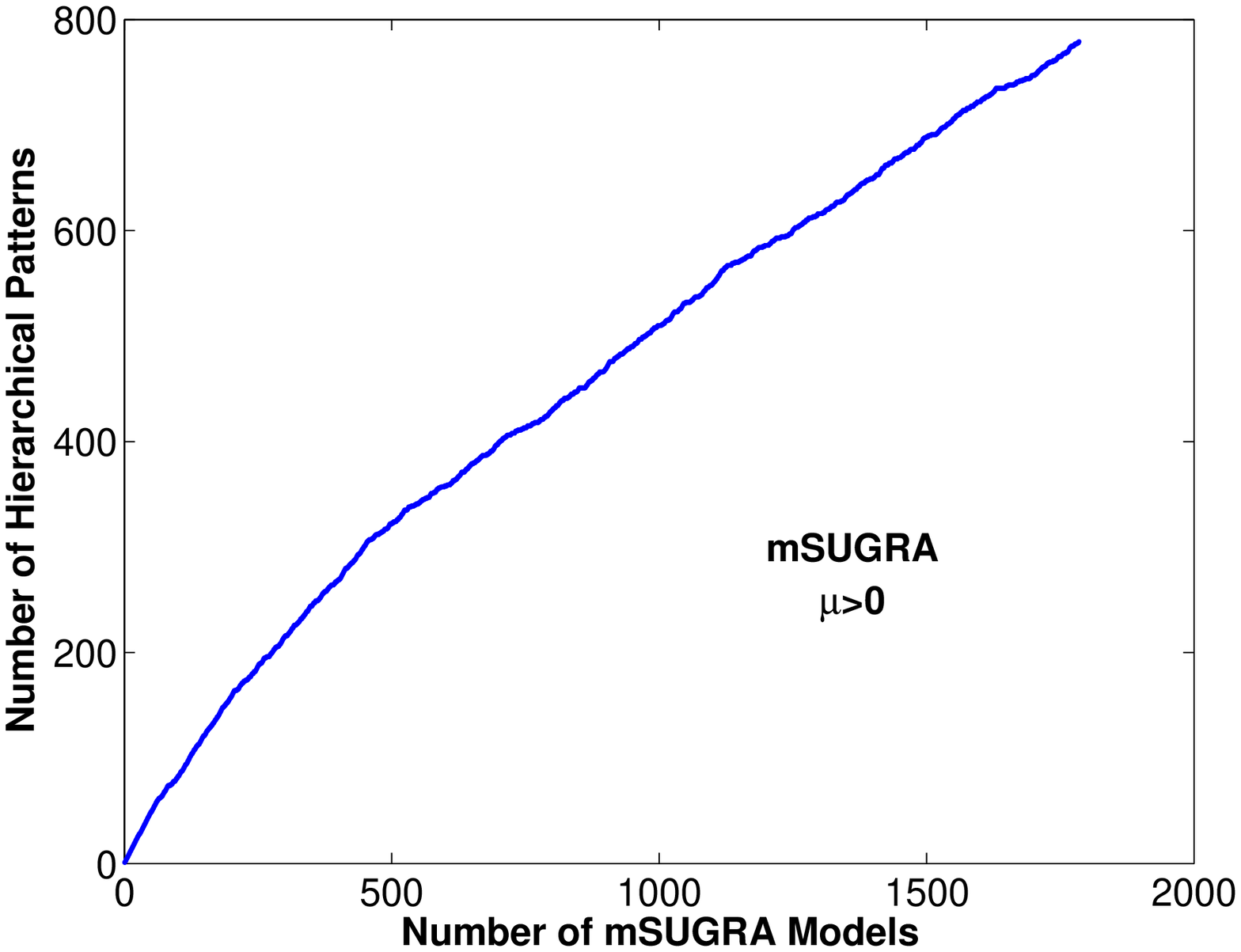}
\hspace*{.2in}\includegraphics[width=6.5cm,height=5cm]{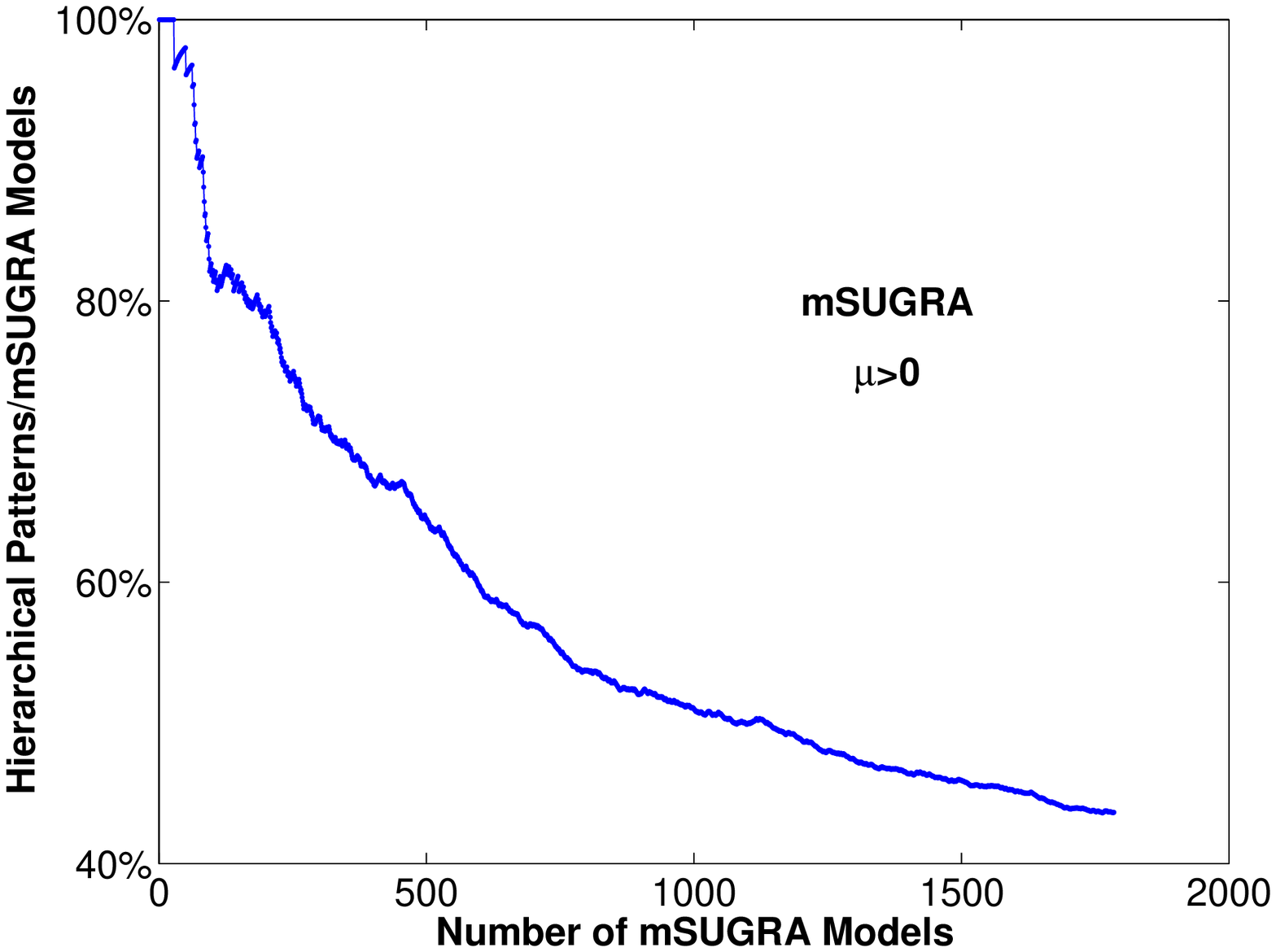}
\caption{Saturation of Sparticle patterns.}
 \label{fig:saturation}
\end{figure*}

We discuss now the number of hierarchical mass patterns  for the full  set of
 32 sparticles
in  SUGRA models when the constraints of electroweak symmetry, relic
density, and other experimental constraints are imposed.  The result of
the analysis is given in Fig.~(\ref{fig:saturation}) and Table (\ref {tab:wholespectrum}).
\begin{table}[htbp]
    \begin{center}
    \scriptsize{
\begin{tabular}
{|r|c|c|}
\hline
Models [No.]    & No. after constraints  &  No. of patterns\\
\hline
mSUGRA($\mu>0$) [$10^6$]    &   902 &   505    \\\hline
mSUGRA($\mu<0$) [$10^6$]   &   487 &   268    \\\hline
NUH($\mu>0$) [$10^6$]  &   724 &   517   \\ \hline
NU3($\mu>0$) [$10^6$]    &   650 &   528     \\\hline
NUG($\mu>0$) [$10^6$]    &   1294    &   1092      \\\hline
All Above[$5 \times 10^6$]    &   4057    &   2557      \\
\hline
\end{tabular}
} \caption{Exhibition of mass patterns and models with various constraints. 
Column 1 shows  one million input parameter points for each
of the models investigated,  and the
number surviving all the constraints are exhibited in column 2,
while column 3 gives the number of  hierarchical patterns.
}
\label{tab:wholespectrum}
    \end{center}
 \end{table}
The left panel of Fig.~(\ref{fig:saturation}) shows the number of  hierarchical mass
patterns for 32 sparticles vs the number of trial  points for mSUGRA
models which survive the electroweak symmetry breaking constraints,
the relic density and all other experimental constraints. The number
of  hierarchical mass patterns show a trend towards saturation.
In the right panel, a similar phenomenon is seen in the ratio between the
number of patterns over the number of surviving trial points  in
mSUGRA models.
Here one finds that increasing the number of model points in the scan
does increase the number of patterns. However, the ratio of the
number of  patterns to the total number of models that survive all
the constraints  from  the scan decreases sharply as shown in the right
panel of Fig.~(\ref{fig:saturation}). This means  that although
saturation is not yet  achieved one is moving fast  towards
achieving saturation with a relatively small number of allowed
patterns for all the 32  sparticles within SUGRA models consistent
with the various experimental constraints.
The analysis of Table (\ref{tab:wholespectrum}) shows
that the number of allowed patterns for the 32 sparticles,
which in the MSSM without the SUGRA framework  can be as large as $O(10^{25})$
or larger, reduces rather drastically when
various constraints are applied in supergravity models.

The Table (\ref{tab:wholespectrum}) exhibits a dramatic reduction of the
landscape from upward of $\sim O(10^{25})$
hierarchical mass patterns for
the 32 sparticle masses  to a much smaller number when the
electroweak symmetry breaking constraints, the relic density
constraints, and other experimental constraints are applied.
We note that some patterns are repeated as we move across different model types listed in the
first column  of Table (\ref{tab:wholespectrum}). Thus the total number of patterns listed at the
bottom of the last column of this table is smaller than the sum of patterns listed above in that
column.  We note that the precise number and nature of the patterns are dependent on the
input parameters such as the top mass and a significant shift in the input values could modify the
pattern structure.

\section{Sparticle Patterns \& Nature of Soft Breaking\label{B}}

\begin{figure*}[htb]
\centering
\hspace*{-.1in}\includegraphics[width=7.0cm,height=6.0cm]{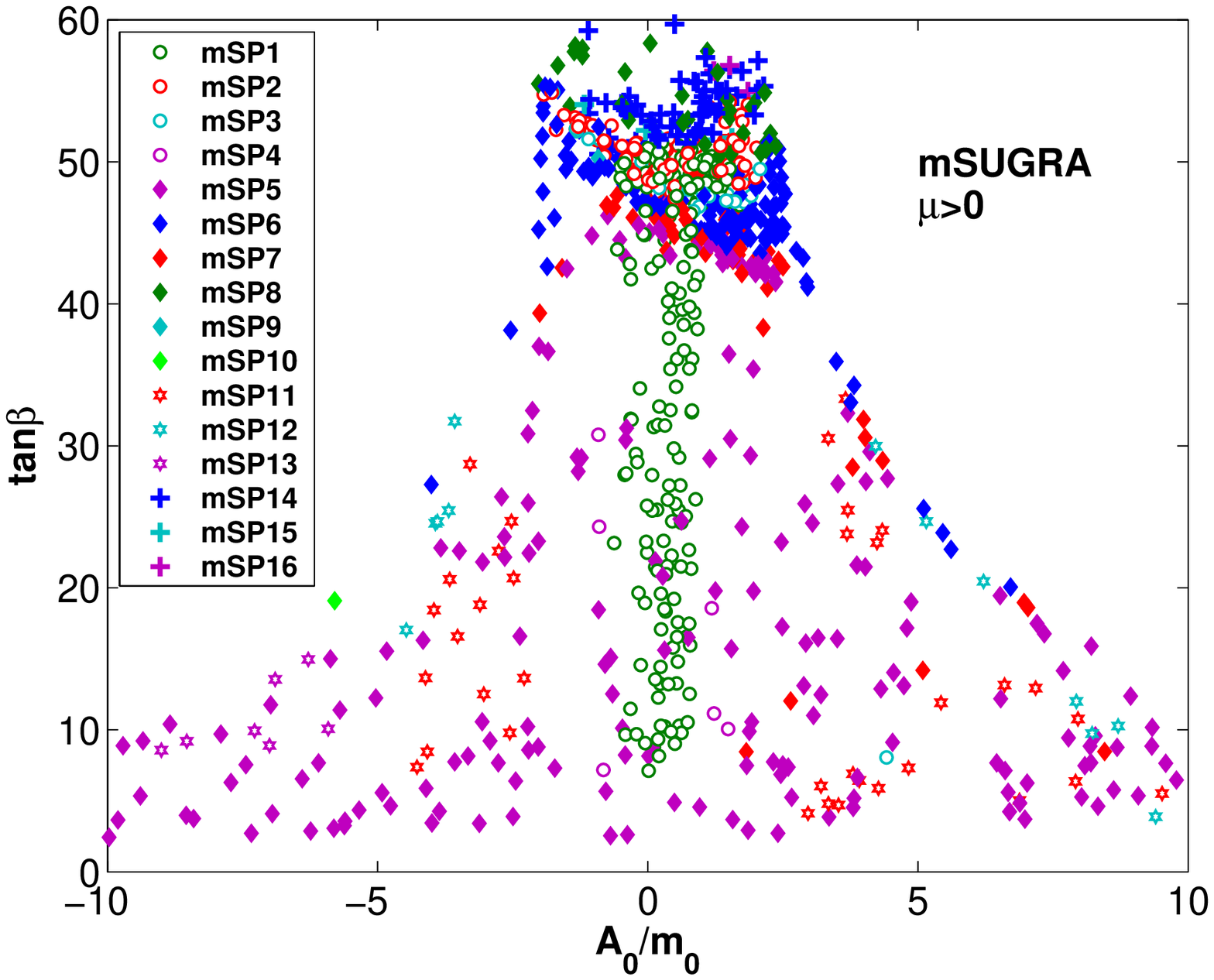}
\hspace*{.2in}\includegraphics[width=7.0cm,height=6.0cm]{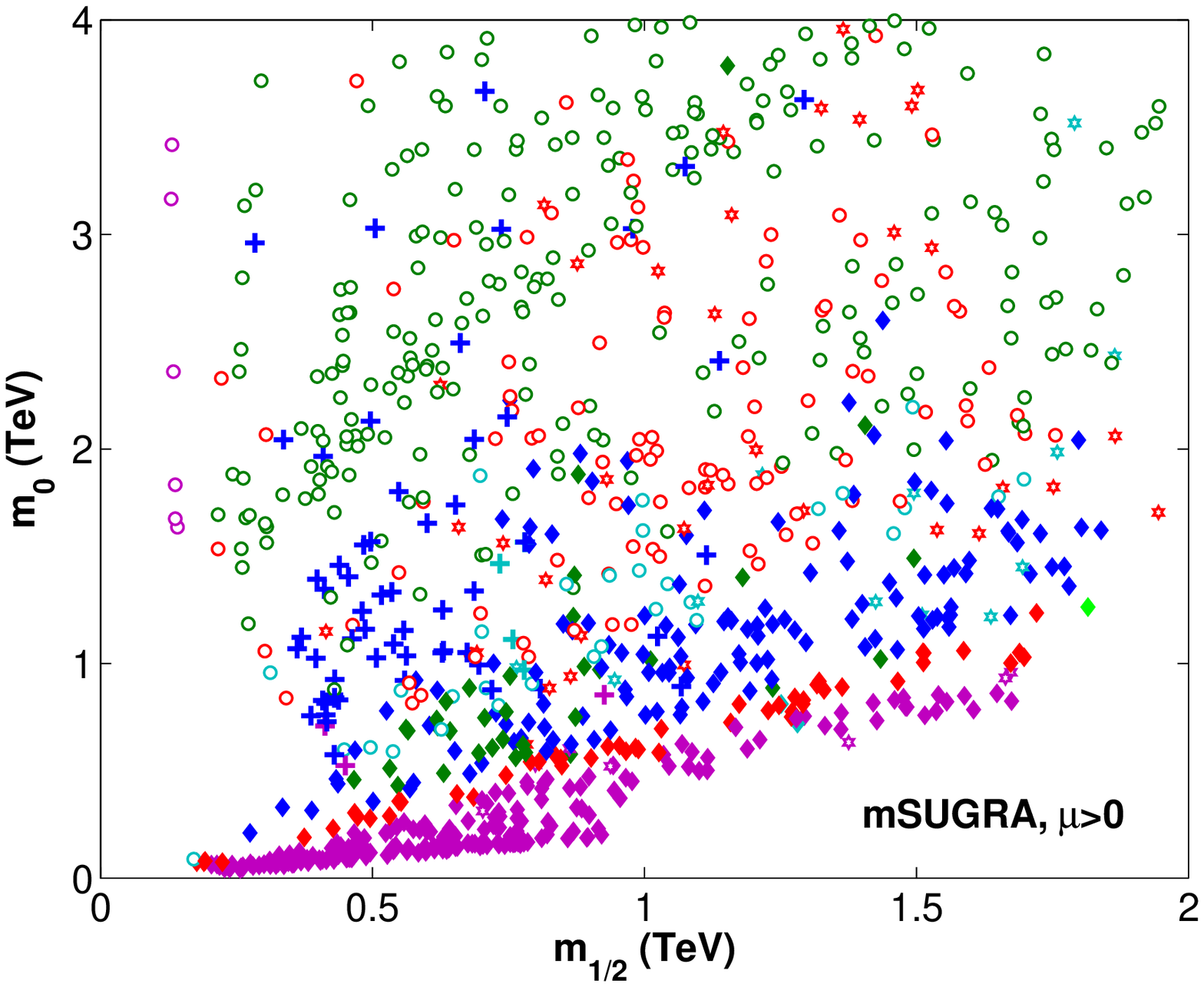}
\hspace*{-.1in}\includegraphics[width=7.0cm,height=6.0cm]{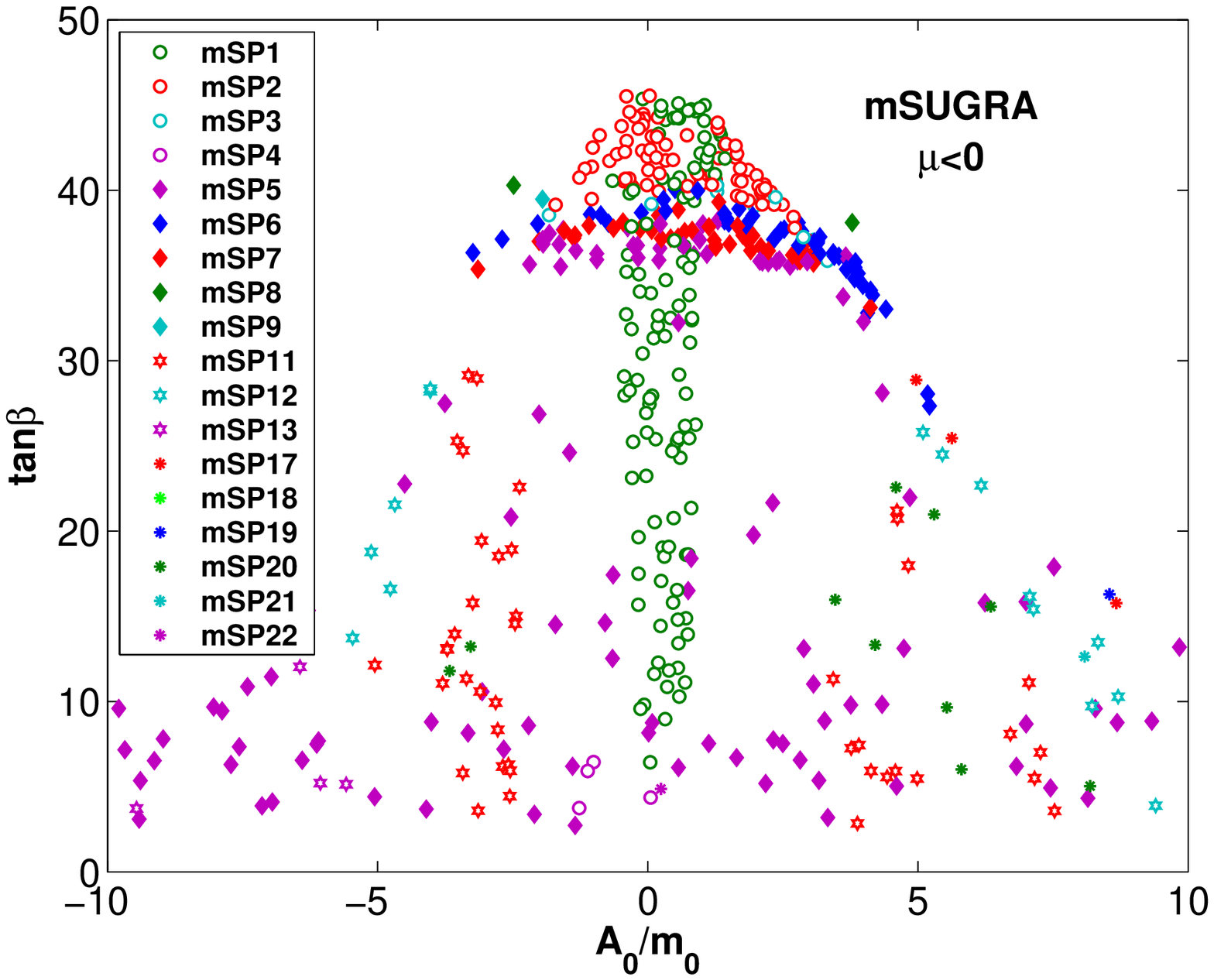}
\hspace*{.2in}\includegraphics[width=7.0cm,height=6.0cm]{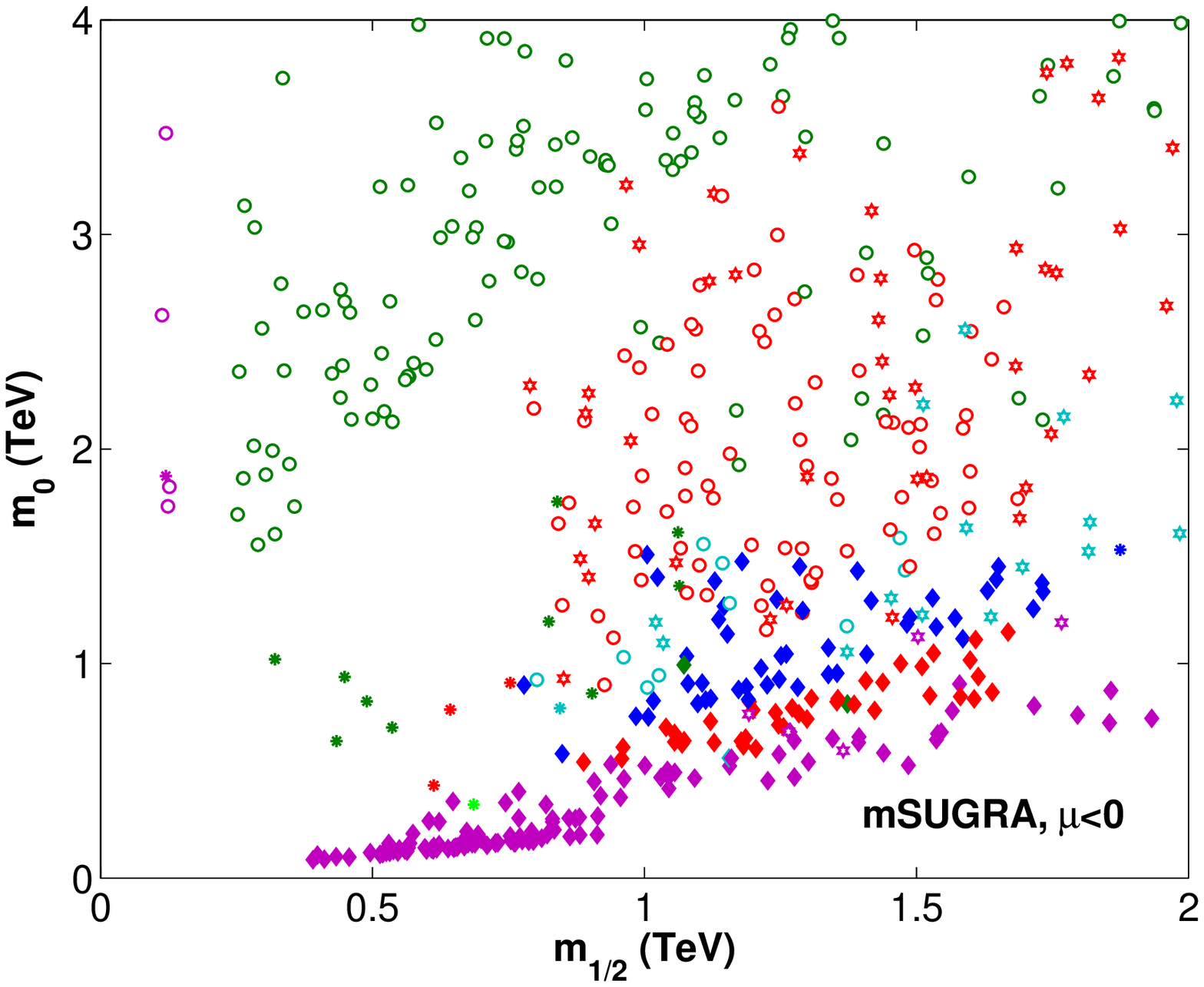}
\caption{Spectrum decomposition of soft parameter space. 
Exhibited are the landscape of sparticle mass spectra in the planes  (I)
$\tan\beta$ vs $A_0/m_0$ and (II) $m_0$ vs $m_{1/2}$, when the soft
parameters  are allowed to vary in the  ranges given in
Eq.~(\ref{softranges}). 
} \label{fig:spec}
\end{figure*}

It is interesting to ask if the patterns can be traced back to some specific regions
of the parameter of soft breaking from where they originate. This indeed is the case,
at least, for some of the patterns. The analysis illustrating the origin of the patterns
in the parameter space is given in  Fig.~(\ref{fig:spec}).
The dispersion of mSPs arising in mSUGRA
in the $\tan\beta$ vs $A_0/m_0$ plane (left panels), and in the
$m_0$ vs $m_{{1}/{2}}$ plane (right panels) for the $\mu>0$ case
(upper panels) and $\mu<0$ case (lower panels). The analysis is
based on a  scan of $10^6$ trial model points with flat priors in
the ranges $m_0<4{\rm~TeV}$, $m_{1/2}<2 {\rm~TeV}$, $1<\tan\beta<60$, and
$|A_0/m_0|<10$. mSP1 is confined to the region where $|A_0/m_0|<2$.
For the case $\mu<0$, no HPs are seen, and also, no model points
survive in the region where $\tan\beta > 50$ in contrast to the
$\mu>0$ case where there is a significant number for $\tan\beta
\gtrsim 45$. 
Many interesting observations can be made
from these spectral decompositions. For example, a significant set
of the mSP1 (CP) models lie in the region $|A_0/m_0|<2$ and correspond to
the Hyperbolic branch/Focus Point (HB/FP) \cite{hb/fp} regions, while most of the
SOPs have a rather large ratio of $A_0/m_0$ with the satisfaction of REWSB.
In this analysis we require
that there be no charge or color breaking (CCB) \cite{Frere:1983ag,Casas:1995pd}
at the electroweak scale.
We note in passing that it has been argued that even if the true minimum is not
color or charge preserving, the early universe is likely to occupy the CCB preserving
minimum and such minima may still be acceptable if the tunneling lifetime from the false
to the true vacuum is much greater than the present age of the universe \cite{Kusenko:1996jn}.
Next, we note that for the mSUGRA $\mu>0$ case,
the region around
$\tan\beta= 50$  has a large number of models that can be realized, while
the region around $\tan\beta= 30$  has far less model points. We also note
that most of the HPs reside only in the very high $\tan\beta$
region in mSUGRA, but this situation can be changed significantly
in
the NUH case where HP points can be realized in the $\tan\beta$
region as low as $\tan\beta \sim 20$.
In the $m_0$ vs $m_{1/2}$ plane,
one finds  that most of CPs and HPs
have a larger universal scalar mass  than most of the
SUPs and  SOPs.

\begin{figure*}[htb]
\centering
\hspace*{-.1in}\includegraphics[width=7.0cm,height=6.0cm]{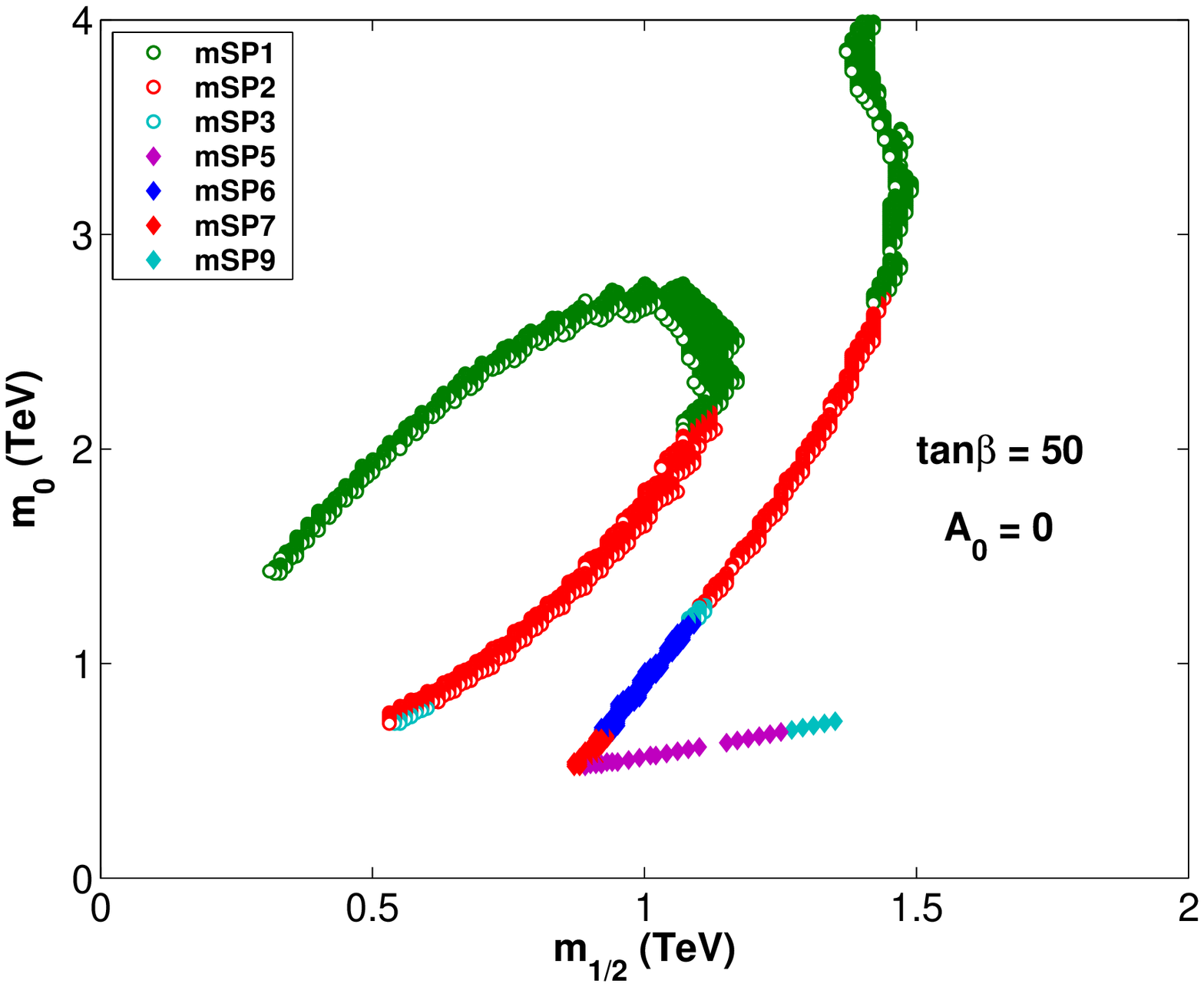}
\hspace*{.2in}\includegraphics[width=7.0cm,height=6.0cm]{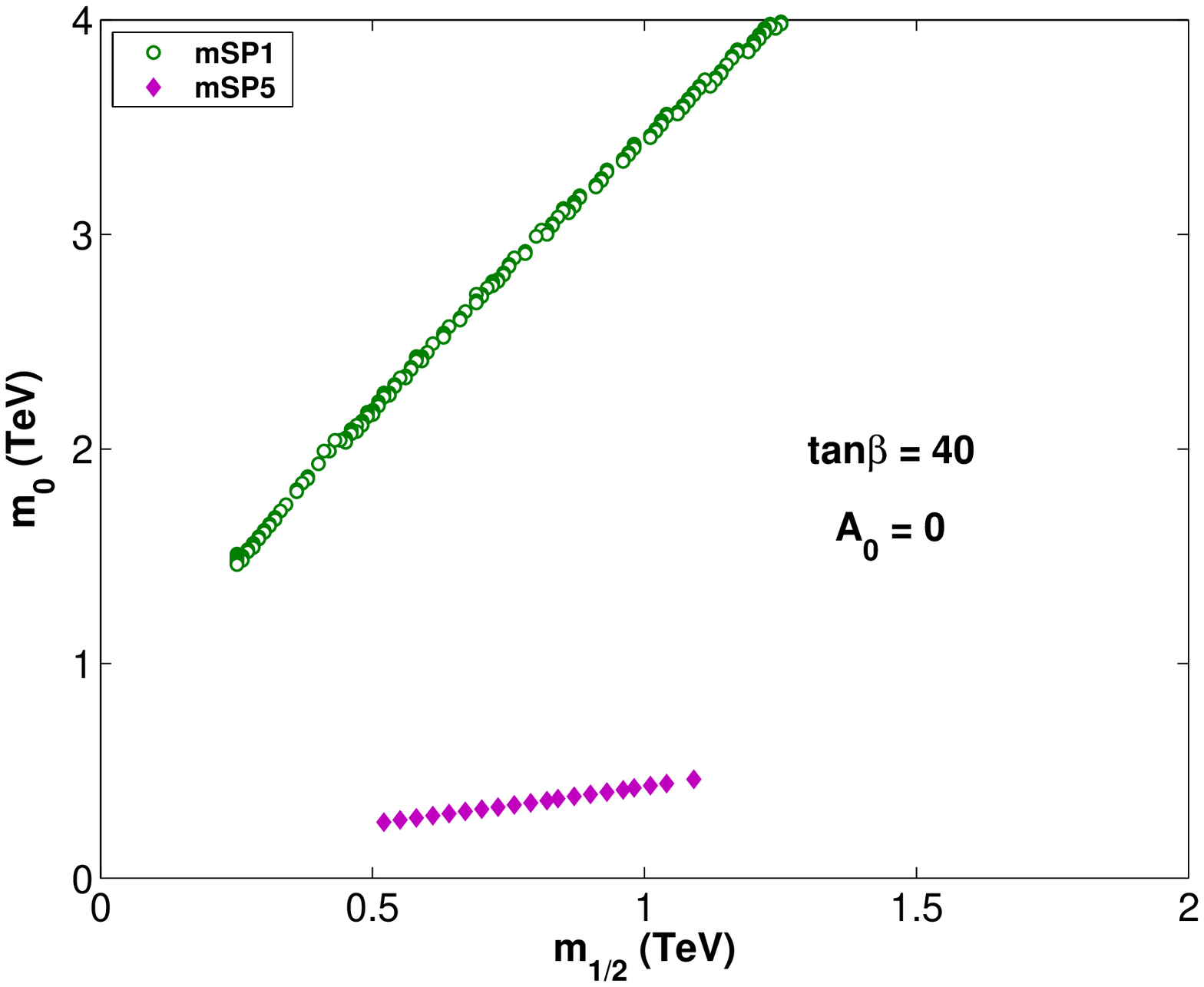}
\hspace*{-.1in}\includegraphics[width=7.0cm,height=6.0cm]{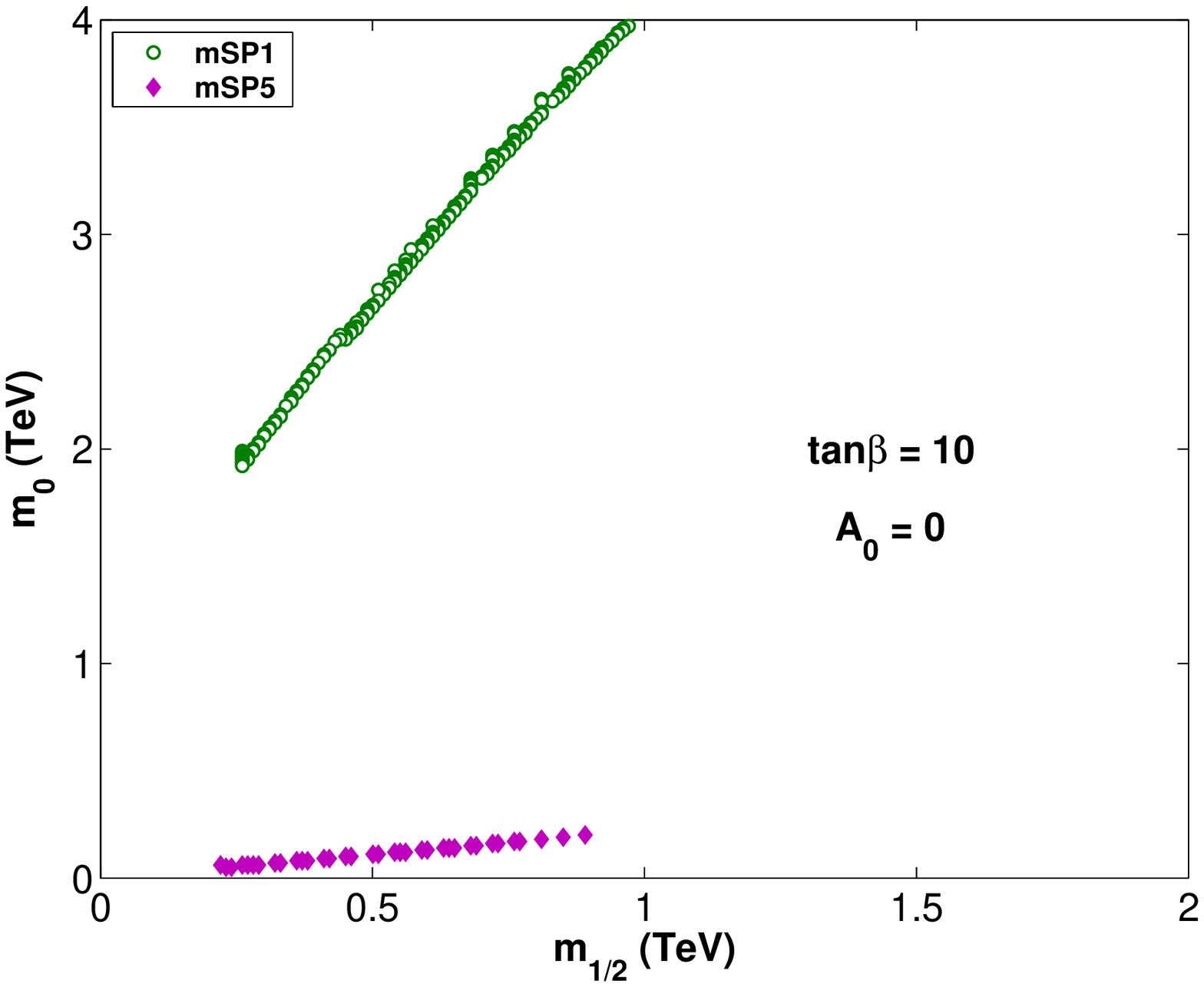}
\hspace*{.2in}\includegraphics[width=7.0cm,height=6.0cm]{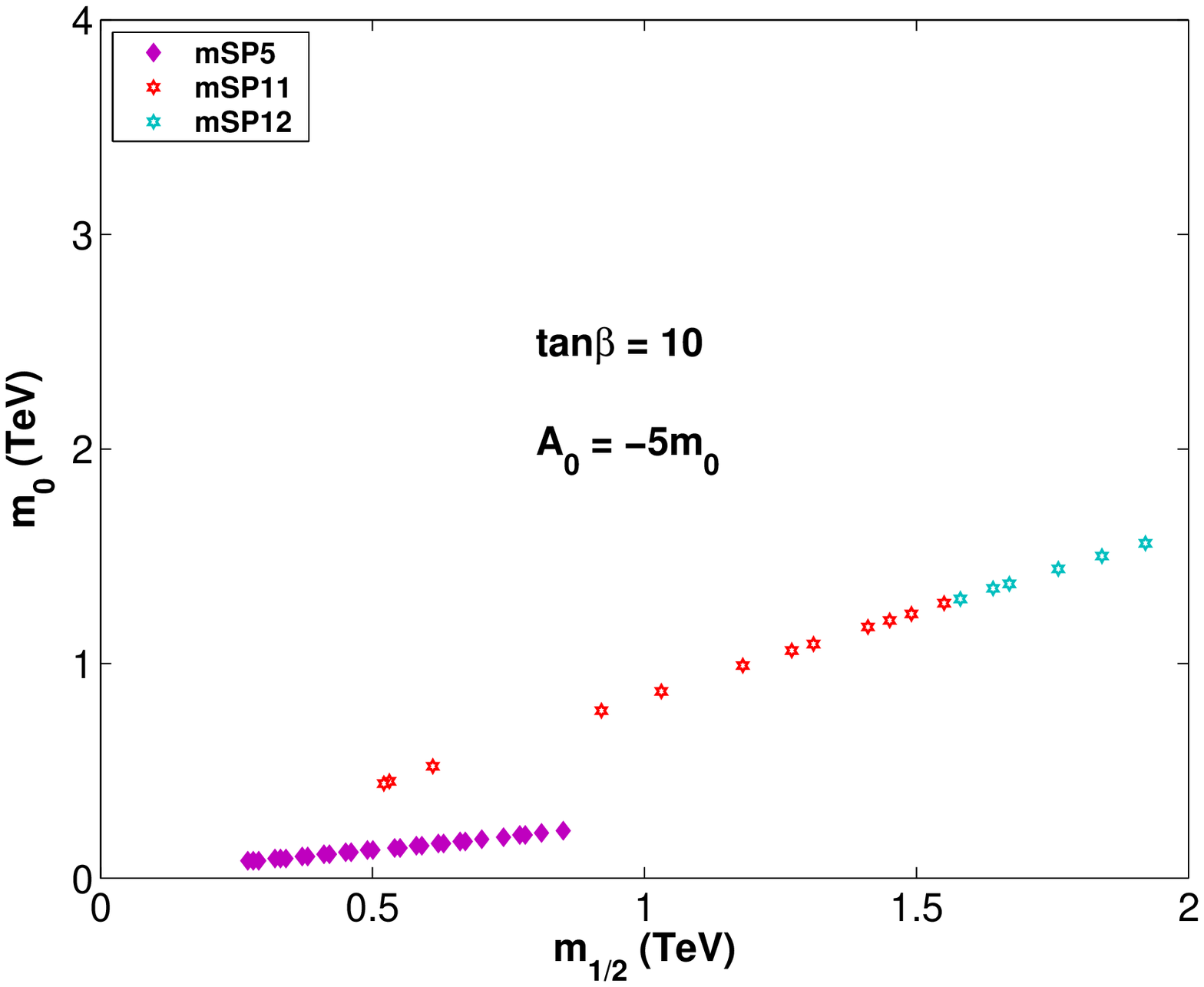}
\caption{Dispersion of patterns in some special 2D surface of mSUGRA space.}
\label{fig:grid}
\end{figure*}

Often in the literature one limits the analysis by fixing specific values of
$A_0$ and $\tan\beta$.  For $A_0$ the value most investigated is
$A_0=0$. However, constraining the values of $A_0$ or of $\tan\beta$
 artificially eliminates a very significant part of the allowed
 parameter space where all the relevant constraints (the REWSB constraint as well as
 the relic density and the experimental constraints) can be satisfied
 as seen in Fig.~(\ref{fig:spec}).
One can extract the familiar plots one finds in the literature
where $A_0$ and $\tan\beta$ are constrained from a reduction of the
top-right  panel of  Fig.~(\ref{fig:spec}). The results of this reduction
are  shown in  Fig.~(\ref{fig:grid}) with a focused scan in specific regions
of the soft parameter space. Fig.~(\ref{fig:grid}) shows a dispersion of patterns 
in the $m_0$ vs $m_{1/2}$ plane for fixed values of  $\tan\beta$
and $A_0/m_0$. The region scanned is  in the range $m_0 < 4$~{\rm TeV} and
$m_{1/2} < 2$~{\rm TeV}  with a 10~{\rm GeV} increment for each mass.
Only a subset of the allowed parameter points relative to Fig.~(\ref{fig:spec}) remain,
since the scans are on constrained surfaces in the mSUGRA parameter space.
Specifically the bottom-left and  top-right panels of Fig.~(\ref{fig:grid})
show the familiar stau coannihilation \cite{Ellis:1999mm,Gomez:2000sj,ArnowittTexas}
regions and the HB/FP branch,
 the bottom-right panel gives the stau coannihilation region and
the stop coannihilation region because of the relatively large $A_0$ value,
and the top-left panel is of the form seen in the works of Djouadi
et al.~\cite{Djouadi:2006be} where the Higgs funnel plays an important
role in the satisfaction of the relic density.

\begin{figure*}[htb]
\centering
\hspace*{-.1in}\includegraphics[width=7.0cm,height=6.0cm]{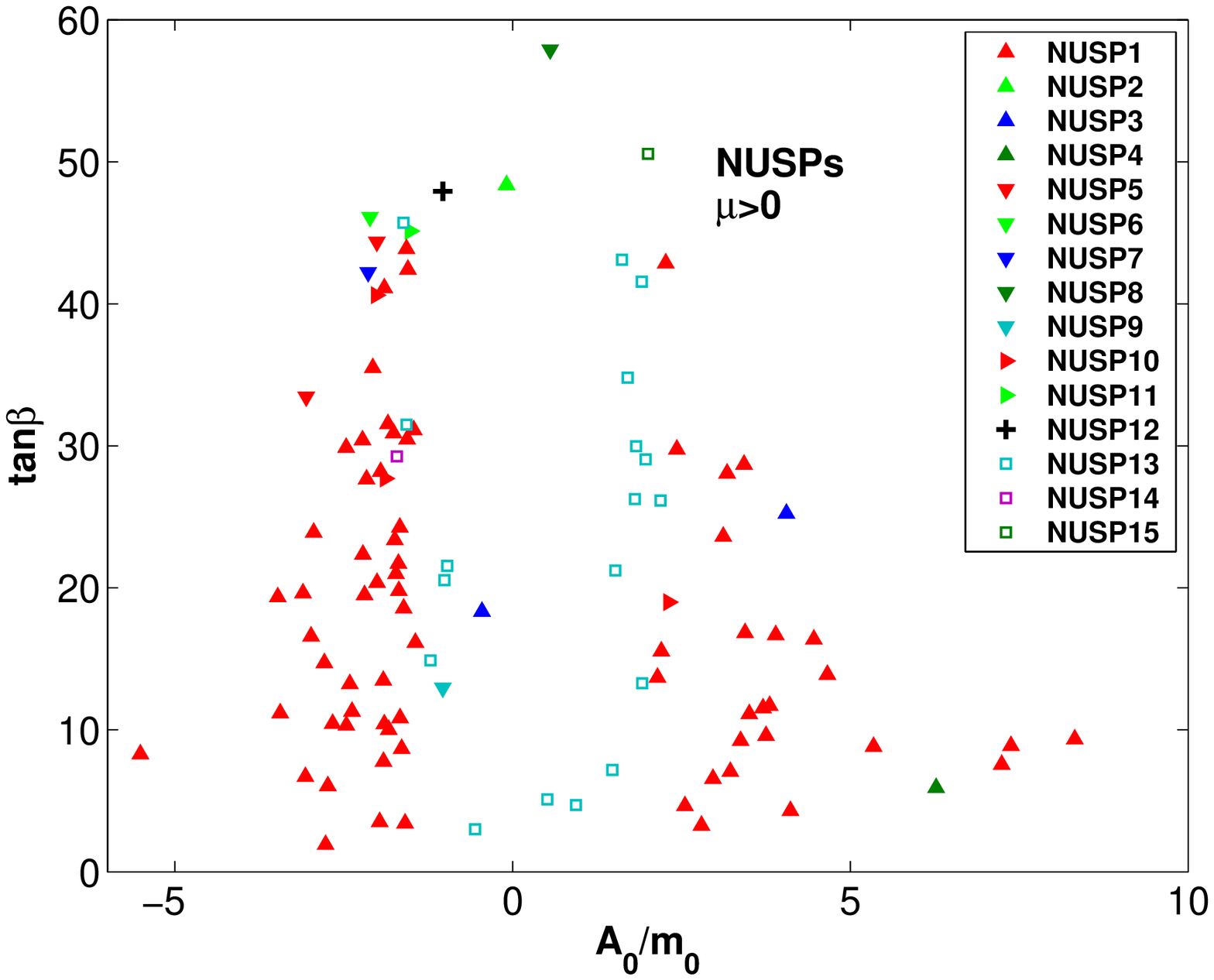}
\hspace*{.2in}\includegraphics[width=7.0cm,height=6.0cm]{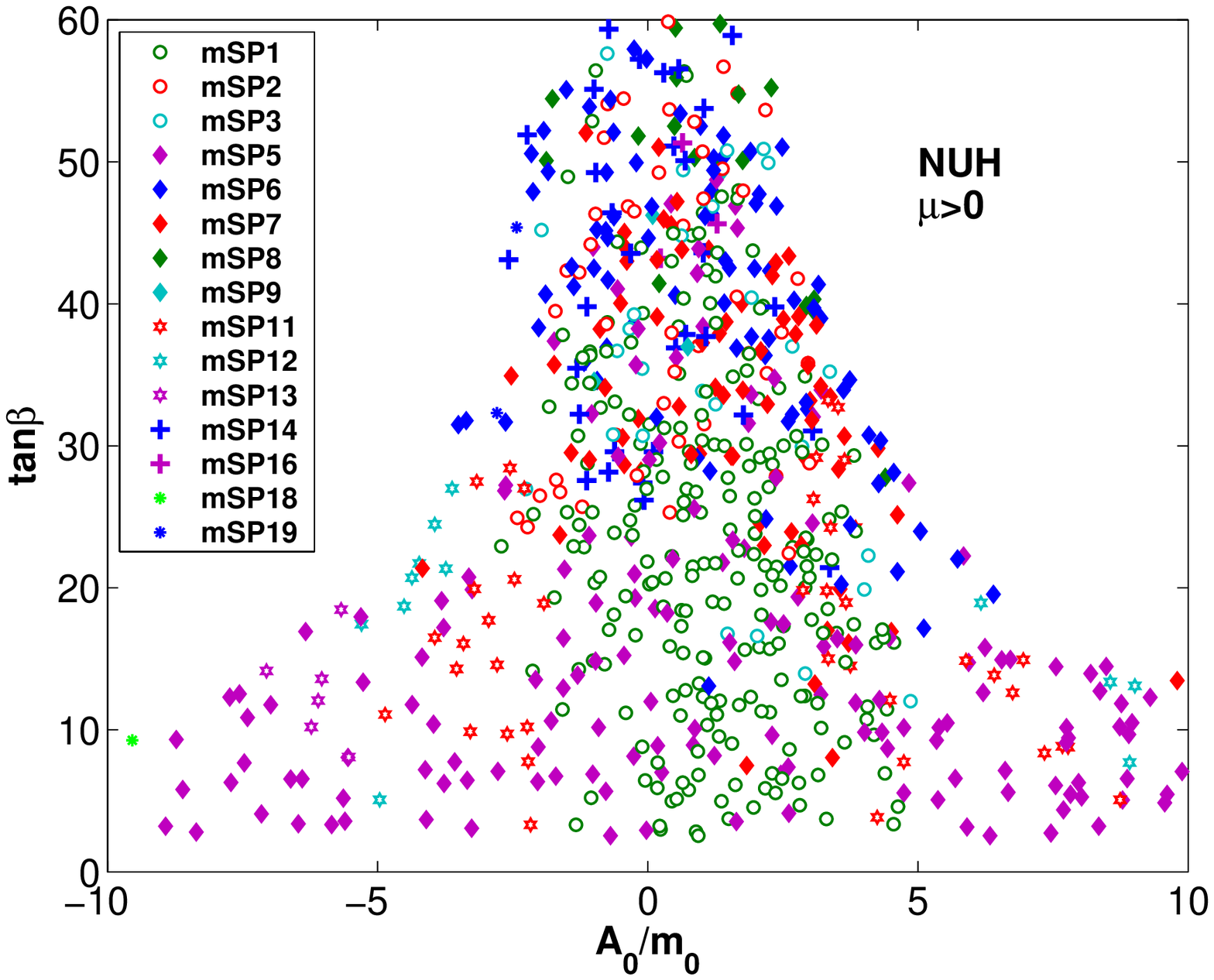}
\hspace*{-.1in}\includegraphics[width=7.0cm,height=6.0cm]{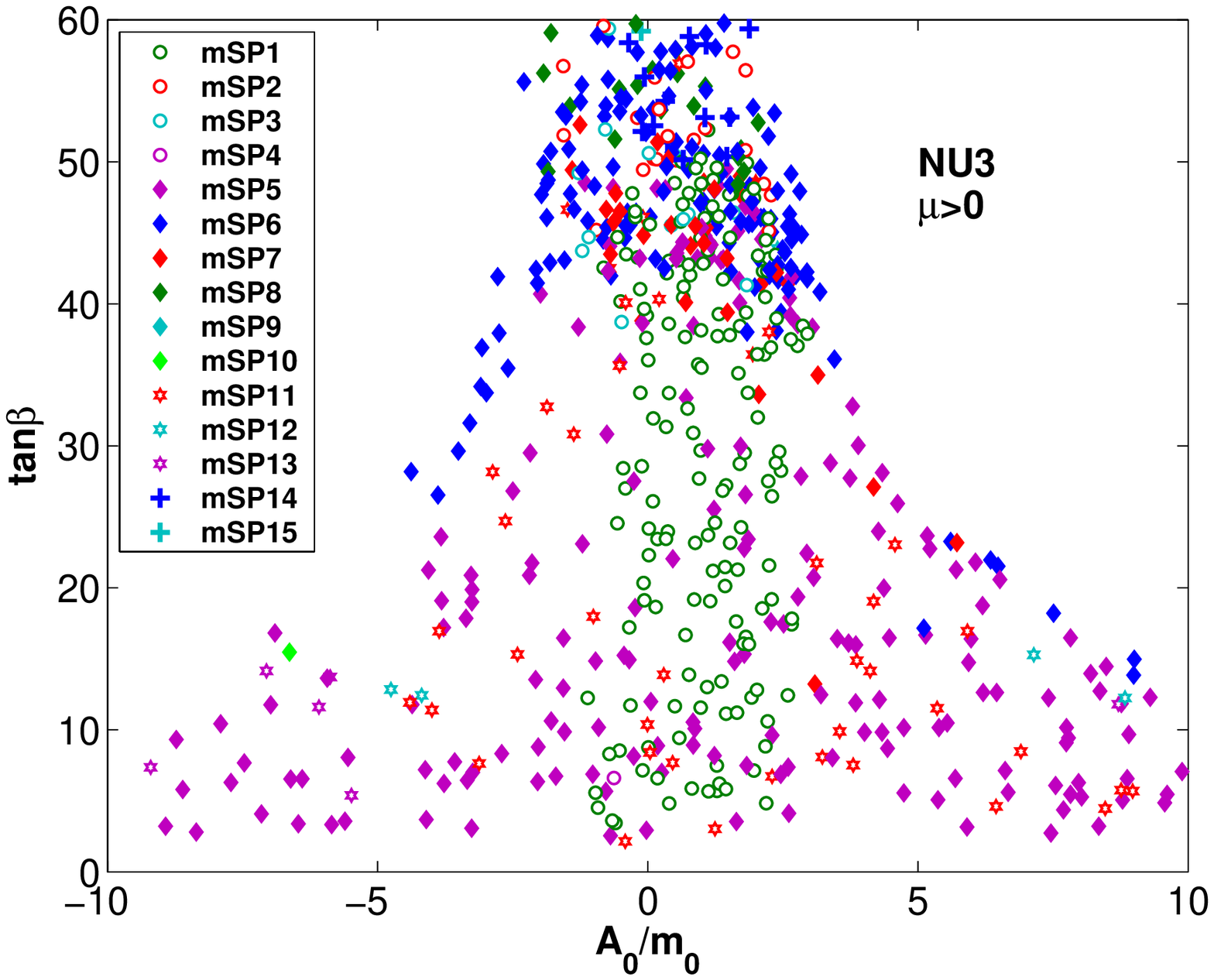}
\hspace*{.2in}\includegraphics[width=7.0cm,height=6.0cm]{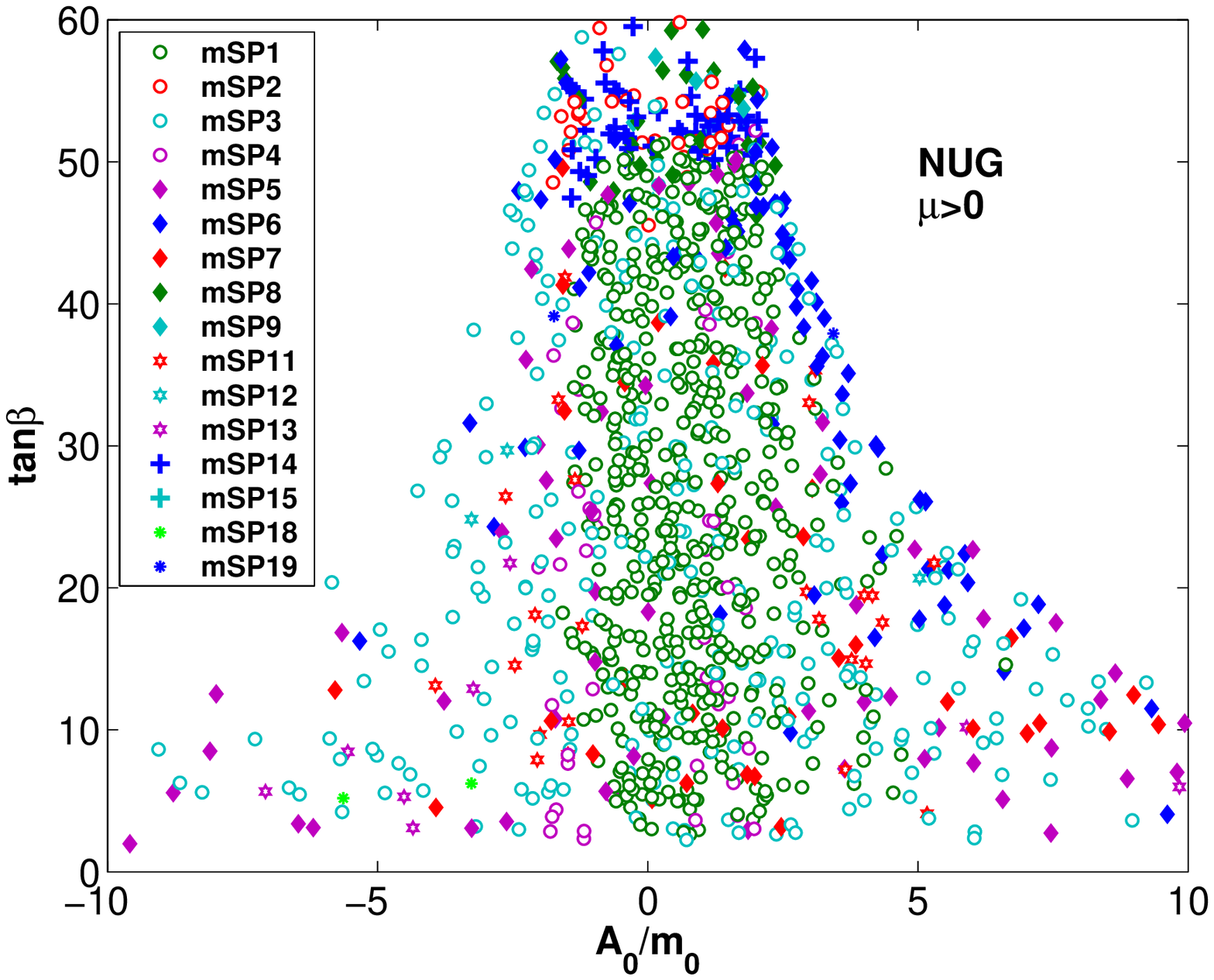}
\caption{Exhibition of NUSPs and mSPs arising from NUSUGRA models.}
 \label{fig:nuspec}
\end{figure*}
A similar analysis for the nonuniversal case is given in Fig.~(\ref{fig:nuspec}) which 
exhibits the NUSPs and mSPs for the NUH, NU3, and NUG models in the
$\tan\beta$ vs $A_0/m_0$ plane. The range of SUGRA parameters are the
same as the case mSUGRA ($\mu>0$). One may notice that the mSP1 points
arising from NU models lie in a relatively larger $A_0/m_0$ region.
Most of the models in NU cases are still mSPs, and among the NUSPs, only
two patterns have a relatively large population,  these being  NUSP1 and NUSP13.
One may also notice that in NUH case, the HPs can exist in a low
$\tan\beta$ region as opposed to the mSUGRA case where
HPs can either exist in the large $\tan\beta$ region ($\mu>0$)
or are totally eliminated  ($\mu<0$). 
Among the NUSPs  the  dominant patterns are NUSP1 (CP) and NUSP13 (GP), which
 are seen to arise from the model with \non in the gaugino sector, i.e., the NUG model.

\section{Benchmarks for Sparticle Patterns}

As discussed already, many of the sparticle mass patterns
discussed in this analysis do not appear in the Snowmass, Post-WMAP,
and CMS benchmark points. With some of these mSP and NUSP having a
significant probability of occurrence, we therefore provide a larger set of
benchmark points for the various patterns in different SUGRA
scenarios. In Table (\ref{tab:benchmark}), we give one benchmark point 
for each mSP pattern that are discovered in mSUGRA with $\mu>0$. 

\begin{table}[htb]
\begin{center}
\begin{tabular}{c|c |c |c |c| c |c c   }
\hline
mSUGRA  &    $m_0$   &   $m_{1/2}$   &   $A_0$ &  $\tan\beta$  & $\mu(Q)$  &   LSP$~|~$LCP             \\
Pattern &     $({\rm GeV})$   &   $({\rm GeV})$ & $ ({\rm GeV})$     &  $v_u/v_d$  &   $({\rm GeV})$    &   $({\rm GeV})$        \\
\hline
mSP1:   &     2001&411&0&30     & 216     &156.1$~|~$202.6     \\
mSP2:   &    1125&614&2000&50   & 673      &256.7$~|~$483.1  \\
mSP3:   &    741&551&0&50        & 632    &230.5$~|~$434.7   \\
mSP4:   &     1674&137&1985&18.6 & 533   &~54.3$~|~$106.9\\\hline
mSP5:   &     111&531&0&5       & 679     &217.9$~|~$226.3  \\
mSP6:   &      245&370&945&31    & 427    &148.6$~|~$156.8  \\
mSP7:   &      75 &201& 230& 14   & 246    &~74.8$~|~$100.2 \\
mSP8:   &      1880&877&4075&54.8 & 1141     &373.1$~|~$379.6 \\
mSP9:   &     667& 1154&-125&51   & 1257     &499.2$~|~$501.8  \\
mSP10:  &      336&772&-3074&10.8  & 1695 &329.2$~|~$331.7\\\hline
mSP11:  &      871&1031&-4355 &10 & 2306 &447.1$~|~$491.5 \\
mSP12:  &      1371&1671&-6855&10 & 3593 &741.2$~|~$791.8  \\
mSP13:  &     524&800&-3315 & 15 & 1782 &342.7$~|~$383.8  \\\hline
mSP14:  &     1036&562&500 & 53.5 & 560&236.2$~|~$399.1  \\
mSP15:  &       1113&758&1097&51.6 & 724 &321.1$~|~$595.9  \\
mSP16:  &       525&450&641&56     & 484 &184.6$~|~$257.9  \\
\hline
\end{tabular}
\caption{Benchmarks  using SUSPECT 2.3  with one point for each
mass pattern mSP1-mSP16. Also given are the neutralino LSP
(Lightest SUSY (R parity odd) Particle), and the Lightest Charged
Particle (LCP) masses. We take ${\mu}
> 0$, ${m_b}^{\overline{\rm MS}}(m_b) = 4.23$ {\rm GeV},
${\alpha_s}^{\overline{\rm MS}}(M_Z)=.1172$, and $m_t({\rm pole}) =
170.9$ ${\rm GeV}$. At least five LCP from these benchmarks will be
accessible  at the International  Linear Collider (ILC).}
\label{tab:benchmark}
\end{center}
 \end{table}
 
 We also provide a large collection of benchmark points which are exhibited in
Tables (\ref{b1}, \ref{b2}, \ref{b3}, \ref{b5}, \ref{b4}) in  the
Appendix.   Each of these benchmarks satisfies the relic density and
other experimental constraints with SuSpect  linked to MicrOMEGAs. We
have explicitly checked that the first mSP benchmark point in each
of the tables can be reproduced by using  SPheno, and SOFTSUSY
by allowing minor variations on the input parameters.
The benchmarks are chosen to  cover wide parts  of the SUGRA
parameter space. We give these benchmarks, several for each mass pattern,
as the search for SUSY from the point of view of mass patterns
has important consequences for LHC experimental searches.
 Some of the patterns are correlated  with certain
well investigated phenomena such as the HB/FP branches of REWSB
and the stau-neutralino co-annihilation regions.
However, many of the patterns arise from multiple annihilation processes.

\chapter{Sparticle Signatures at the Large Hadron Collider}
\label{ch:lhc}

In this chapter, first we will give the detailed description of our LHC simulations. 
And then we will give our post trigger cuts and all the signatures investigated. 

\section{LHC Simulations}
After the imposition of all the constraints mentioned in the previous sections, 
such as the relic density constraints from WMAP data, the constraints on the FCNCs, 
as well as mass limits on the sparticle spectrum, we are left with the candidate 
model points for the signature analysis. For each of these model points, a SUSY
Les Houches Accord (SLHA) file \cite{SKANDS} is interfaced to PYTHIA 6.4.11 
\cite{PYTHIA} through  PGS4 \cite{PGS} for the computation of SUSY production cross 
sections and branching fractions. In this analysis, for signals, we have generated all
of PYTHIA's $2 \to 2$ SUSY production modes using MSEL~$=$~39. 
More specifically this choice generates 91 SUSY production modes including 
gaugino, squark, slepton, and SUSY Higgs pair production but leaves out singly produced 
Higgs production. For further details, see \cite{PYTHIA}. A treatment of singly produced 
Higgs production in the context of sparticle mass hierarchies was included in the 
analysis of Ref. \cite{Feldman:2007fq}. Leading order cross sections from PYTHIA and 
leading order cross sections from PROSPINO 2.0 \cite{PROSPINO} were cross checked against 
one another for consistency over several regions of the soft parameter space. TAUOLA 
\cite{TAUOLA} is called by PGS4 for the calculation of tau branching fractions as 
controlled in the PYTHIA parameter card (.pyt) file.

With PGS4 we use the Level 1 (L1) triggers based on  the Compact Muon Solenoid detector
(CMS) specifications \cite{CMS,Ball:2007zza} and the LHC detector card. Muon isolation 
is controlled by employing the cleaning script in PGS4. We take the experimental 
nomenclature of lepton being defined only as electron or muon and thus distinguish 
electrons and muons from tau leptons. SM backgrounds have been generated with
QCD multi-jet production due to light quark flavors, heavy flavor jets 
($b \bar b$,  $t \bar t$), Drell-Yan, single $Z/W$ production in association with quarks 
and gluons ($Z$/$W$+ jets), and $ZZ$, $WZ$, $WW$ pair production resulting 
in multi-leptonic backgrounds. Extraction of final state particles from the PGS4 event 
record is accomplished with a code SMART (= SUSY Matrix Routine) written by us 
\cite{Feldman:2007zn} which provides an optimized processing of PGS4 event data files.
The standard criteria for the discovery limit of new signals is that the SUSY signals
should exceed either $5\sqrt{N_{\rm SM}}$ or 10 whichever is larger, i.e.,
${\rm N_{SUSY}}>{\rm Max}\left\{5\sqrt{N_{\rm SM}},10\right\}$ and such a criteria 
is imposed where relevant. We have also  cross checked various results of our analysis 
with three CMS notes \cite{CMSnote1,CMSnote2,CMSnote3} and we have found agreement with 
these works using SMART and PGS4 for signal and backgrounds.

We note that several works where sparticle signatures are discussed have appeared recently 
\cite{chameleon,Kane:2006yi,Conlon:2007xv,Baer:2007ya,Mercadante:2007zz,Kitano:2006gv}.
However, the issue of  hierarchical mass patterns and the correlation of signatures
with such patterns has not been discussed which is what the analysis of this work investigates.
\begin{table*}[htb]
\begin{center} 
\small{
\begin{tabular}{|c|l||c|l|}
\hline Signature   &   Description &   Signature   &
Description \\  \hline 0L  &   0 Lepton    &   0T  &
0 $\tau$    \\  \hline 1L  &   1 Lepton    &   1T  &   1 $\tau$
\\  \hline 2L  &   2 Leptons    &   2T  &   2 $\tau$    \\  \hline 3L
&   3 Leptons    &   3T  &   3 $\tau$    \\  \hline 4L  &   4 Leptons
and more   &   4T  &   4 $\tau$ and more   \\  \hline 0L1b    &   0
Lepton + 1 b-jet  &   0T1b    &   0 $\tau$ + 1 b-jet  \\  \hline
1L1b    &   1 Lepton + 1 b-jet  &   1T1b    &   1 $\tau$ + 1 b-jet
\\  \hline 2L1b    &   2 Leptons + 1 b-jet  &   2T1b    &   2 $\tau$
+ 1 b-jet  \\  \hline 0L2b    &   0 Lepton + 2 b-jets  &   0T2b    &
0 $\tau$ + 2 b-jets  \\  \hline 1L2b    &   1 Lepton + 2 b-jets  &
1T2b    &   1 $\tau$ + 2 b-jets  \\  \hline 2L2b    &   2 Leptons + 2
b-jets  &   2T2b    &   2 $\tau$ + 2 b-jets  \\  \hline ep  &   $e^+$
in 1L &   em  &   $e^-$ in 1L \\  \hline mp  &   $\mu^+$ in 1L   &
mm  &   $\mu^-$ in 1L   \\  \hline tp  &   $\tau^+$ in 1T  &   tm  &
$\tau^-$ in 1T  \\  \hline OS  &   Opposite Sign Di-Leptons &   0b  &
0 b-jet \\  \hline SS  &   Same Sign Di-Leptons &   1b  &   1 b-jet
\\  \hline OSSF    &   Opposite Sign Same Flavor Di-Leptons &   2b  &
2 b-jets \\  \hline SSSF    &   Same Sign Same Flavor Di-Leptons &
3b  &   3 b-jets \\  \hline OST &   Opposite Sign Di-$\tau$ &   4b  &
4 b-jets and more    \\  \hline SST &   Same Sign Di-$\tau$ & TL  &
1 $\tau$ plus 1 Lepton  \\  \hline
\end{tabular}
}
\caption{The table gives  a list of 40 counting signatures for each  point in the SUGRA model
parameter space. $L=e,\mu$ signifies only electrons and muons.} \label{tab:counting}
\end{center}\end{table*}

\begin{table*}[htb]
    \begin{center}
\small{
\begin{tabular}{|l|}
\hline Kinematical signatures\\  \hline 1. $P_T^{miss}$ \\  \hline
2. Effective Mass = $P_T^{miss}$ + $\sum_j P_T^j$\\  \hline 3.
Invariant Mass of all jets\\  \hline 4. Invariant Mass of $e^+e^-$
pair\\  \hline 5. Invariant Mass of $\mu^+\mu^-$ pair\\  \hline 6.
Invariant Mass of $\tau^+\tau^-$ pair\\  \hline
\end{tabular}
}
\caption{The table give a list kinematical signatures analyzed.} \label{tab:kin}
    \end{center}
 \end{table*}

\section{Post Trigger Level Cuts and LHC Signatures \label{D1}}

Generally speaking, there are two kinds of LHC signatures: (i) event  counting
signatures,  and (ii) kinematical signatures. We have investigated both of
these for the purpose of discriminating the sparticle mass patterns.
We list our  event counting signatures in Table (\ref{tab:counting}), where we have
carried out analyses of a large set of lepton + jet signals.
In our counting procedure, only electron and muon are
counted as leptons, while tau jets are counted independently.
For clarity, from here on, our use of `jet(s)' will exclude tau jets. Thus, for jet
identification, we divide jets into two categories: b-tagged jets
and jets without b-tagging, which we simply label as b-jets and
non-b-jets (see also \cite{chameleon}).
There are some counting signatures that only concern one
class of measurable events, for example,
the number of events containing one tagged b-jet and any other final state
particles. There are also types of signatures of final state particles with
combinations of two or three different species.
For instance, one such example would be the number of events in which there is
a single lepton and a single tau.

When performing the analysis of event counting, for  each  SUGRA model point,
we impose  global post trigger cuts to analyze most of our PGS4 data.   Below we give
our default post trigger cuts which are used throughout this thesis unless stated otherwise.
\begin{enumerate}
\item In an event, we only select photons, electrons, and muons
that have transverse momentum $P^{p}_T>10$ GeV and $|\eta^{p}|<2.4$, $p=(\gamma,e,\mu)$.
\item Taus which satisfy $P^{\tau}_T>10$ GeV and $|\eta^{\tau}|<2.0$ are selected.
\item For hadronic jets, only those satisfying $P^{j}_T>60$ GeV and $|\eta^{j}|<3$ are selected.
\item We require a large amount of missing transverse momentum, $P^{miss}_T>200$ GeV.
\item There are at least two jets that satisfy the $P_T$ and $\eta$ cuts.
\end{enumerate}
Our default post trigger level cuts are standard
and are designed to  suppress the
Standard Model background, and highlight the SUSY events over a broad class of models.

The different kinematical signatures we investigated for the purpose
of discriminating among sparticle mass patterns are exhibited in Table
(\ref{tab:kin}). One may further divide the kinematical
signatures into two classes: namely those involving transverse momentum $P_T$ and
those which involve invariant
mass. For those involving  $P_T$, we have investigated missing $P_T$ distributions
and the effective mass, the latter being the sum of missing
$P_T$ and $P_T$ of all jets contained  within an event.
For the kinematical variables using  invariant
mass, we reconstruct such quantities for four different cases, i.e.,
 the invariant mass for all jets, for $e^+e^-$ pair, for
$\mu^+\mu^-$ pair, and for $\tau^+\tau^-$ pair. The reconstruction of the
invariant mass of $\tau^+\tau^-$ pair is  based on hadronically decaying
taus (for recent analyses see \cite{ArnowittTexas}).

\chapter{Event Counting Signatures for Sparticles at the LHC}
\label{ch:sig}

\section{Discrimination among mSPs in mSUGRA\label{D2}}

We turn now to a discussion of how one may distinguish among different
patterns. The analysis begins by considering the 902  model points that survive our
mSUGRA scan with $10^6$ trial points, and simulating  their LHC signals
with PGS4 using, for illustration, 10 fb$^{-1}$ of integrated luminosity at the LHC.
In our analysis we will focus mostly on the counting signatures.
Here the most useful counting signature is the total number of SUSY
events after trigger level cuts and post trigger level cuts are imposed.
All other counting signatures are normalized with respect to the
total number of SUSY events passing the cuts
and thus appear as fractions lying
between (0,1) in our figures.
To keep the analysis statistically significant, we admit only those points in the
parameter space that generate at least 500 total SUSY events.

\begin{figure*}[htb]
\centering
\includegraphics[width=7.0cm,height=6.0cm]{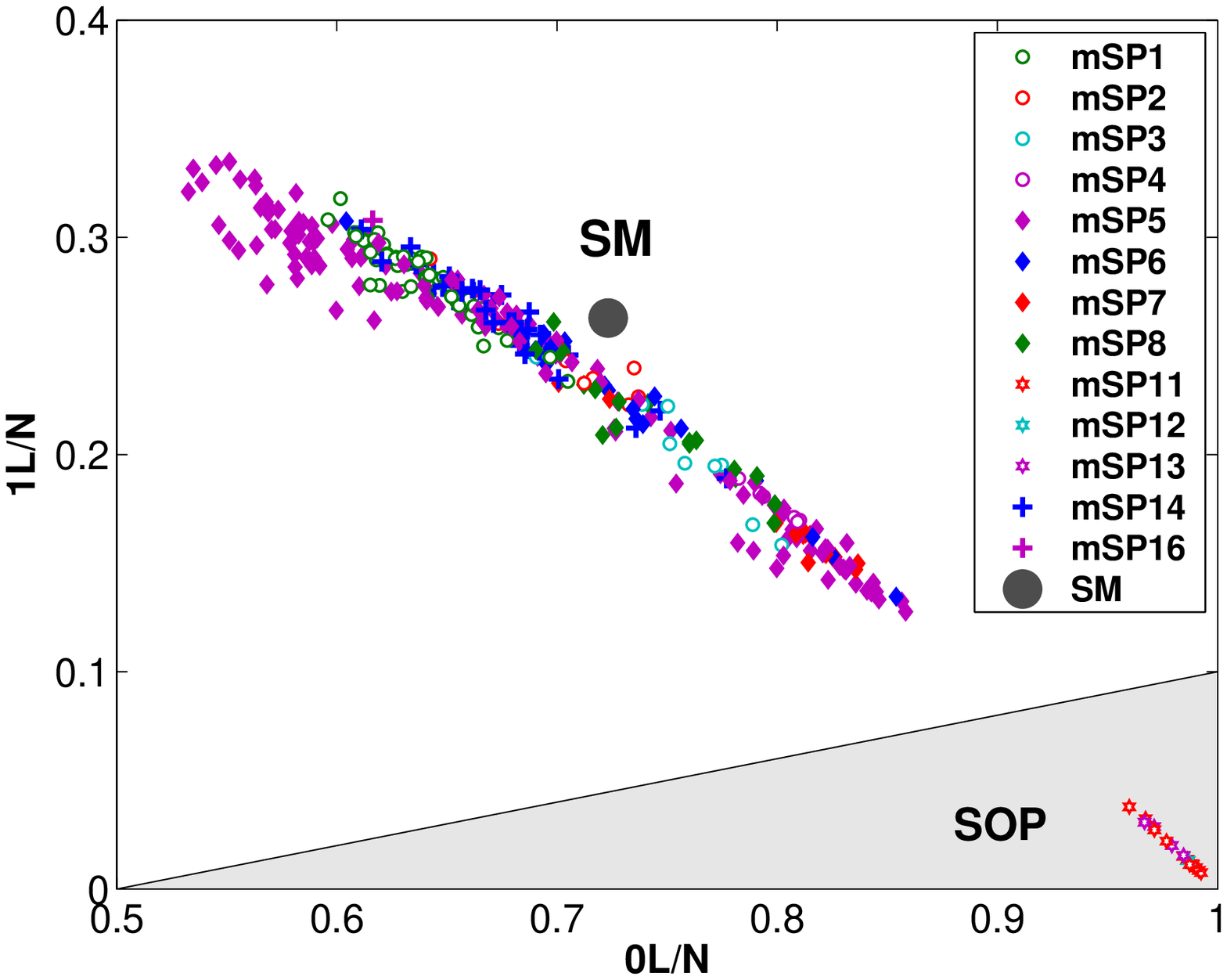}
\includegraphics[width=7.0cm,height=6.0cm]{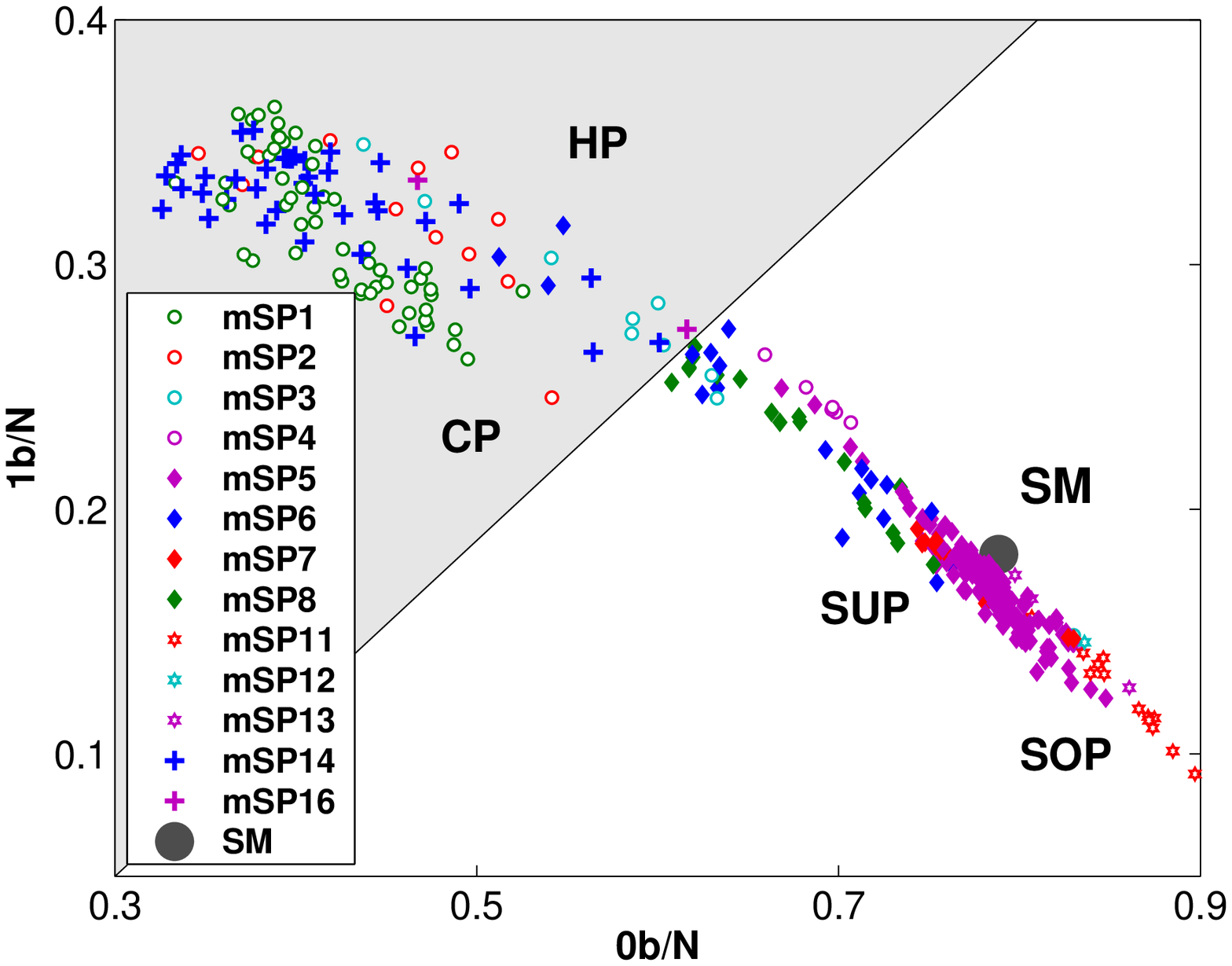}
\includegraphics[width=7.0cm,height=6.0cm]{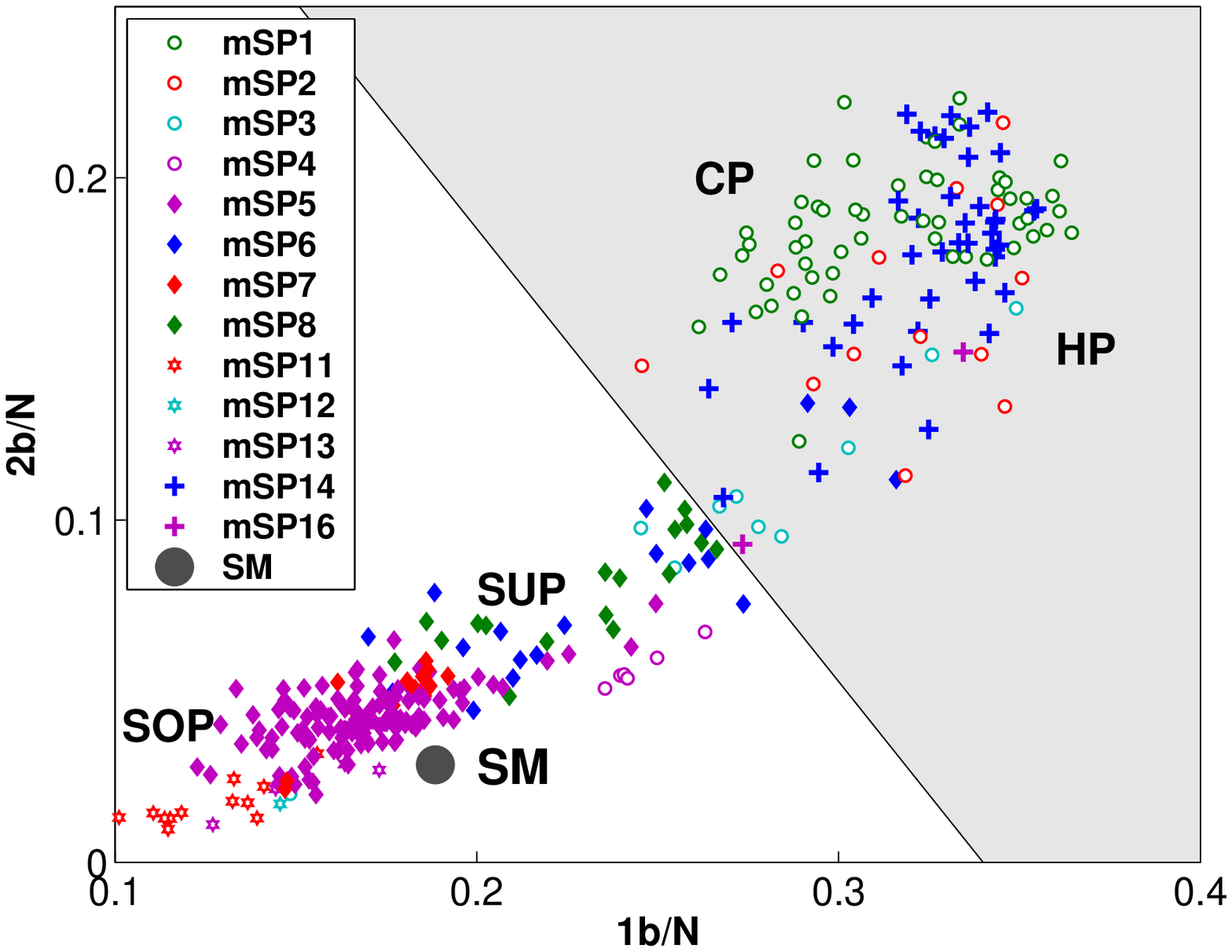}
\includegraphics[width=7.0cm,height=6.0cm]{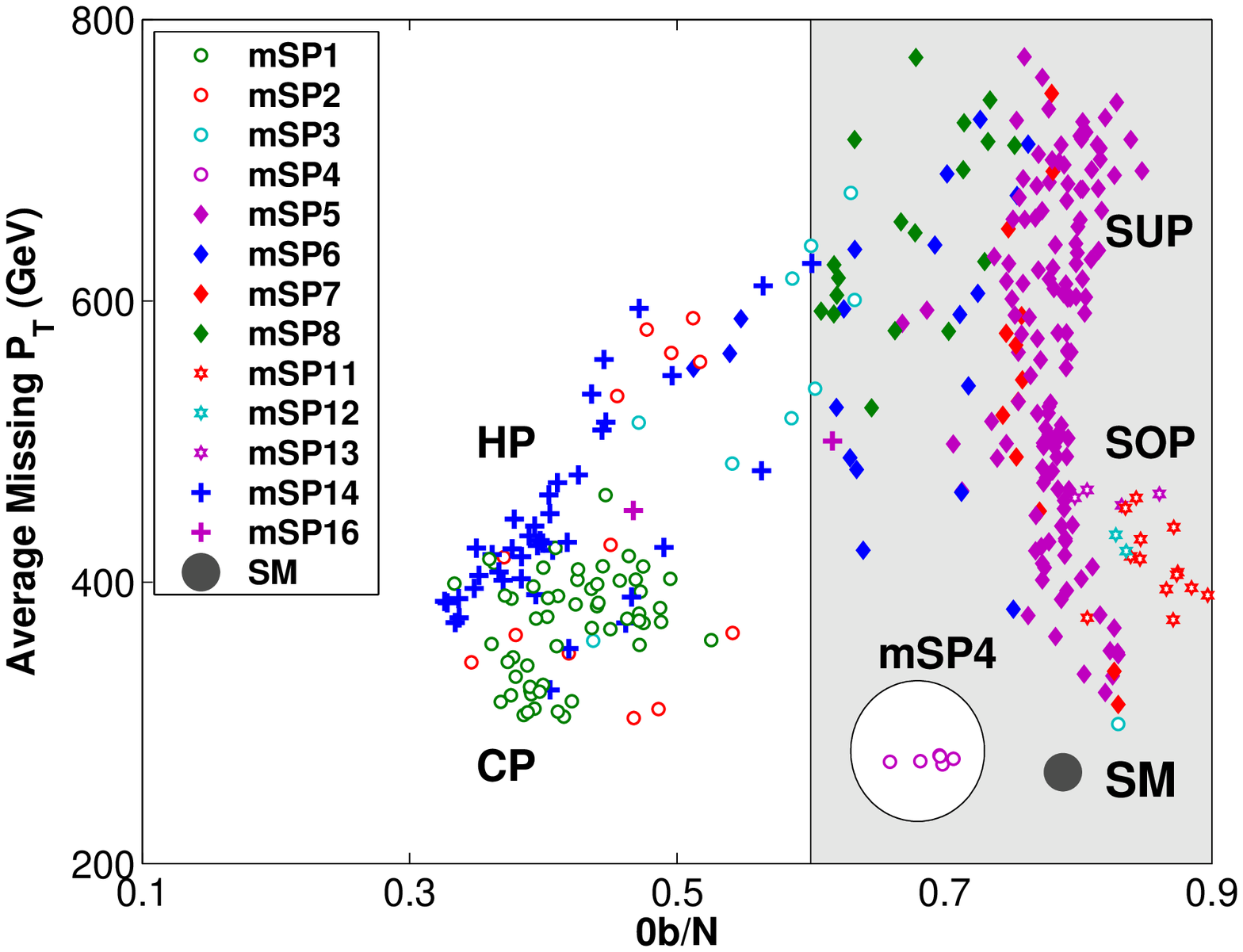}
\caption{Exhibitions of the mSPs in mSUGRA with $\mu > 0$ in various signature channels.}
\label{fig:bjet}
\end{figure*}

\begin{figure*}[htb]
\centering
\includegraphics[width=7.0cm,height=6.0cm]{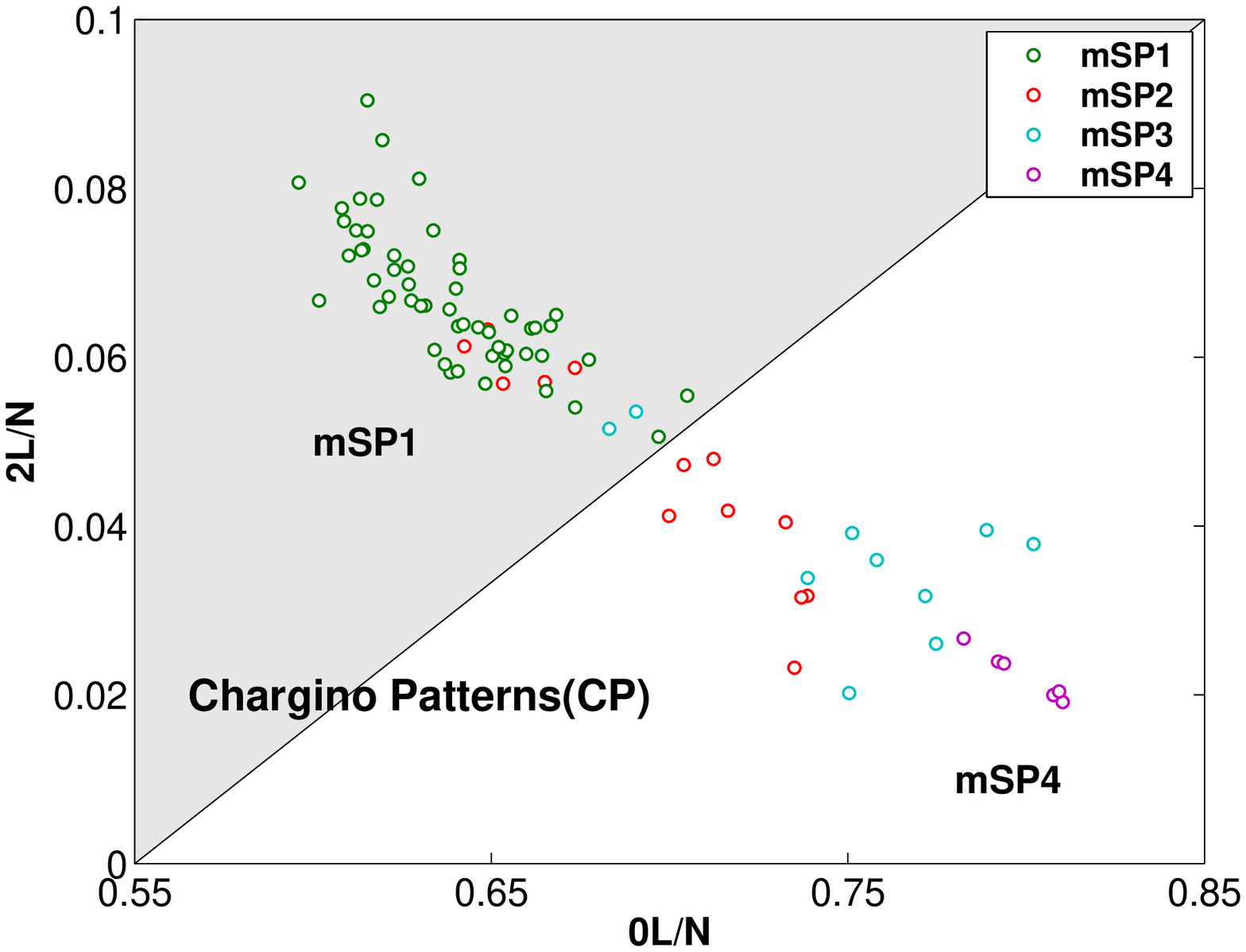}
\includegraphics[width=7.0cm,height=6.0cm]{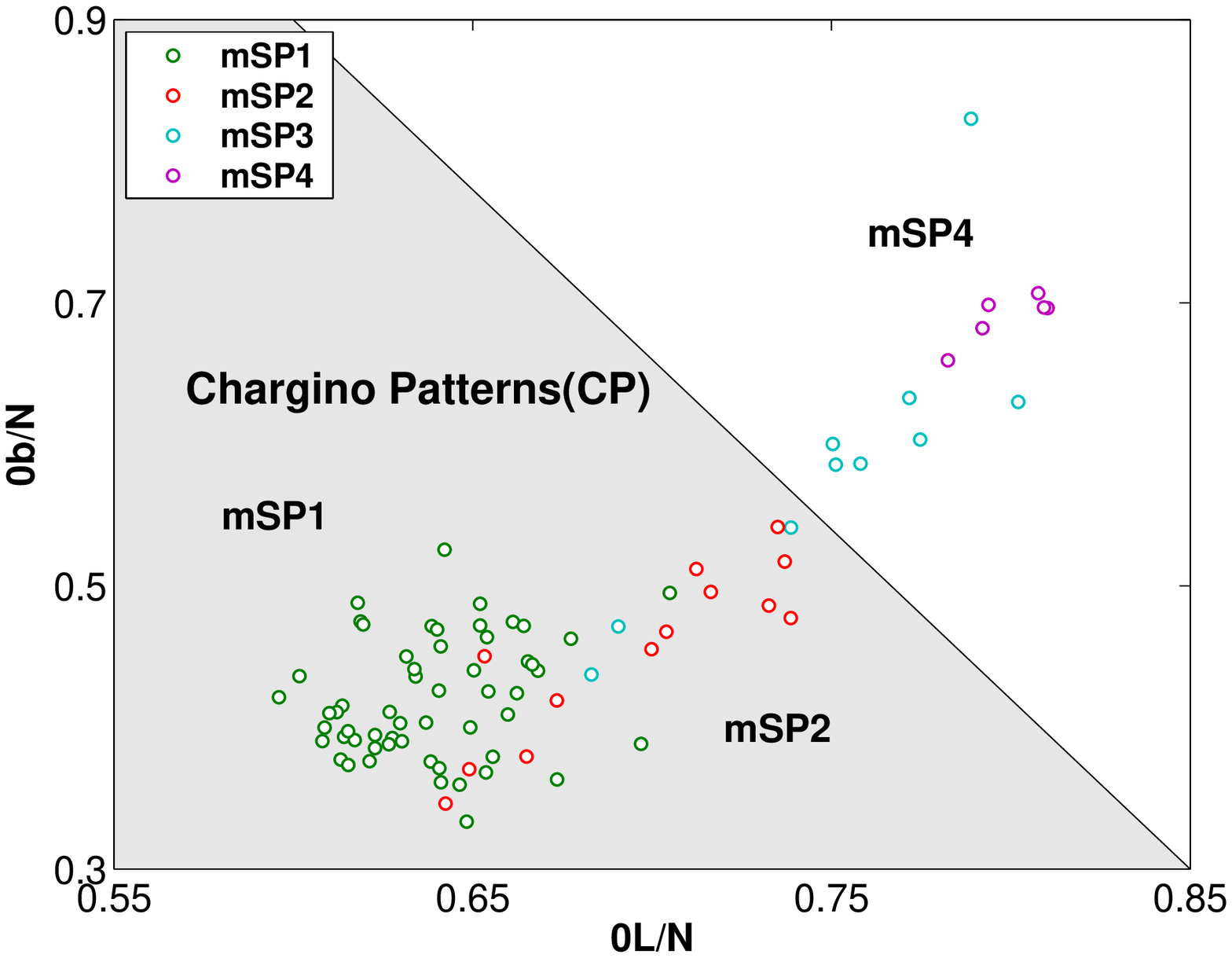}
\includegraphics[width=7.0cm,height=6.0cm]{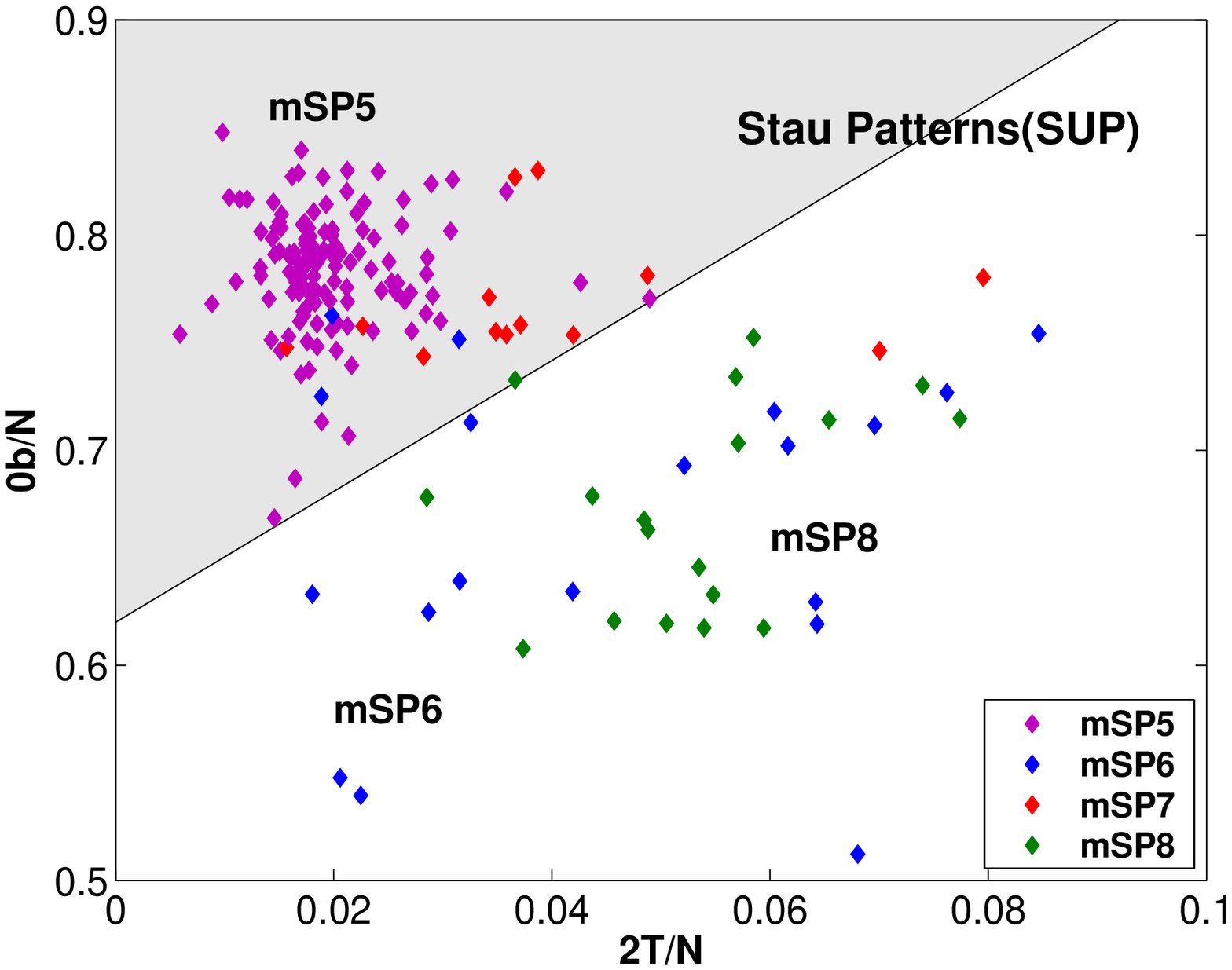}
\includegraphics[width=7.0cm,height=6.0cm]{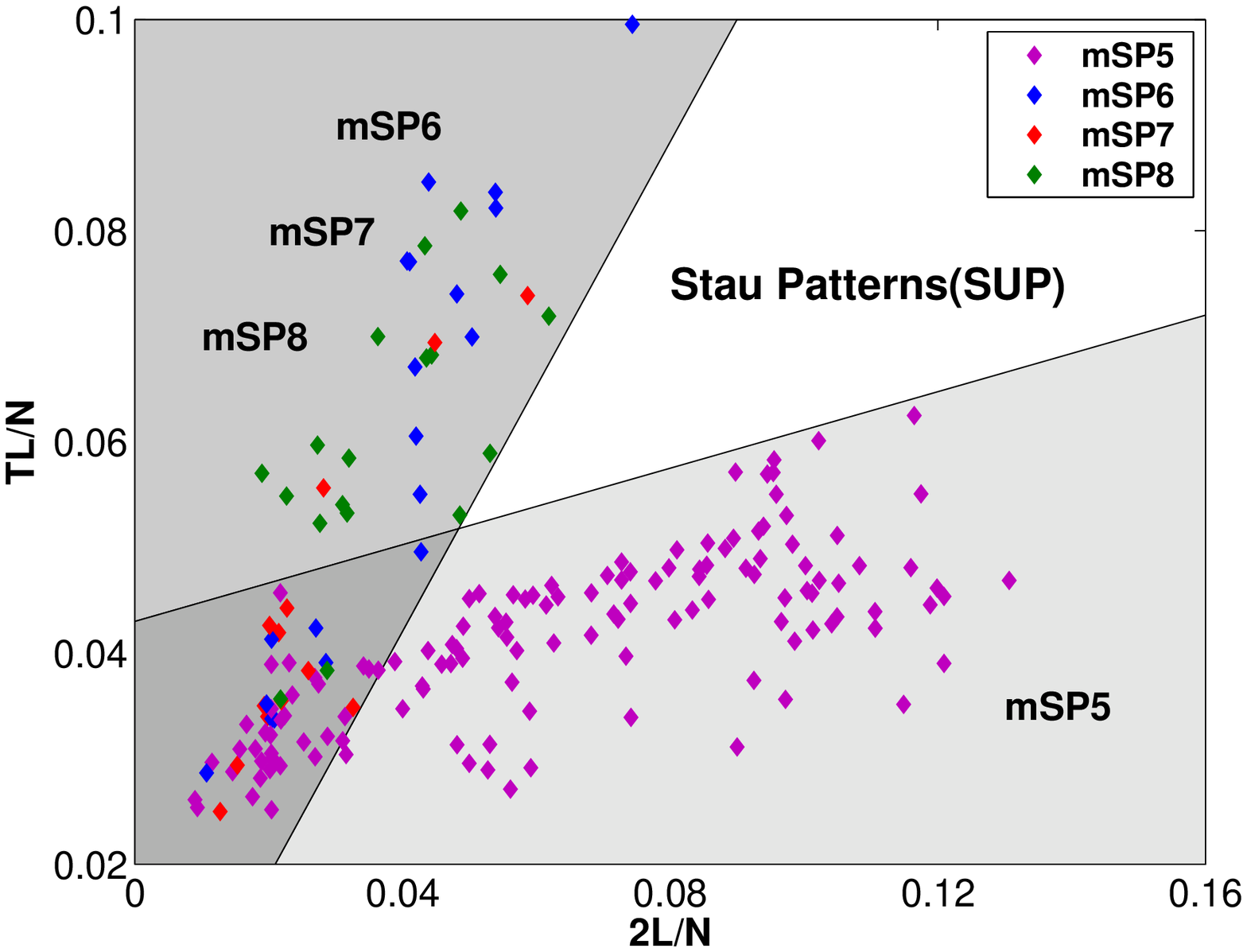}
\caption{An exhibition of how the mSPs can be  discriminated within CPs, SUPs.}
\label{fig:sups}
\end{figure*}

\begin{figure*}[htb]
\centering
\includegraphics[width=7.0cm,height=6.0cm]{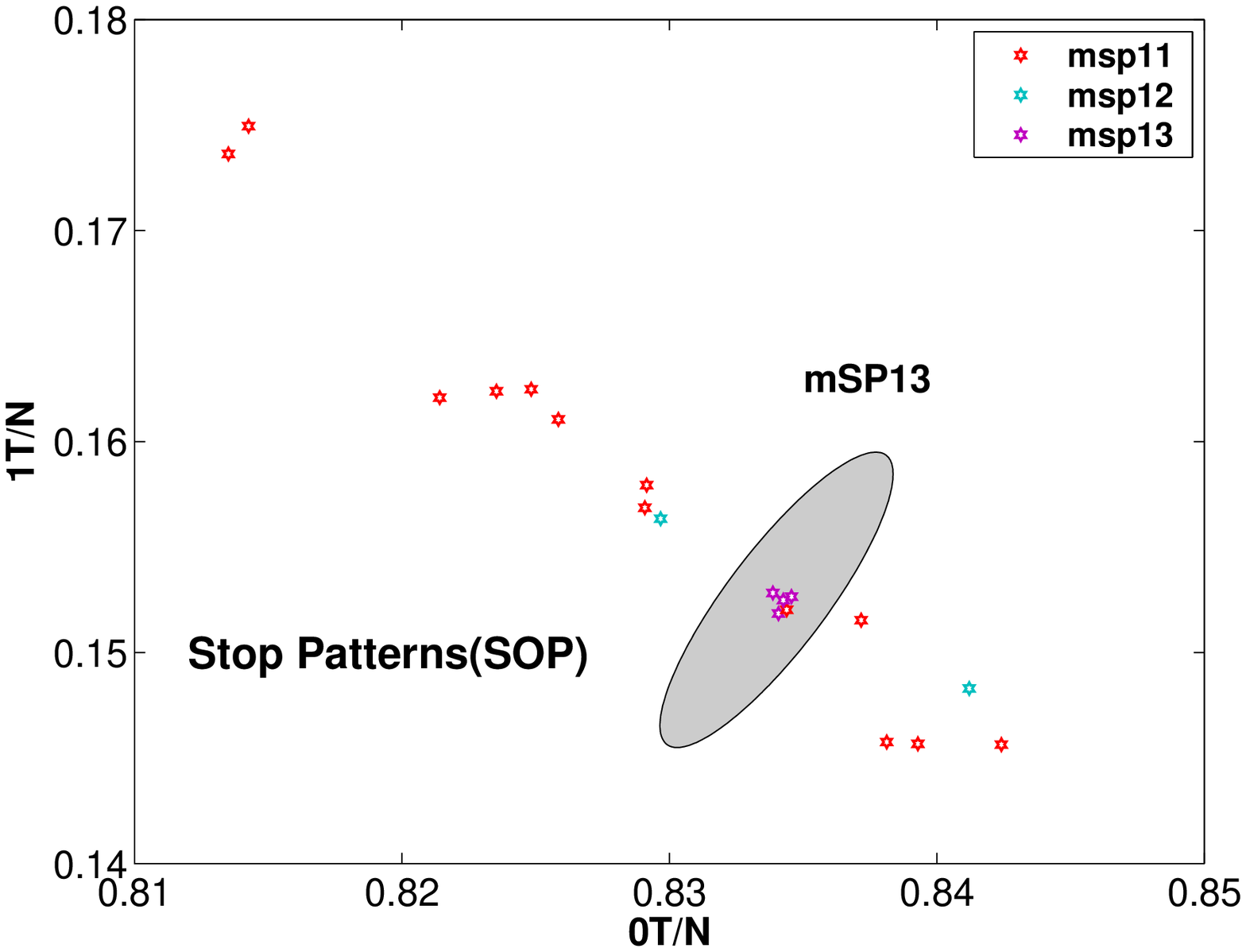}
\includegraphics[width=7.0cm,height=6.0cm]{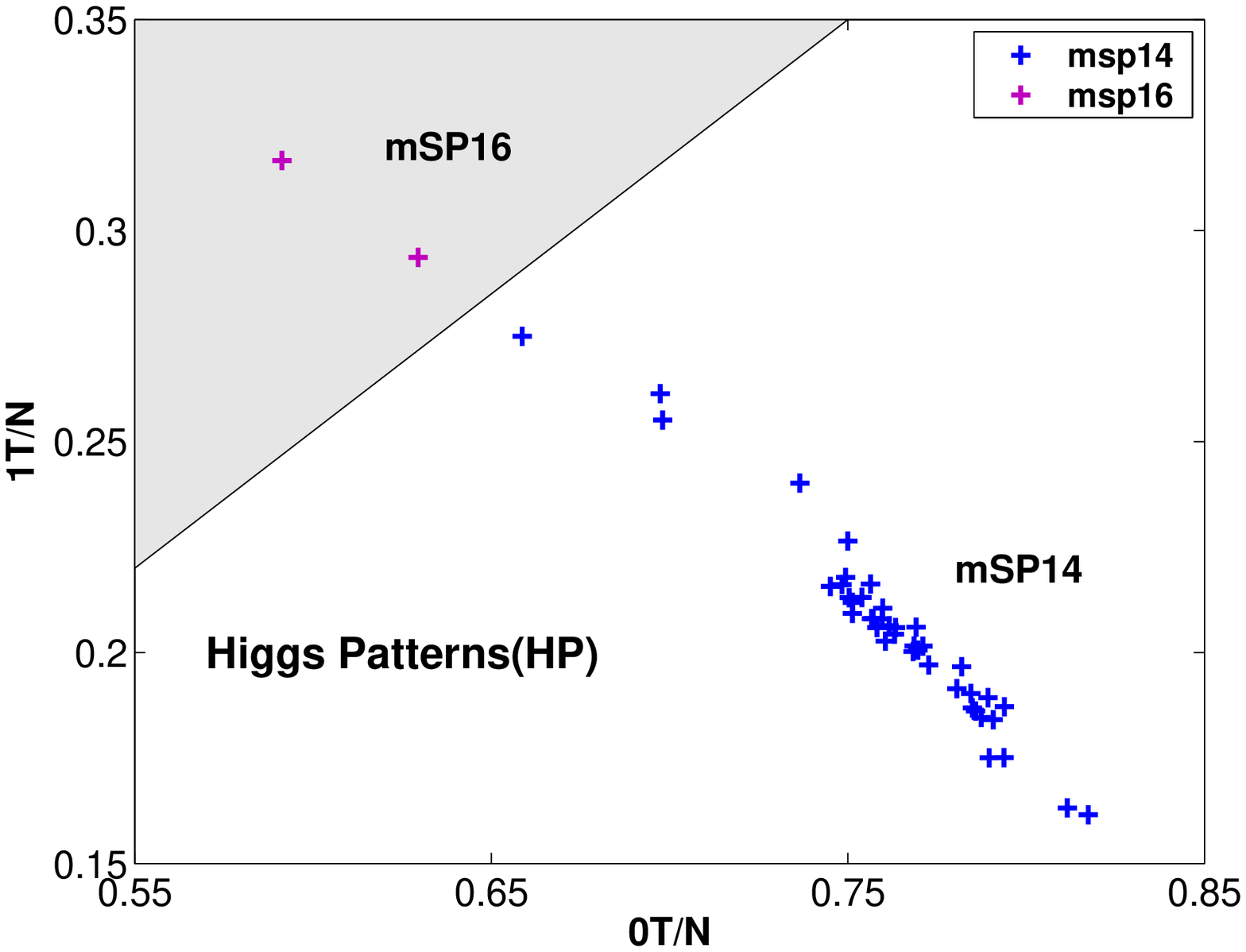}
\caption{An exhibition of how the mSPs can be  discriminated within SOPs and HPs.}
\label{fig:sups2}
\end{figure*}

We give now the details of the analysis.
In Fig.(\ref{fig:bjet}), we investigate the signature space spanned
by a variety of signature channels. Top Left: An exhibition of the mSPs in the 1L
vs 0L where the fraction of events to the
total number of events  in each case  is plotted.
 The analysis shows that the Stop Patterns (SOP)
appearing on the right-bottom corner are easily distinguished from
other patterns. The analysis shows that SOP has few lepton signals.
Top Right and Bottom Left: Plots in the signature space with fraction of events with
 1b vs 0b and 2b vs 1b  exhibiting the separation of CPs and HPs from SOPs
and SUPs, with CPs and HPs occupying one region, and SOPs and SUPs occupy another in
this signature space except for a very small overlap.
Bottom Right: An exhibition of the mSPs in the signature space with the average
missing $P_T$ for each parameter point in the mSUGRA parameter
space along the y-axis and the fraction of events with $0b$ along the
x-axis. The plot shows a separation of the CPs and HPs from SOPs and
SUPs. Further, mSP4 appears isolated in this plot.
Most of the CPs and HPs have less than 60\% events without b-jet content.
The ratios for the SUSY models refer to the SUSY signal only.  The SM point is purely
background.
The top left panel gives a plot with
one signature consisting of events with one lepton and the second
signature consisting of event with no leptons.  It is seen that
the stop patterns (SOPs) that survive the cuts
are confined in a  small region at the right-bottom corner
and have a significant separation from all other mSPs.
The panel illustrates the negligible leptonic content in stop decays.
The top-right panel is a plot between two signatures where one
signature contains a tagged b-jet while the other signature has no
tagged b-jets.  In this case one finds a significant separation of the
CPs and HPs  from SUPs and SOPs.
The lower-left panel gives a plot where one signature has two tagged b-jets
and the other signature has only one tagged b-jet. One again finds that
the CPs and HPs are well separated from the SOPs and the SUPs for much
the same reason as in upper-right panel.  Finally,  a plot is given in the lower-right
panel where one signature is the average missing $P_T$ while the other
signature involves events with no tagged b-jets. Again in this plot the CPs
(which include mSP4)  and HPs are well separated from the SOPs and SUPs.

The analysis of  Fig.(\ref{fig:bjet}) exhibits that
for some cases, e.g., for the patterns CP and HP in the upper right
hand corner of Fig.(\ref{fig:bjet}),  the separation between the SUGRA
prediction and the Standard Model background is strikingly clear, allowing for the
 identification not only of new physics but also of the nature of the pattern that
leads to such a signature.

We discuss now the possibility of discriminating sub-patterns within a given pattern class.
 An analysis illustrating this possibility
is given in Fig.~(\ref{fig:sups}). Here the top two panels illustrate
how  the sub-patterns mSP1, mSP2, mSP4 within the chargino class (CP)
are distinguishable  with appropriate choice of the signatures.
A similar analysis regarding the discrimination for the sub-patterns in the stau class  (SUP)
is given in the two bottom panels. 
The left panel in Fig.~(\ref{fig:sups2}) gives an analysis of how one may
discriminate the stop sub-patterns mSP11, mSP12, mSP13 in the stop class (SOP),
and finally the right panel shows the plots that allows one to discriminate
the Higgs patterns mSP14 and mSP16 from each other.
There are a variety of other plots which allow one to discriminate
among patterns. With 40 counting signatures one  can have 780
such plots and it is not possible to display all of them.
A global analysis where the signatures are simultaneously considered for
a large collection of mSPs and NUSPs.

 As mentioned in the above analysis we have included models which can produce at
least 500 SUSY events with 10 fb$^{-1}$  which is  lower than
our estimated discovery limits for total SUSY events which are about 2200 in this case. The reason for
inclusion of points below the discovery limit in the total SUSY events is
 that some of them can be detected in other
channels such as in the trileptonic channel while others will be detectable 
as the luminosity goes higher.
We note in passing that reduction of admissible points makes
separation of patterns easier.

\section{Sparticle Signatures including Nonuniversalities   \label{D3}   }

\begin{figure}[htb]
\centering
\includegraphics[width=7.0cm,height=6.0cm]{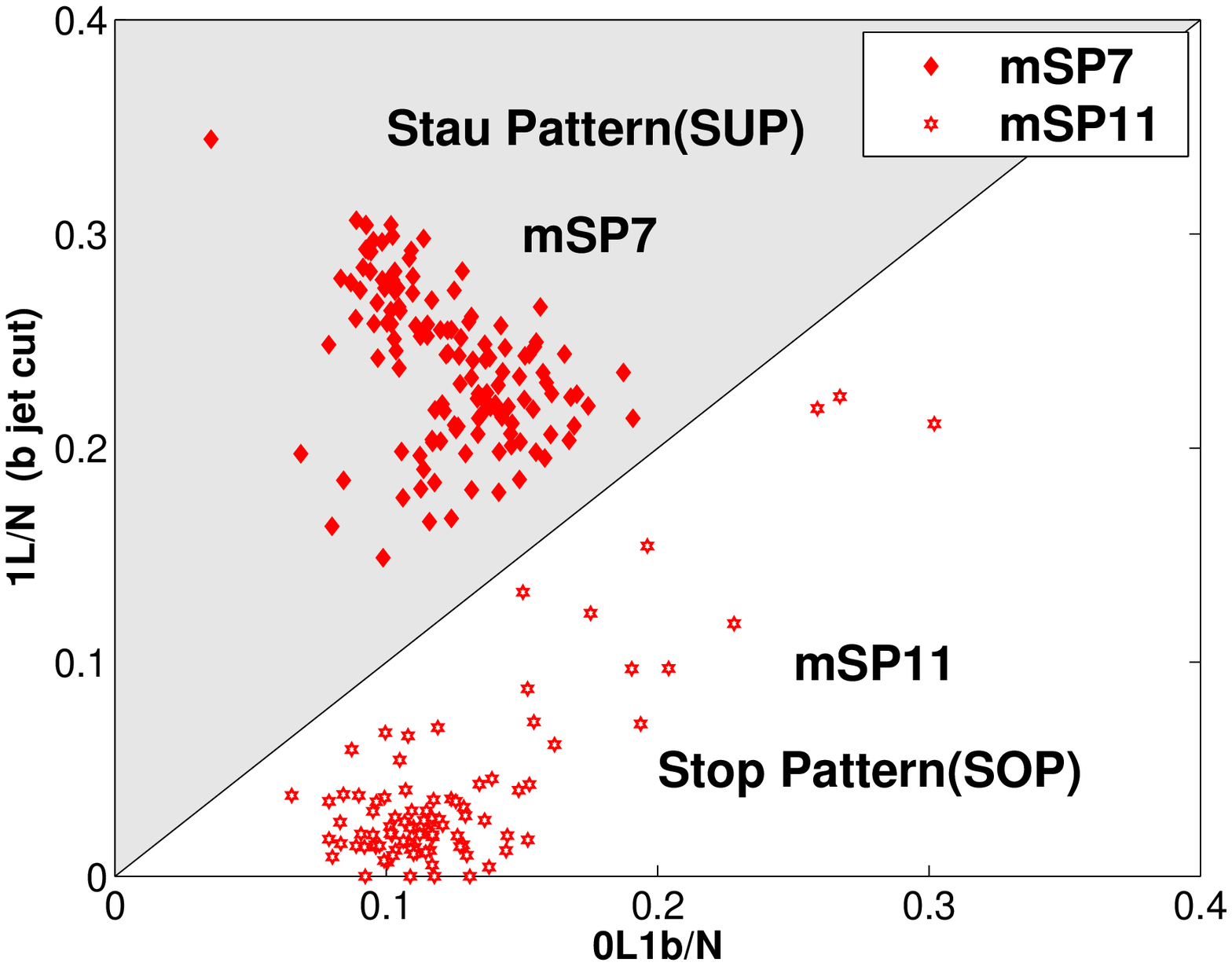}
\includegraphics[width=7.0cm,height=6.0cm]{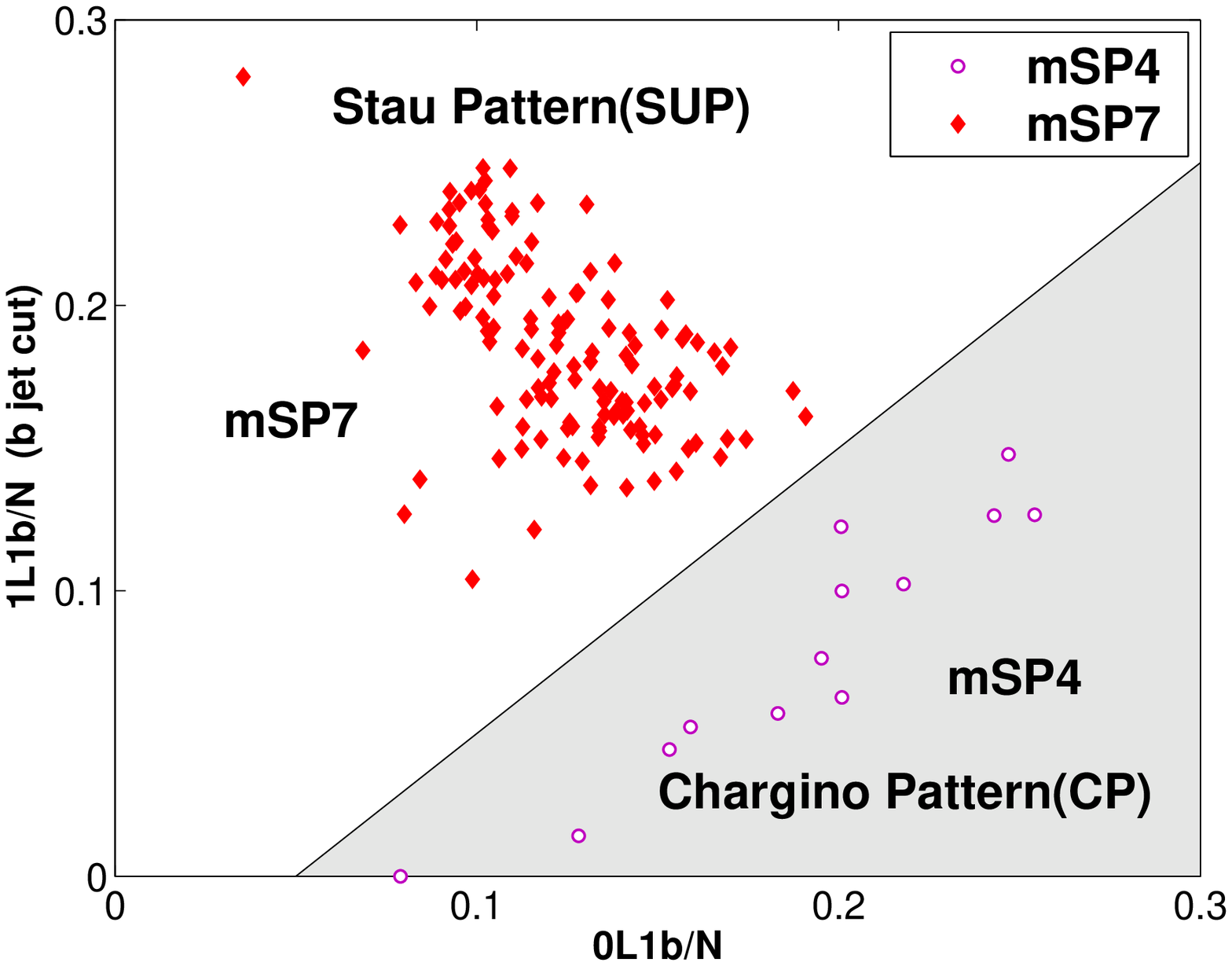}
\includegraphics[width=7.0cm,height=6.0cm]{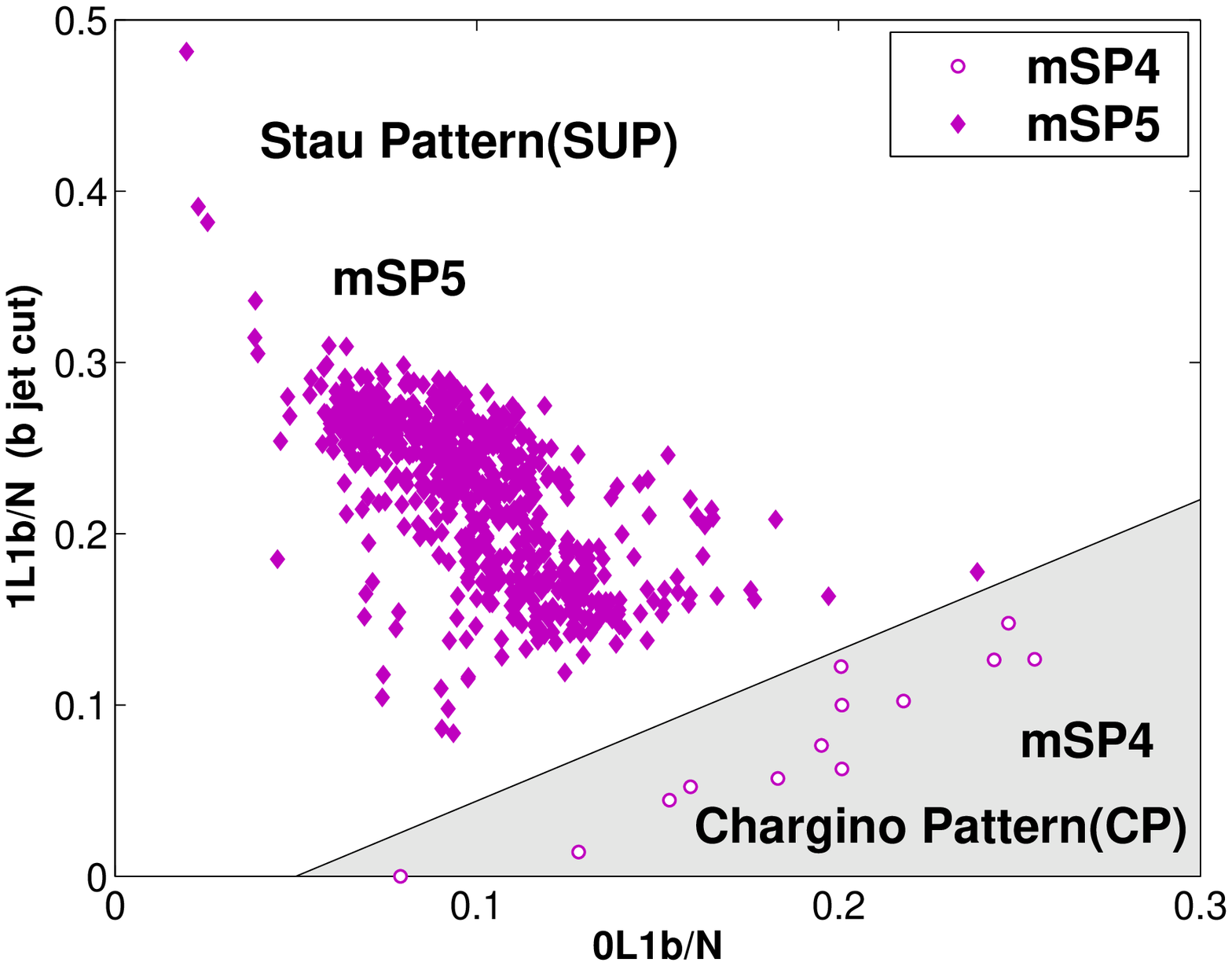}
\includegraphics[width=7.0cm,height=6.0cm]{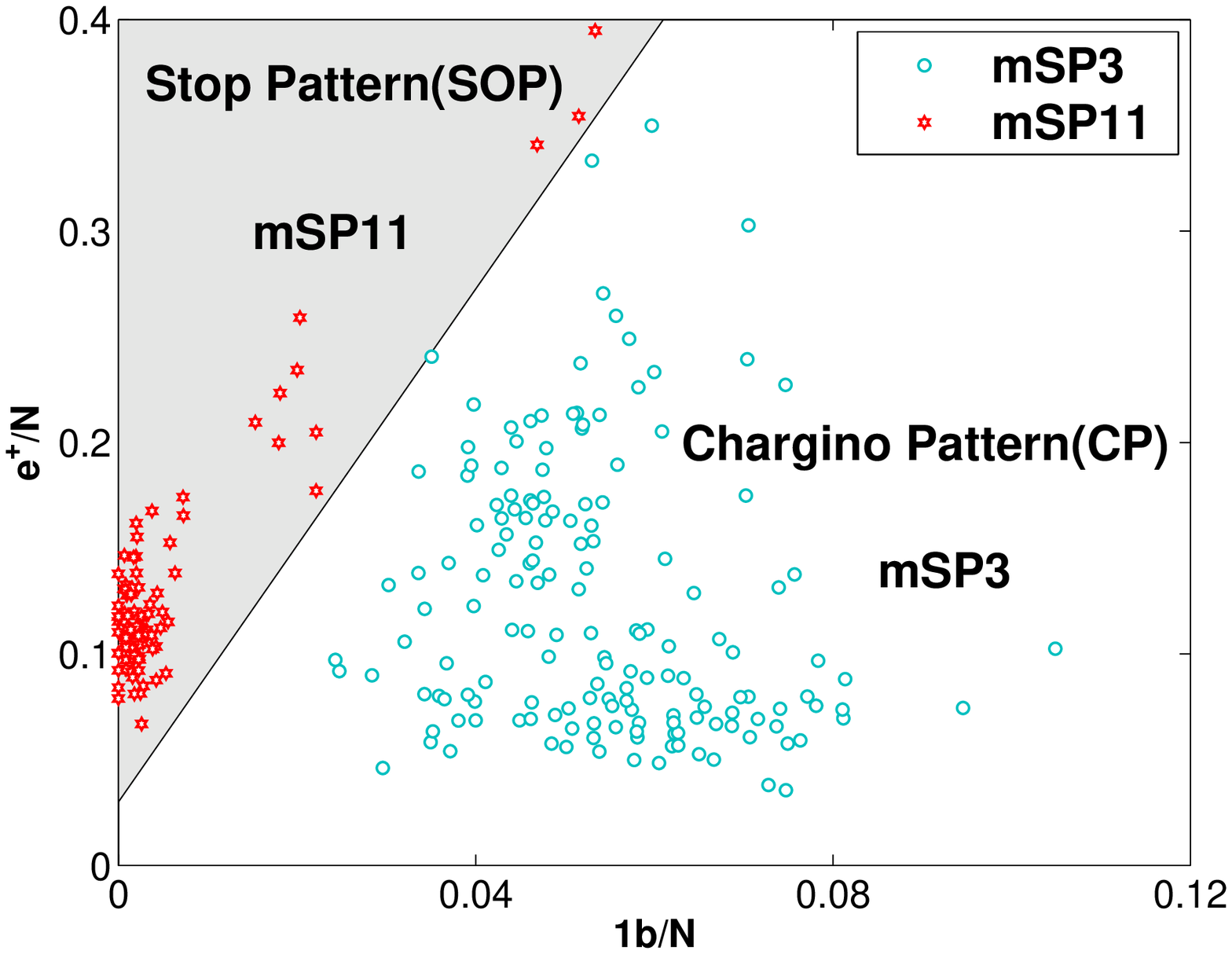}
\caption{Discrimination among mSPs within both mSUGRA and
NUSUGRA models.} \label{fig:sugrafig}
\end{figure}

In this section, we give an analysis including \non in three different sectors:
NUH, NU3, and NUG.  In our analysis we simulate various models with the
same constant number of events N which we take, as an example to be N$= 10^4$.
To discriminate among the  patterns in the signature space,
we introduce another set of post trigger cuts, which we denote as `b jet cuts',
in addition to  the default post trigger cuts specified in previous chapter.
The criteria in the b-jet cuts are the same as the default post trigger cuts, except that
we change the condition `at least two hadronic jets in the event' to
`specifically at least one b-tagged jet in the event'.
We exhibit our analysis utilizing both the default cuts and the b jet cuts
in  Fig.(\ref{fig:sugrafig}). Two  mSPs are presented in each figure in
different signature spaces to show the separation for each case.
Signals are simulated with constant number of events in PGS4
for each pattern. One can see that even with inclusion of
a variety of soft breaking scenarios, some mSPs still have
very distinct signatures in some specific channels.

Thus in the top-left panel of Fig.~(\ref{fig:sugrafig}) we give a plot of mSP7 (SUP) and
mSP11 (SOP) in the signature space 1L/N (b jet cuts) vs 0L1b/N,
where 0L1b/N is obtained with the default post trigger cuts.
Here we find that these two model types are clearly distinguishable
as highlighted by shaded and unshaded regions. A similar analysis with
signatures consisting of 1L1b/N (b jet cuts) vs 0L1b/N for
mSP4 (CP) and mSP7 (SUP) is given in the top-right panel.
The lower-left panel gives an analysis of mSP4 (CP) and mSP5 (SUP)
also in the signature space consisting of 1L1b/N (b jet cuts) vs 0L1b/N.
Finally, in the lower-right panel we give an analysis of mSP3 (CP) and mSP11 (SOP)
in the signature plane $e^+$/N vs 1b/N.  These analyses
illustrate that the patterns and often even the sub-patterns can be discriminated
 with the appropriate choice of signatures for a general class of SUGRA models
including \non.


\section{The Trileptonic Signal as a Pattern Discriminant \label{D6} }

\begin{figure*}[htb]
  \begin{center}
\includegraphics[width=9.0cm,height=7.0cm]{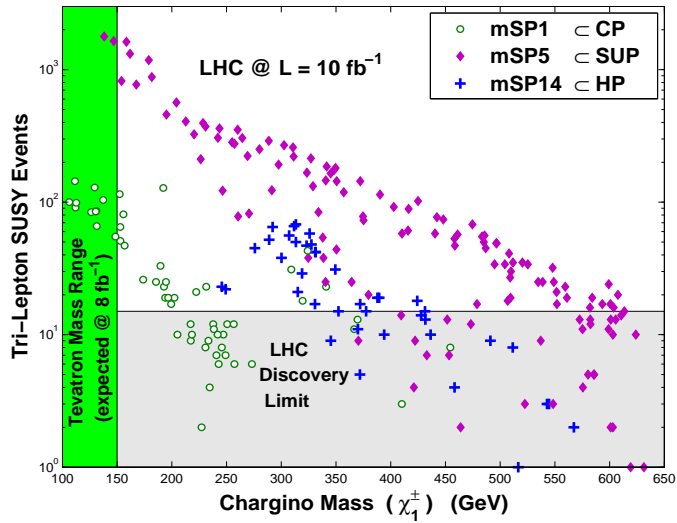}
\caption{
A plot of the number of trilepton events versus  the light chargino mass for three patterns,
one from each class, CP, SUP and HP. The SUP pattern gives the largest trileptonic signal
followed by the HP and CP patterns.
}
\label{fig:sigmass2}
  \end{center}
\end{figure*}

 \begin{figure*}[htb]
  \begin{center}
\includegraphics[width=7.0cm,height=6.0cm]{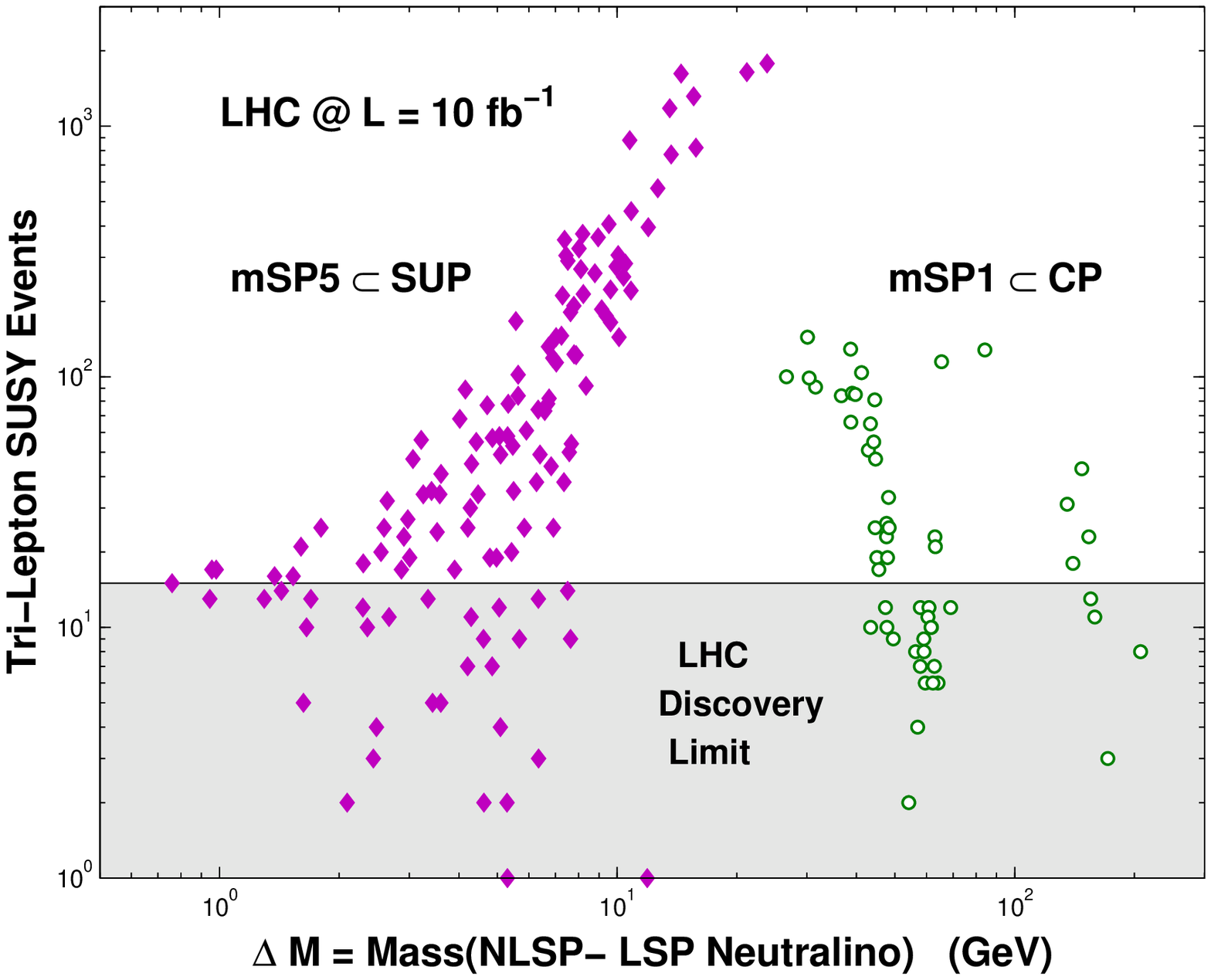}
\includegraphics[width=7.0cm,height=6.0cm]{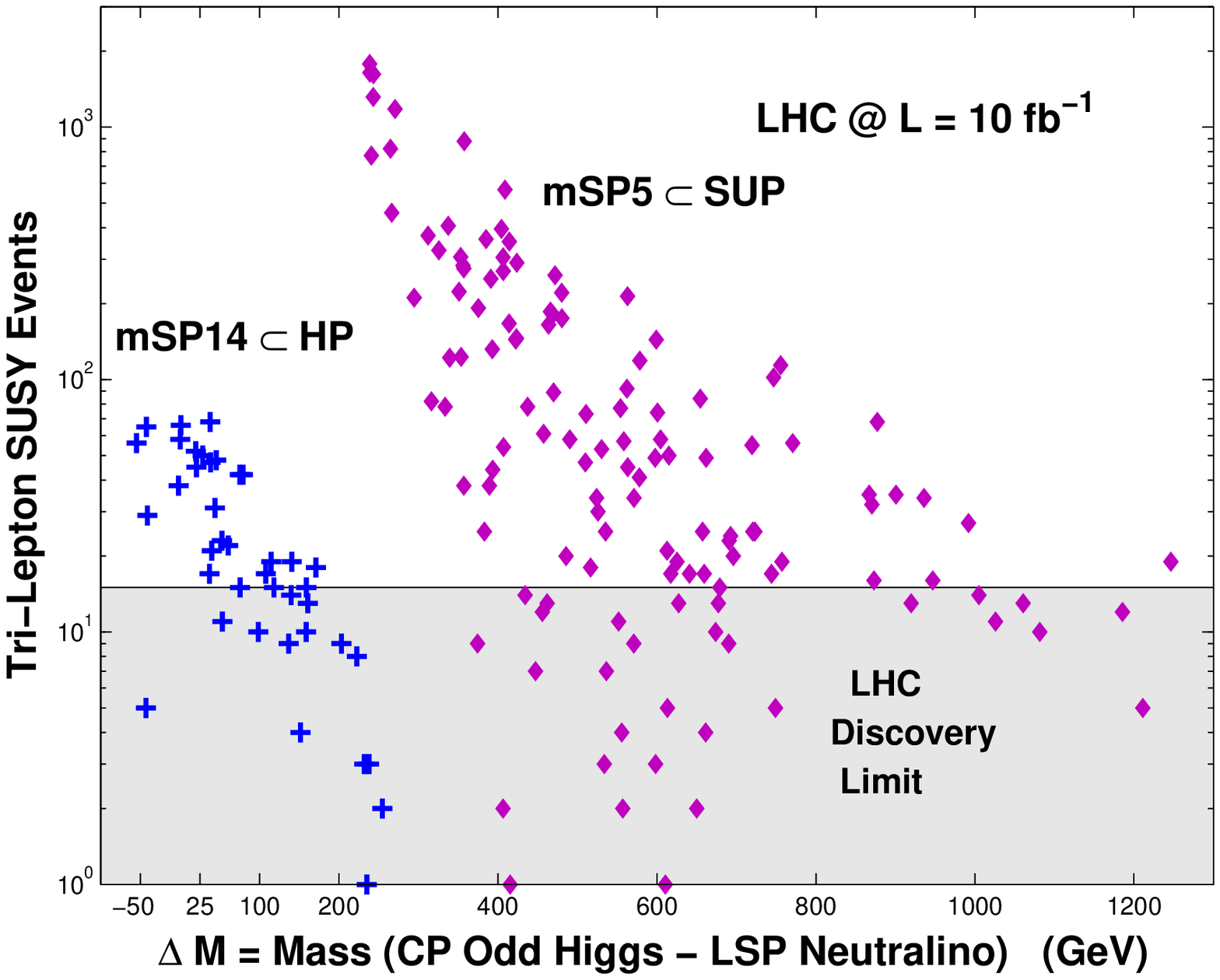}
\caption{The number of tri-lepton events versus  the sparticle mass splittings. }
\label{fig:sigmass1}
  \end{center}
\end{figure*}

The trileptonic signal is an important signal for the discovery of supersymmetry.
For on-shell decays the trileptonic signal was discussed in the early days in
\cite{earlypheno,Baer:1986vf} and for off-shell decays in \cite{Nath:1987sw}.
(For a recent application see \cite{CMSnote3}).
Here we discuss the trileptonic signal in the context of discrimination
of hierarchical  patterns.
In Fig.~(\ref{fig:sigmass2}) we exhibit the dependency of the trilepton
signal on the chargino mass. It is seen  that mSP5 gives
the largest number of events in this channel while
 the CP pattern (mSP1) and the HP pattern (mSP14)
can also produce a large number of trilepton events above
the discovery limit,  while the chargino mass reach is extended for the mSP5
as opposed to the mSP1 and mSP14. The above observations
hold for some of the other SUP patterns as well.
Thus the trileptonic signal is strong enough to be probed up to  chargino
masses of  about 500~\rm GeV in the SUP pattern.
Another interesting display of the trileptonic signal is when this signal is
plotted against some relevant mass splittings.
The left panel of Fig.~(\ref{fig:sigmass1}) shows clear separations
for hierarchical mass patterns in the number of trilepton events
produced with  $10~\rm fb^{-1}$ as a function of the NLSP and the  LSP
mass splitting for the chargino (CP) pattern mSP1 and Stau (SUP) mSP5.
The plot on the right shows a similar effect for the case where the mass splitting is taken
to be the difference of the CP odd Higgs boson mass and the LSP
for both the Higgs pattern mSP14 and the stau pattern mSP5.
The Standard Model background is highly suppressed in this channel.
Thus  the left-panel of Fig.~(\ref{fig:sigmass1}) gives an analysis for the trileptonic signal for
two patterns:  the Chargino pattern mSP1
and the Stau pattern mSP5 plotted against the  NLSP-LSP mass
 splitting  with  $10~\rm fb^{-1}$ of data.
 
The analysis of  the left-panel of Fig.~(\ref{fig:sigmass1}) shows that
the SUP pattern
presents an excellent opportunity for discovering SUSY through the 3
lepton mode.  The analysis also shows a clear separation  among mass patterns
and further a majority of the model points stand above the discovery limit which
in this channel is  $\approx 15$  events under the post trigger level
cuts discussed previously. 
The right-panel of
Fig.~(\ref{fig:sigmass1}) gives an analysis of the trileptonic signal
 vs the mass splitting of the CP odd Higgs and the
lightest neutralino LSP for patterns mSP5  and mSP14.
Again, we see a clear separation of model points.
We note that CP odd Higgs can sometimes be even
 lighter than the LSP, and thus the quantity $\Delta M= M_A-M_{\tilde \chi_1^0}$ plotted
 on the x-axis can sometimes become  negative.

\chapter{Kinematic Signatures of Sparticles at the LHC}
\label{ch:kin}

In this chapter, we discuss various kinematical variables that are useful for 
pattern discrimination. Usually, the kinematic distributions require much 
higher luminosity than the counting signatures, so it is unlikely that SUSY 
is first discovered with kinematical distributions. But the kinematic distributions 
carry more precise information which is essential for determining the detailed 
structure of SUSY models.

Typically, there are two types of kinematic distributions, 
$P_T$ distributions and invariant mass distributions. 
The $P_T$ distributions we investigate include missing $P_T$ distribution 
and the effective mass distributions. 
The invariant mass distribution analyzed here is  
the opposite sign same flavor dilepton invariant mass distribution.

\section{Transverse Momentum Distributions}

\begin{figure}[htb]
  \begin{center}
\includegraphics[width=9.0cm,height=8.0cm]{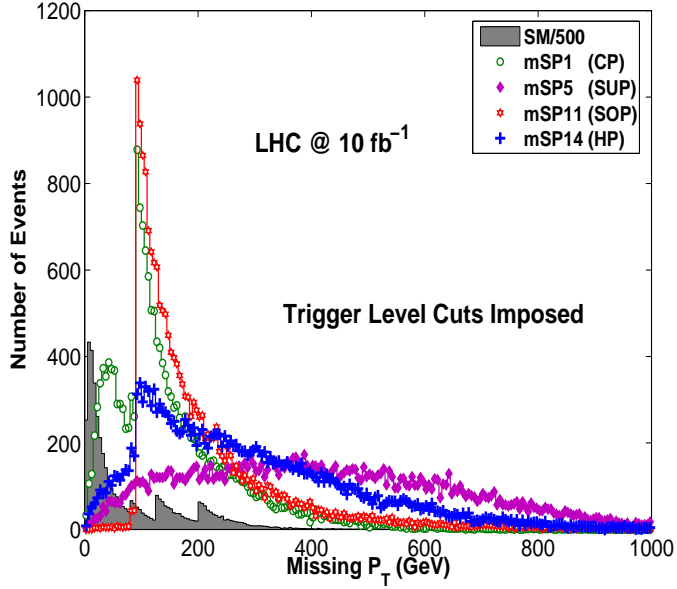}
\caption{An exhibition of the missing $P_T$ 
distributions for 4 different mSUGRA models
with each corresponding to one class of mSPs, and for the Standard Model.
Only trigger level cuts are employed here.}
\label{fig:ptmiss}
\end{center}
\end{figure}

The kinematical signatures are important for pattern discrimination in addition to the event 
counting signatures discussed previously. We illustrate this using the kinematical variables consisting of
missing $P_T$ and the effective mass (see Table (\ref{tab:counting}) for their definitions)
and an illustration is given in Fig.(\ref{fig:ptmiss}).
\begin{figure}[htb]
  \begin{center}
\includegraphics[width=9.0cm,height=8.0cm]{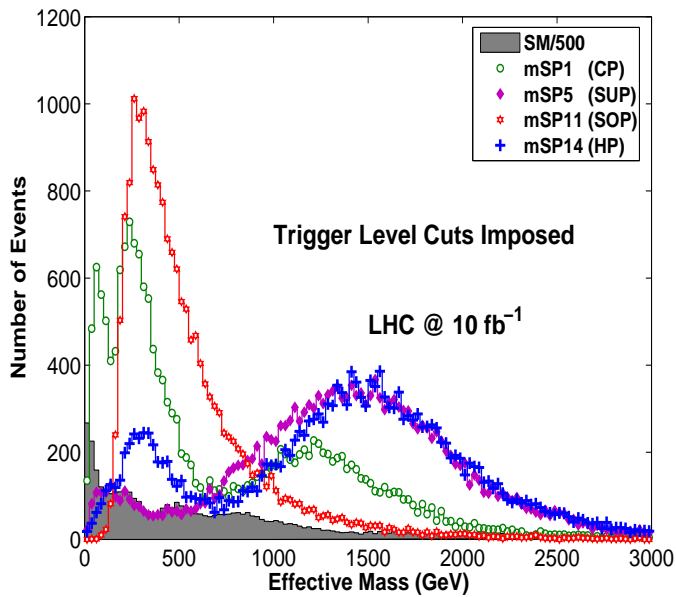}
\caption{An exhibition of the effective mass
 distributions for the same mSUGRA models
as shown in Fig.~(\ref{fig:ptmiss}). 
Only trigger level cuts are employed here.}
\label{fig:meff}
\end{center}
\end{figure}
\begin{figure}[htb]
  \begin{center}
\includegraphics[width=9.0cm,height=8.0cm]{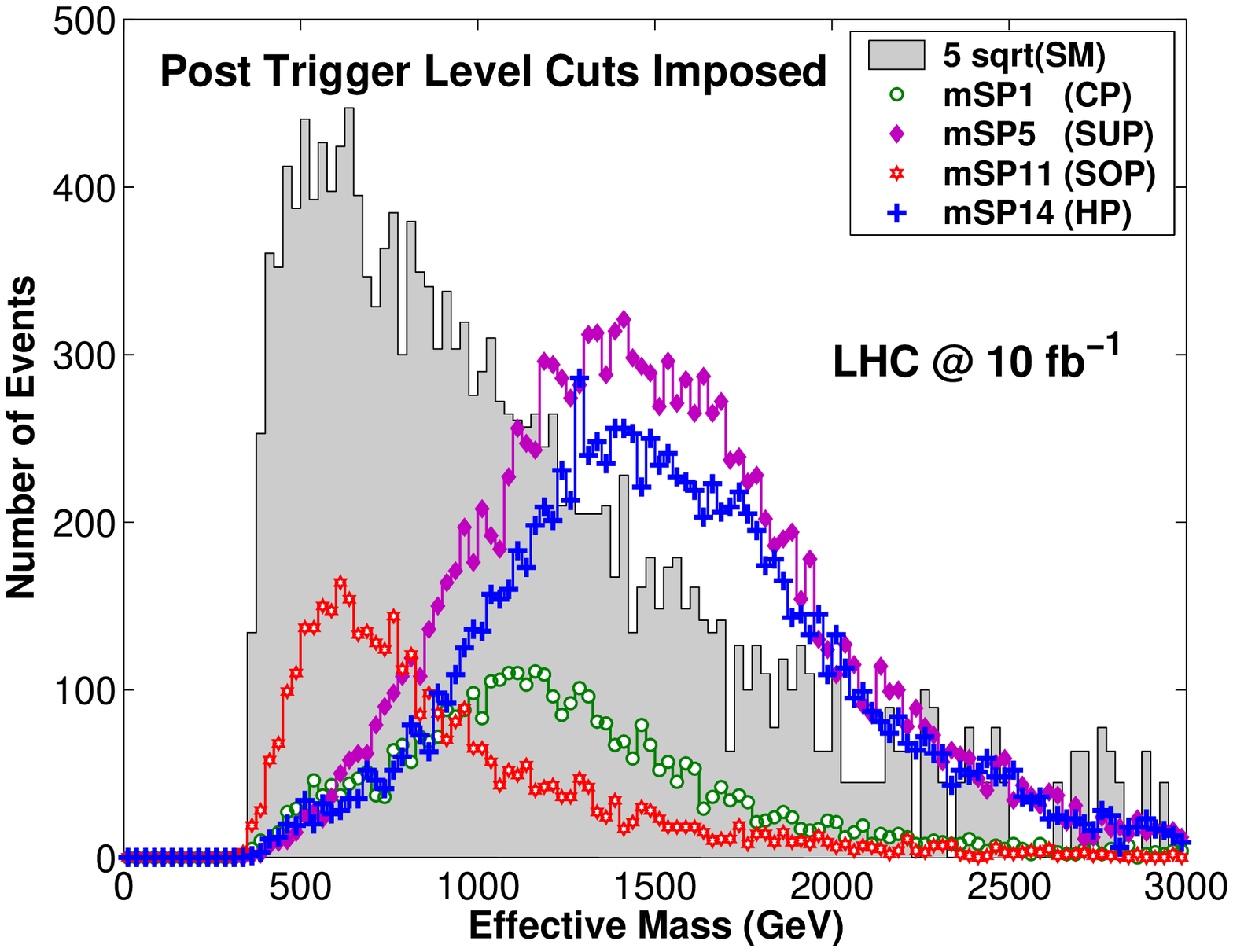}
\caption{
The effective mass distributions for 4 different mSUGRA models
with each corresponding to one class of mSPs, and for the Standard Model.
Post trigger level cuts are imposed here. The bin size used here is 25 GeV.
}
\label{fig:meff-cut}
\end{center}
\end{figure}
Specifically the analysis of Fig.(\ref{fig:ptmiss}) uses  four mSUGRA points one each 
in the patterns CP, SUP, SOP and HP. We exhibit the mSUGRA points used here in
the order ($m_0$, $m_{1/2}$, $A_0$, $\tan\beta$, sign$\mu$):
\begin{equation}
\begin{array}{rrrrrrr}
{\rm CP~~Point}   &   ( 3206.9, & 285.3,  & -1319.8, & 9.7, & +1) ,\\
{\rm SUP~~Point} &  ( 92.6,     & 462.1,  & 352.2,    & 4.5, & +1) ,\\
{\rm SOP~~Point} &  (  2296.9, & 625.0,     &-5254.9,  & 13.6, & +1) ,\\
{\rm HP~~Point}   &  ( 756.8,     & 387.0,     &1144.9,   & 56.5, & +1) .\\
\end{array}\label{4msps}
\end{equation}
In the missing $P_T$ distribution the Standard Model tends to produce events with a
lower missing  $P_T$ relative to the  mSUGRA case which
generates events at relatively higher missing $P_T$. 
Further,  there is  a large variation between different mSUGRA models, as  can be seen 
in Fig.~(\ref{fig:ptmiss}). Thus, for example mSP5 (a stau pattern) 
and mSP14 (a Higgs pattern) have peaks at larger values of missing $P_T$ 
relative to mSP1 (a chargino pattern) and mSP11 (a stop pattern).
Additionally, the shapes of the distributions are also different.

The analysis of effective mass distribution in Fig.~(\ref{fig:meff}) 
is carried out with the same 
mSUGRA model points as in Fig.~(\ref{fig:ptmiss}). 
And it is found that in the effective mass distribution, 
the Standard Model tends to produce events with a lower 
effective mass relative to the mSUGRA models, and the variation 
between mSUGRA models remain similar to the case as in Fig.~(\ref{fig:ptmiss}).

The analysis of Fig.~(\ref{fig:ptmiss}) and Fig.~(\ref{fig:meff})  shows  that
the distributions for the CP, HP, SOP and SUP are substantially different.
It is  interesting  to note that
in the missing $P_T$ distribution, the HP and SUP model points have
a relatively flat distribution compared to the CP and SOP model points.
The missing $P_T$ distribution and the effective
mass distribution are useful when designing post trigger level
cuts to optimize the signal over the background.
For instance, one can take a 1 TeV effective mass cut to analyze the
SUP and HP signals shown in Fig.(\ref{fig:meff}), but this method 
will not work well when it comes to the CP and SOP points since most of their events
have a rather small effective mass.  To illustrate that different models
have different effective mass distributions, and consequently
different effective mass cuts
are needed for different patterns, an analysis is given in Fig.(\ref{fig:meff-cut})
for the same set of points in Fig.(\ref{fig:meff}) with post trigger level cuts imposed.

\section{Invariant Mass Distributions}

\begin{figure}[htb]
  \begin{center}
\includegraphics[width=9.0cm,height=8.0cm]{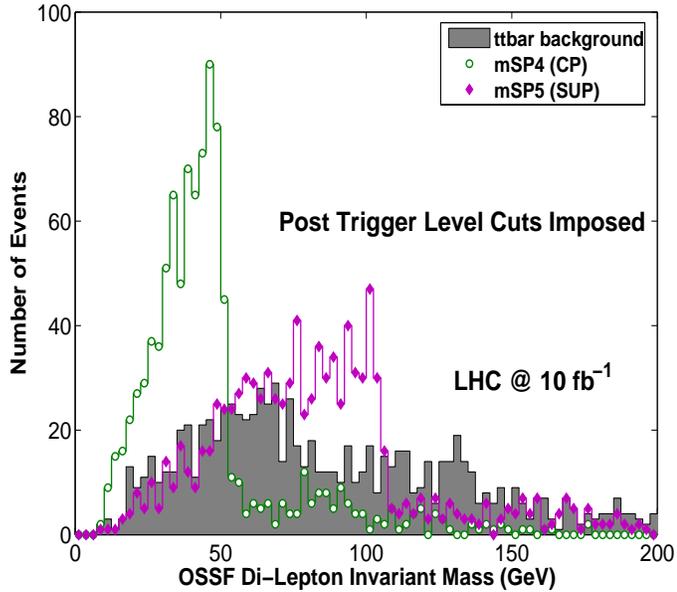}
\caption{
A plot of the opposite sign same flavor (OSSF) di-lepton invariant
mass distribution at LHC with 10 fb$^{-1}$ with the default post trigger cuts
imposed for two different mSP points on top of the SM $t\bar t$ background.}
\label{kinmass}
  \end{center}
\end{figure}

We also investigate the invariant mass distribution
for the opposite sign same flavor (OSSF) di-leptons ($e^+e^-, \mu^+\mu^-$)
in Fig.(\ref{kinmass}).
We applied the default post trigger cuts as discussed previously to suppress
the SM background. As a comparison  the dominant Standard Model
$t\bar t$ background is also exhibited. We have cross checked our work with the CMS Note
\cite{CMSnote1}, and found good agreement regarding the SUSY signals and
the Standard Model background. It is seen that the two mSP points,  
\begin{equation}
\begin{array}{rrrrrrr}
{\rm mSP4 ~~Point}   &   ( 1674.9, & 137.6,  & 1986.5, & 18.6, & +1) ,\\
{\rm mSP5 ~~Point}   &   ( 84.4, & 429.3,  &  -263, & 3.4, & +1) ,\\
\end{array}\label{2msps}
\end{equation}
plotted in Fig.~(\ref{kinmass}) are clearly distinguishable from each other in the distribution.
An analysis of invariant mass distribution is given in the Appendix. 
Here, we apply the general formula given in the Appendix 
 to one specific interesting SUSY 
decay chains $\nb \to \ell^{\pm}\sellpm \to\ell^{\pm}\ell^{\mp}\na$
\begin{equation}
M_{\ell\ell}^{\rm max}=M_{\nb}\sqrt{1-\frac{M_{\sell}^2}{M_{\nb}^2}}
\sqrt{1-\frac{M_{\na}^2}{M_{\sell}^2}}.
\end{equation}\label{mll}
For mSP5 model point plotted here, the relevant branching ratios are
\begin{equation}
BR(\nb\to\sell+\ell) \simeq 23.5\%~~{\rm and}~~~BR(\sell\to\na+\ell) \simeq 100\%
\end{equation}
where $\sell$ are $\el$ and $\ml$, and $\ell$ are electron and muon.
The relevant sparticle masses are $M_{\sell}=300.9$ GeV, 
$M_{\nb}=327.4$ GeV, and $M_{\na}=181.9$ GeV. 
Therefore, the maximum value of the invariant mass of dilepton is 
\begin{equation}
M_{\ell\ell}^{\rm max}=327.4\sqrt{1-\frac{300.9^2}{327.4^2}}
\sqrt{1-\frac{181.9^2}{300.9^2}}=102.8~~~{\rm GeV}\label{msp5mass}
\end{equation}
which is consistent with the result of Fig.~(\ref{kinmass}).

For the mSP4 model point plotted here, 
since the sfermion masses are quite large,  
the SUSY production is dominated by the 
Ino-production, especially $\g$, $\nb$, and $\cha$. The decay mode which is 
responsible for most of the production of the dilepton events is 
\begin{equation}
BR(\nb\to\na+\ell^++\ell^-) = 5.66\%.
\end{equation}
For this model point, the lightest neutralino masses are 
$M_{\na}=54.6$ GeV and $M_{\nb}=107.4$ GeV. 
Here the decay process is realized through the off-shell slepton decay, 
since the on-shell slepton masses are above TeV. 
When the off-shell mass of the corresponding slepton is 
$m_{\rm offshell}=\sqrt{M_{\nb}M_{\na}}$, 
the invariant mass edge value achieves its maximum value
\begin{equation}
(M_{\ell\ell}^{\rm max})^{\rm max}=M_{\nb}-M_{\na}
=107.4-54.6=52.8~~~{\rm GeV}\label{msp4mass}
\end{equation}
which is again consistent with the result of Fig.~(\ref{kinmass}).
\begin{figure}[htb]
\begin{center}
 	\begin{fmffile}{offshell}
		\begin{fmfgraph*}(160,80)
			\fmfstraight
			\fmfleft{i1}
			\fmfright{o1,o2,o3}
			\fmf{fermion,tension=3,label=$\ell$}{v1,i1}
			\fmf{dashes,tension=3,label=$\sell_{\rm offshell}$}{v1,v2}
			\fmfblob{.05w}{v1}
			\fmfblob{.05w}{v2}
			\fmf{fermion,label=$\na$}{v2,o1}
			\fmf{dashes}{v2,o2}
			\fmf{fermion,label=$\ell$}{v2,o3}	
			\fmfv{label=$\nb$,label.angle=90,label.dist=10}{v1}
			\fmfv{label=$\theta$,label.angle=22,label.dist=20}{v2}
			\fmflabel{$\hat{Z}$}{o2}
		\end{fmfgraph*}
	\end{fmffile}
	\caption{Decay process via off-shell slepton.}
	\label{fig:offshell}
	\end{center}
\end{figure}
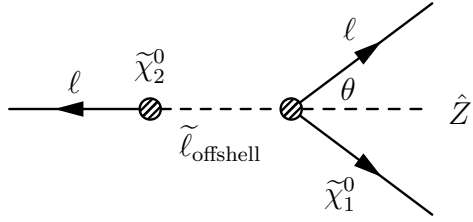
The off-shell mass of the slepton $m=\sqrt{M_{\nb}M_{\na}}
=\sqrt{107.4*54.6}=76.6~{\rm GeV}$ gives rise to the edge of the OSSF 
dilepton invariant mass distribution. 
When the off-shell slepton has an off-shell mass roughly the half way 
between the $\na$ and $\nb$, both leptons are likely to gain 
enough transverse momentum in order to pass the detector cuts on 
lepton $P_T$.

The mSP4 model point shown here has recently been investigated  \cite{Gounaris:2008pa} in
the context of helicity  amplitudes as a discovery mechanism for supersymmetry.

\chapter{Global Analysis of Sparticle Patterns: Fuzzy Signature Vectors}
\label{ch:fuzzy}

In this chapter, we discuss how one may distinguish sparticle patterns utilizing 
all signature channels that are available. 
We have given specific examples of how  patterns can be differentiated  from each other. 
In the  previous analysis we used only  a few of the 40 signatures exhibited in Table (\ref{tab:counting}). 
Here we want to examined all of them. 

Thus for each parameter point we have analyzed 40 signatures. 
We now define correlations among these  signatures.  
Thus consider an ordered set where the signatures are labeled $S_1, S_2, .., S_{40}$
and let the number of events in each  signature be $N_1, N_2, .., N_{40}$. 
Define a signature vector for a given point $x_{\alpha}$ ($\alpha =1, 2, .., p$) in the parameter space
\begin{equation}\xi^a=(\xi_1^a, \xi_2^a, .., \xi_{40}^a)\end{equation}
where $\xi_i=N_i^a/N$ and $N$ is the total number of SUSY events. As the parameter point 
$x_{\alpha}$ varies  over the allowed range within a given pattern it generates a signature vector 
where the elements trace out a given range. Thus  for a pattern X one generates a  fuzzy pattern 
vector $\Delta\xi^X$ so that
\beqn\Delta \xi^X=(\Delta\xi_1^X, \Delta\xi_2^X, .., \Delta\xi_{40}^X),\eeqn
where $\Delta \xi_i^X$ is the range traced  out by the element $\xi_i^X$ as the parameter point 
$x_{\alpha}$ moves in the allowed parameter space of the pattern X. What makes the vector 
$\Delta \xi^X$ fuzzy is that its elements are not single numbers but a set which cover a range.
We define now the inner product of two  such fuzzy pattern vectors  so that
 \beqn C_{XY}\equiv (\Delta\xi^X|\Delta\xi^Y)=0(1)\label{inner}\eeqn
where the inner product is 0 if the element $\Delta\xi_i^X$ and $\Delta\xi_i^Y$ overlap 
for all $i$  $( i=1,.,40)$, and 1 if at least one of the elements of pattern X, $\Delta \xi_j^X$
does not overlap with $\Delta\xi_j^Y$,  the element for pattern Y. Therefore,  if for two patterns
X and Y one finds there is no overlap at least for one signature component
$\Delta\xi_j$, then these two patterns can be distinguished in this
specific signature and one  obtains  $C_{XY}=1$. Otherwise $C_{XY}=0$
which means that all components of
$\Delta\xi^X$  and $\Delta\xi^Y$  have an overlap  and cannot be distinguished under
this criteria.

We can generalize the above procedure for the signatures
\begin{equation}
\zeta_{i,j}=\frac{N_i}{N_j}, ~~ (i,j=1,...,40). \label{tensor}
\end{equation}
Repeating the previous analysis, one can construct another fuzzy signature
vector for pattern X as
\begin{equation}
\Delta\zeta^X=(\Delta\zeta_{1,2}^X, ..,\Delta\zeta_{i,j}^X, ..,
\Delta\zeta_{39,40}^X)
\label{tensor1}
\end{equation}
where the elements have a range corresponding to the range spanned by
the soft parameters $x_{\alpha}$ as they move over the parameter space
specific to the pattern.
Further, the definition of the inner product Eq.~(\ref{inner}) still holds
for this new fuzzy signature vector.
We have carried out a full signature analysis of such comparisons, using 40
different signatures, and their combinations as defined in
Eq.~(\ref{tensor}) and Eq.~(\ref{tensor1}).
\begin{table}[htbp]
    \begin{center}
\begin{tabular}
{|p{0.8cm}|p{0.8cm}|p{0.8cm}|p{0.8cm}|p{0.8cm}|p{0.8cm}
|p{0.8cm}|p{0.8cm}|p{0.8cm}|p{0.8cm}|p{0.8cm}|p{0.8cm}|}\hline
       &m5&m1&m3&m7&m11&m6&m12&m13&N1&m4&m18\\\hline
m5  &0&0&0&0&1&0&1&1&0&1&0 \\\hline
m1  &0&0&0&0&0&0&1&0&0&1&0\\\hline
m3  &0&0&0&0&1&0&1&1&0&0&0\\\hline
m7  &0&0&0&0&1&0&1&1&0&1&0\\\hline
m11&1&0&1&1&0&0&1&1&1&0&1\\\hline
m6  &0&0&0&0&0&0&1&0&0&0&0\\\hline
m12&1&1&1&1&1&1&0&0&1&1&1\\\hline
m13&1&0&1&1&1&0&0&0&1&1&0\\\hline
N1  &0&0&0&0&1&0&1&1&0&1&1\\\hline
m4  &1&1&0&1&0&0&1&1&1&0&1\\\hline
m18&0&0&0&0&1&0&1&0&1&1&0 \\\hline
\end{tabular}
\caption{A table exhibiting the discrimination of patterns using the criterion of  Eq.(\ref{inner})
where various signatures with both the default post trigger cuts and b jet cuts are utilized.
If the element of $i^{\rm th}$ row and $j^{\rm th}$ column is 1, i.e., $C_{i j} = 1$,
one can distinguish the $i^{\rm th}$ mass pattern from the $j^{\rm th}$ one. Here 
m stands for mSP, and N for NUSP. Thus shown here is the discrimination table between the 
patterns: mSP5, mSP1, mSP3, mSP7, mSP11, mSP6, mSP12, mSP13, NUSP1, mSP4, mSP18. 
The order of the patterns indicates how often these patterns appear in our Monte Carlo scan. 
} 
\label{tab:spdis-1}
\end{center}
\end{table}

An illustration of the global analysis is given in 
Tables (\ref{tab:spdis-1}, \ref{tab:spdis-2}, \ref{tab:spdis-3}). 
We carry out this analysis with a large collection of SUGRA model 
points which belong to 22 different hierarchical mass patterns. 
The complete set of the LHC signatures are obtained with 
the default post trigger cuts as well as the b jet cuts as specified in 
chapter (\ref{ch:sig}). We have roughly divided the 22 sparticle patterns 
into two equal size sets, and classified the more probable patterns into the first set:  
mSP5, mSP1, mSP3, mSP7, mSP11, mSP6, mSP12, mSP13, NUSP1, mSP4, mSP18; 
and the less probable patterns into the second set: 
NUSP13, mSP20, mSP10, mSP17, NUSP3, mSP19, 
NUSP5, NUSP8, NUSP10, NUSP4, NUSP9. 
The global analysis within these two sets is exhibited in  
Tables (\ref{tab:spdis-1}, \ref{tab:spdis-2}), 
and the analysis between these two sets is exhibited in Tables (\ref{tab:spdis-3}). 
Altogether, Tables (\ref{tab:spdis-1}, \ref{tab:spdis-2}, \ref{tab:spdis-3}) show 
whether or not one can distinguish any pair of patterns chosen from the 22 different 
patterns utilizing the signatures investigated here. 
\begin{table}[htbp]
    \begin{center}
\begin{tabular}
{|p{0.8cm}|p{0.8cm}|p{0.8cm}|p{0.8cm}|p{0.8cm}|p{0.8cm}
|p{0.8cm}|p{0.8cm}|p{0.8cm}|p{0.8cm}|p{0.8cm}|p{0.8cm}|}\hline
       &N13&m20&m10&m17&N3&m19&N5&N8&N10&N4&N9\\\hline
N13  &0&1&1&1&1&1&1&1&1&1&1 \\\hline
m20  &1&0&1&1&1&1&1&1&1&1&1\\\hline
m10  &1&1&0&1&1&1&1&1&1&1&1\\\hline
m17  &1&1&1&0&1&1&1&1&1&1&1\\\hline
N3    &1&1&1&1&0&1&1&1&1&1&1\\\hline
m19  &1&1&1&1&1&0&1&1&1&1&1\\\hline
N5    &1&1&1&1&1&1&0&1&1&1&1\\\hline
N8    &1&1&1&1&1&1&1&0&1&1&1\\\hline
N10  &1&1&1&1&1&1&1&1&0&1&1\\\hline
N4    &1&1&1&1&1&1&1&1&1&0&1\\\hline
N9    &1&1&1&1&1&1&1&1&1&1&0 \\\hline
\end{tabular}
\caption{A discrimination table between the 
patterns: NUSP13, mSP20, mSP10, mSP17, NUSP3, mSP19, 
NUSP5, NUSP8, NUSP10, NUSP4, NUSP9.} 
\label{tab:spdis-2}
\end{center}
\end{table}

The analysis shows that it is possible to often distinguish patterns 
using the criterion of Eq.(\ref{inner}).
We note that the analyses exhibited in  Fig.(\ref{fig:sugrafig}) 
are the special cases of the results in Tables (\ref{tab:spdis-1}, \ref{tab:spdis-2}, \ref{tab:spdis-3}). 
For instance, the clear separation between mSP7 and mSP11 in the signature
space shown in the top-left panel of Fig.(\ref{fig:sugrafig}) gives the
elements $C_{45}=C_{54}=1$ of Table (\ref{tab:spdis-1}). 
As indicated in Table (\ref{tab:spdis-2}) all the patterns analyzed here 
can be discriminated from each other. This is not really surprising, because 
the probability of finding the sparticle patterns shown in Table (\ref{tab:spdis-2}) 
are not big, and each pattern here is not analyzed with enough model points 
in order to gain sufficient statistics. 
As emphasized already the analysis of 
in Tables (\ref{tab:spdis-1}, \ref{tab:spdis-2}, \ref{tab:spdis-3}) 
is for illustrative purposes as we used a random sample of 22 patterns out of 37.
Inclusion of each additional mass pattern brings in a significant set of model
points which need to be simulated, and here one is limited by computing power.
\begin{table}[htbp]
    \begin{center}
\begin{tabular}
{|p{0.8cm}|p{0.8cm}|p{0.8cm}|p{0.8cm}|p{0.8cm}|p{0.8cm}
|p{0.8cm}|p{0.8cm}|p{0.8cm}|p{0.8cm}|p{0.8cm}|p{0.8cm}|}\hline
       &N13&m20&m10&m17&N3&m19&N5&N8&N10&N4&N9\\\hline
m5  &1&1&1&0&1&1&1&1&1&0&1\\\hline
m1  &1&1&1&0&0&1&1&1&1&0&1\\\hline
m3  &1&1&1&1&0&1&1&1&1&1&1\\\hline
m7  &1&1&1&1&0&1&1&1&1&0&1\\\hline
m11&1&1&1&1&1&1&1&1&1&1&1\\\hline
m6  &1&1&1&0&0&1&1&1&1&0&1\\\hline
m12&1&1&1&1&1&1&1&1&1&1&1\\\hline
m13&1&1&1&1&1&1&1&1&1&1&1\\\hline
N1  &1&1&1&1&1&1&1&1&1&1&1\\\hline
m4  &1&1&1&1&1&1&1&1&1&1&1\\\hline
m18&1&1&1&1&1&1&1&1&1&1&1\\\hline
\end{tabular}
\caption{A discrimination table between the two sets of 
patterns exhibited in the previous Tables (\ref{tab:spdis-1}, \ref{tab:spdis-2}).
} 
\label{tab:spdis-3}
\end{center}
\end{table}
The full analysis  including all the patterns can 
be implemented along similar lines with the necessary computing  power.
Finally we note that the analysis 
in Tables (\ref{tab:spdis-1}, \ref{tab:spdis-2}, \ref{tab:spdis-3}) 
is done without statistical uncertainties.
Inclusion of uncertainties in pattern analysis would certainly be worthwhile in a future work.

\chapter{Signature Degeneracies and Resolution of Soft Parameters}
\label{ch:resolution}

\section{Lifting Signature Degeneracies  \label{E1}}
It may happen that two distinct points in the soft parameter space
may lead to the same  set of signatures for a given integrated
luminosity within some predefined notion of indistinguishability.
Thus consider two parameter points $A$ and $B$ and define
the `pulls' in each of their signatures by
\begin{eqnarray}
P_i & = & \frac{|n^A_i-n^B_i|}{\sigma_{AB}},\nonumber\\
\sigma_{AB} & = & \sqrt{(\delta n^A_i)^2+ (\delta n^B_i)^2+ (\delta
n_i^{SM})^2}. \label{pulls1}
 \end{eqnarray}
Here $\delta n_i^A\sim\sqrt{n_i^A}$ is the uncertainty in the signature
events $n_i^A$, and we estimate the SM uncertainty as $\delta
n_i^{SM}\sim\sqrt{y}(\delta n_i^A+\delta n_i^B)/2$. Here the
parameter $y$ parameterizes the effect of the SM events, and for the analysis
in this section, we take $y=1$. In other words, if the pull in each of the
signatures is less than 5, then the two SUGRA 
\begin{table}[htb]\scriptsize{
\begin{center}
\begin{tabular}{|l|l||l|l|l|l|l|l||l|l|l|l|l|l|}\hline\hline
$i$&$S_i$&$A$&$B$&$P_i$&$A'$&$B'$&$P_i$&$A$&$B$&$P_i$&$A'$&$B'$&$P_i$\\\hline
0&N&743&730&0.3&878&817&1.2&35770&35570&0.6&45479&41135&12.1\\\hline
1&0L&430&414&0.4&484&437&1.3&20645&20490&0.6&25897&23427&9.1\\\hline
2&1L&221&230&0.3&294&271&0.8&10565&10410&0.9&13669&12414&6.3\\\hline
3&2L&78&71&0.5&83&96&0.8&3740&3945&1.9&4904&4369&4.5\\\hline
4&3L&10&13&0.5&16&11&0.8&725&675&1.1&927&830&1.9\\\hline
5&4L&4&2&0.6&1&2&0.4&95&50&3.1&82&95&0.8\\\hline
6&0T&620&610&0.2&731&674&1.2&29325&29860&1.8&38213&34138&12.4\\\hline
7&1T&112&104&0.4&137&129&0.4&5710&5125&4.6&6528&6296&1.7\\\hline
8&2T&11&14&0.5&10&14&0.7&685&540&3.4&693&659&0.8\\\hline
9&3T&0&2&1.1&0&0&0.0&45&40&0.4&43&40&0.3\\\hline
10&4T&0&0&0.0&0&0&0.0&5&5&0.0&2&2&0.0\\\hline
11&TL&38&26&1.2&50&45&0.4&1730&1595&1.9&2069&2029&0.5\\\hline
12&OS&59&57&0.2&66&70&0.3&2785&2910&1.4&3665&3285&3.7\\\hline
13&SS&19&14&0.7&17&26&1.1&955&1035&1.5&1239&1084&2.6\\\hline
14&OSSF&40&46&0.5&49&52&0.2&2050&2140&1.1&2710&2389&3.7\\\hline
15&SSSF&7&9&0.4&10&13&0.5&435&480&1.2&537&481&1.4\\\hline
16&OST&7&8&0.2&5&9&0.9&420&340&2.4&428&402&0.7\\\hline
17&SST&4&6&0.5&5&5&0.0&265&200&2.5&265&257&0.3\\\hline
18&0L1b&50&59&0.7&61&56&0.4&2595&2695&1.1&3527&3387&1.4\\\hline
19&1L1b&45&39&0.5&48&53&0.4&1905&1800&1.4&2431&2268&1.9\\\hline
20&2L1b&9&8&0.2&15&21&0.8&585&660&1.7&853&778&1.5\\\hline
21&0T1b&86&88&0.1&100&110&0.6&4095&4260&1.5&5734&5353&3.0\\\hline
22&1T1b&21&15&0.8&22&20&0.3&1005&905&1.9&1150&1106&0.8\\\hline
23&2T1b&3&3&0.0&4&2&0.6&135&95&2.2&111&129&0.9\\\hline
24&0L2b&20&20&0.0&12&13&0.2&590&660&1.6&890&838&1.0\\\hline
25&1L2b&11&12&0.2&15&24&1.2&425&505&2.1&625&598&0.6\\\hline
26&2L2b&3&5&0.6&1&2&0.4&220&165&2.3&251&227&0.9\\\hline
27&0T2b&30&29&0.1&25&32&0.8&995&1120&2.2&1481&1379&1.6\\\hline
28&1T2b&4&6&0.5&4&6&0.5&245&205&1.5&300&297&0.1\\\hline
29&2T2b&0&2&1.1&0&1&0.7&25&35&1.0&28&27&0.1\\\hline
30&ep&71&71&0.0&93&83&0.6&3060&3010&0.5&4251&3957&2.6\\\hline
31&em&47&44&0.3&52&51&0.1&2135&1955&2.3&2618&2358&3.0\\\hline
32&mp&60&70&0.7&103&78&1.5&3360&3415&0.5&4236&3821&3.8\\\hline
33&mm&43&45&0.2&46&59&1.0&2010&2030&0.3&2564&2278&3.4\\\hline
34&tp&60&53&0.5&69&80&0.7&3255&2705&5.8&3564&3504&0.6\\\hline
35&tm&52&51&0.1&68&49&1.4&2455&2420&0.4&2964&2792&1.9\\\hline
36&0b&597&585&0.3&717&642&1.7&29045&28795&0.8&36432&32602&11.9\\\hline
37&1b&110&107&0.2&126&132&0.3&5250&5270&0.2&7003&6593&2.9\\\hline
38&2b&34&37&0.3&29&39&1.0&1265&1360&1.5&1810&1706&1.4\\\hline
39&3b&1&1&0.0&6&2&1.1&195&120&3.5&215&205&0.4\\\hline
40&4b&1&0&0.7&0&2&1.1&15&25&1.3&19&29&1.2\\\hline
\hline
\end{tabular}
\caption{
An exhibition of lifting the degeneracy of two points in the 
mSUGRA parameter space using luminosity.
Two pairs of points ($A$, $B$) and ($A'$, $B'$) are indistinguishable 
under the 2 sigma criteria at 10 fb$^{-1}$ luminosity (column 3-8), 
but can be clearly separated when the luminosity increases to
500 fb$^{-1}$ (column 9-14). 
The Standard Model uncertainty 
is estimated as $\delta n_i^{SM}= (\delta n_i^{A}+\delta n_i^{B})/2$.
}
\label{tbl:degelum}
\end{center}}
 \end{table}
parameter space points are 
essentially indistinguishable in the signature space. In such a situation
one could still distinguish model points  either by including more signatures,
or by an increase in luminosity. Thus, for example,
inclusion of the Higgs production cross sections, $B_s\to \mu^+\mu^-$ constraints,
as well as the inclusion of neutralino proton scattering cross section constraints
tend to discriminate among the model parameter points as shown in
Ref.~\cite{Feldman:2007fq}. Here we point out that in some cases
increasing the luminosity can allow one to lift the degeneracies
enhancing a subset of signatures in one case relative to the other.
For illustration we consider the following two sets of
points in the pattern mSP5 in the mSUGRA parameter space in the
following order ($m_0$, $m_{1/2}$, $A_0$, $\tan\beta$, sign$\mu$).
\begin{equation}
\begin{array}{rl}
{\rm Point ~A} & (192.6, 771.3, ~1791.1, 8.8, +1),\\
{\rm Point ~B} & (163.0, 761.3, -775.8, 4.7, +1);
\end{array}
\end{equation}
\begin{equation}
\begin{array}{rl}
{\rm Point ~A'} & (159.3, 732.3, -783.1, 5.6, +1),\\
{\rm Point ~B'} & (163.5, 753.3, -918.2, 3.3, +1).
\end{array}
\end{equation}
In  Table(\ref{tbl:degelum}) we compare the pulls for the pairs of
points ($A$, $B$) and ($A'$, $B'$) at an integrated luminosity of 10
fb$^{-1}$ and 500 fb$^{-1}$.
For points $A$ and $B$, one finds that  the pulls are all less
than 2 for an integrated  luminosity of 10 fb$^{-1}$.
However,  for an integrated luminosity of 500 fb$^{-1}$, the  pulls
for signatures $(5,7, 8, 34,39)$ increase significantly and the pull for signature
number 34 is in excess of 5 allowing one to discriminate between the
two parameter points $A$ and $B$. A very similar analysis is carried
out for parameter points $A'$ and $B'$. Here one finds that the
signature $(0, 1, 2, 3, 6, 12, 14, 32, 33, 36)$ receive a big
boost as we go from 10 fb$^{-1}$ to 500 fb$^{-1}$, and the
signatures $(0, 1, 2, 6, 36)$ give pulls greater than 5, with the largest
pulls being in excess of 12, allowing one to discriminate between
the parameter points $A'$ and $B'$.
 We note the analysis ignores
systematic errors and also does not consider an ensemble of simulations.
Nonetheless it  does  illustrate the effects of moving from
a  low to a high LHC luminosity
allowing one to discriminate some model pairs, which appear degenerate in
the signature space at one luminosity, but can become distinct from each other
at a larger luminosity.

\section{Resolving Soft Parameters using LHC Data\label{E2}}

\begin{figure*}[htb]
\centering
\includegraphics[width=7.0cm,height=6.0cm]{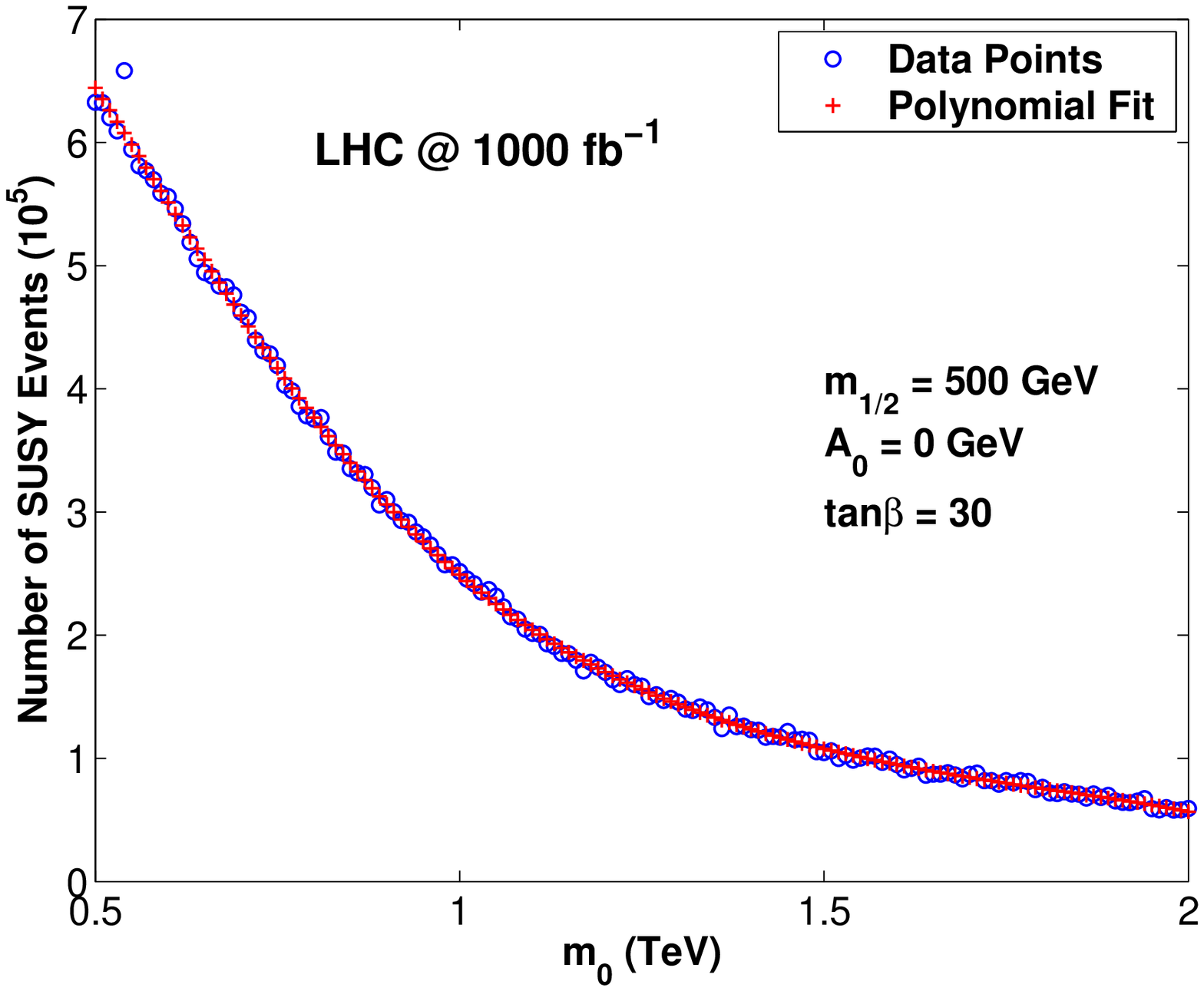}
\includegraphics[width=7.0cm,height=5.8cm]{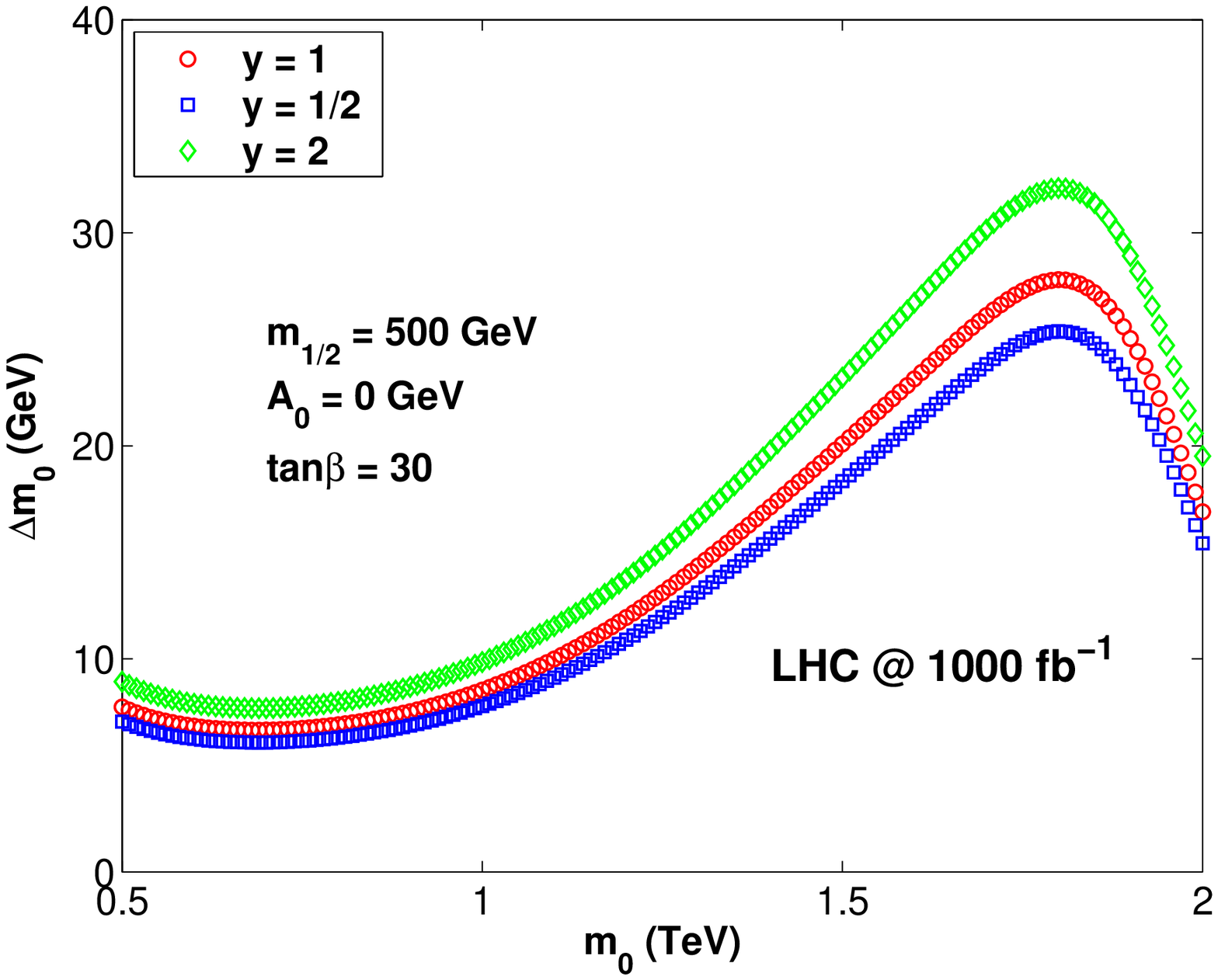}
\includegraphics[width=7.0cm,height=6.0cm]{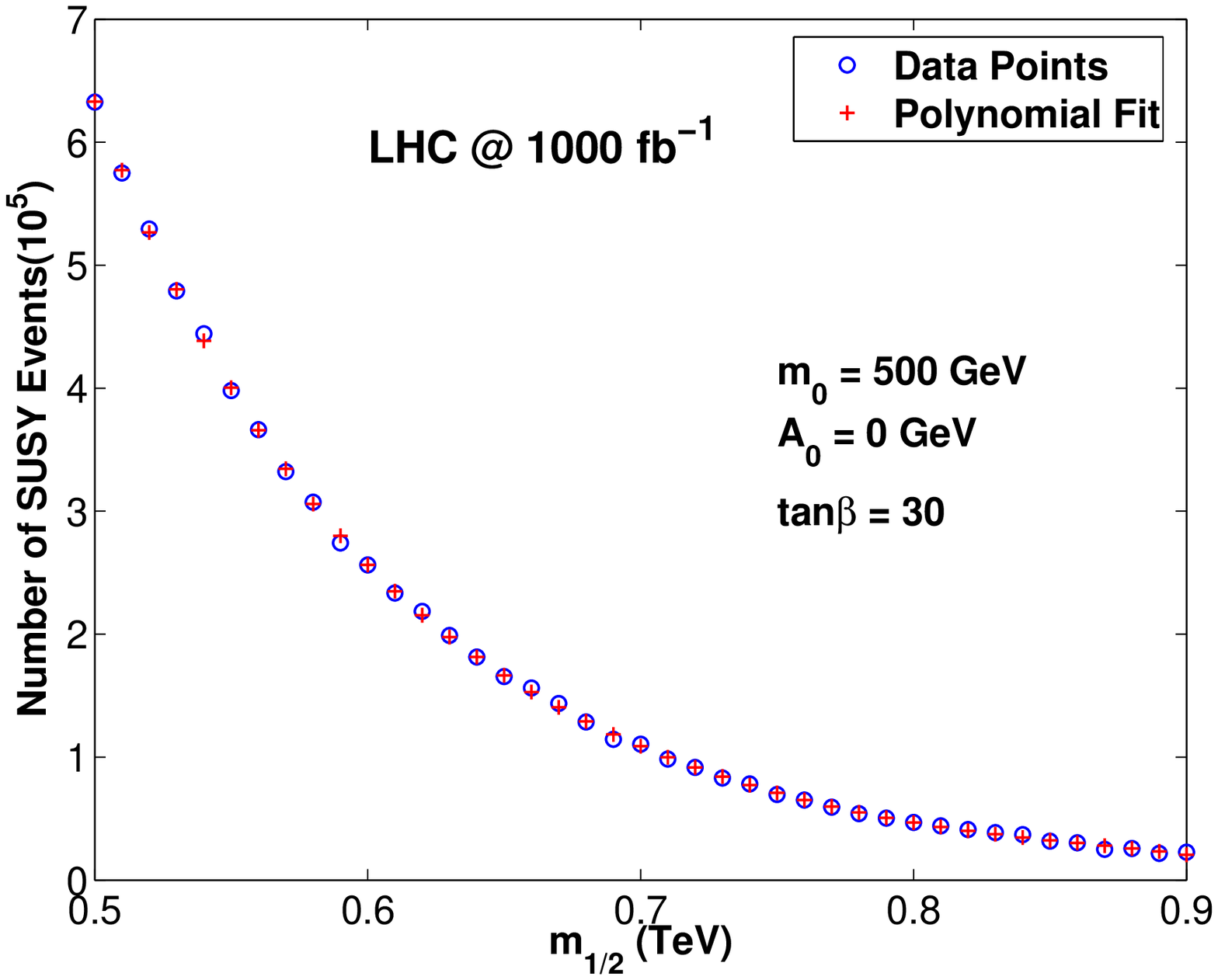}
\includegraphics[width=7.0cm,height=5.8cm]{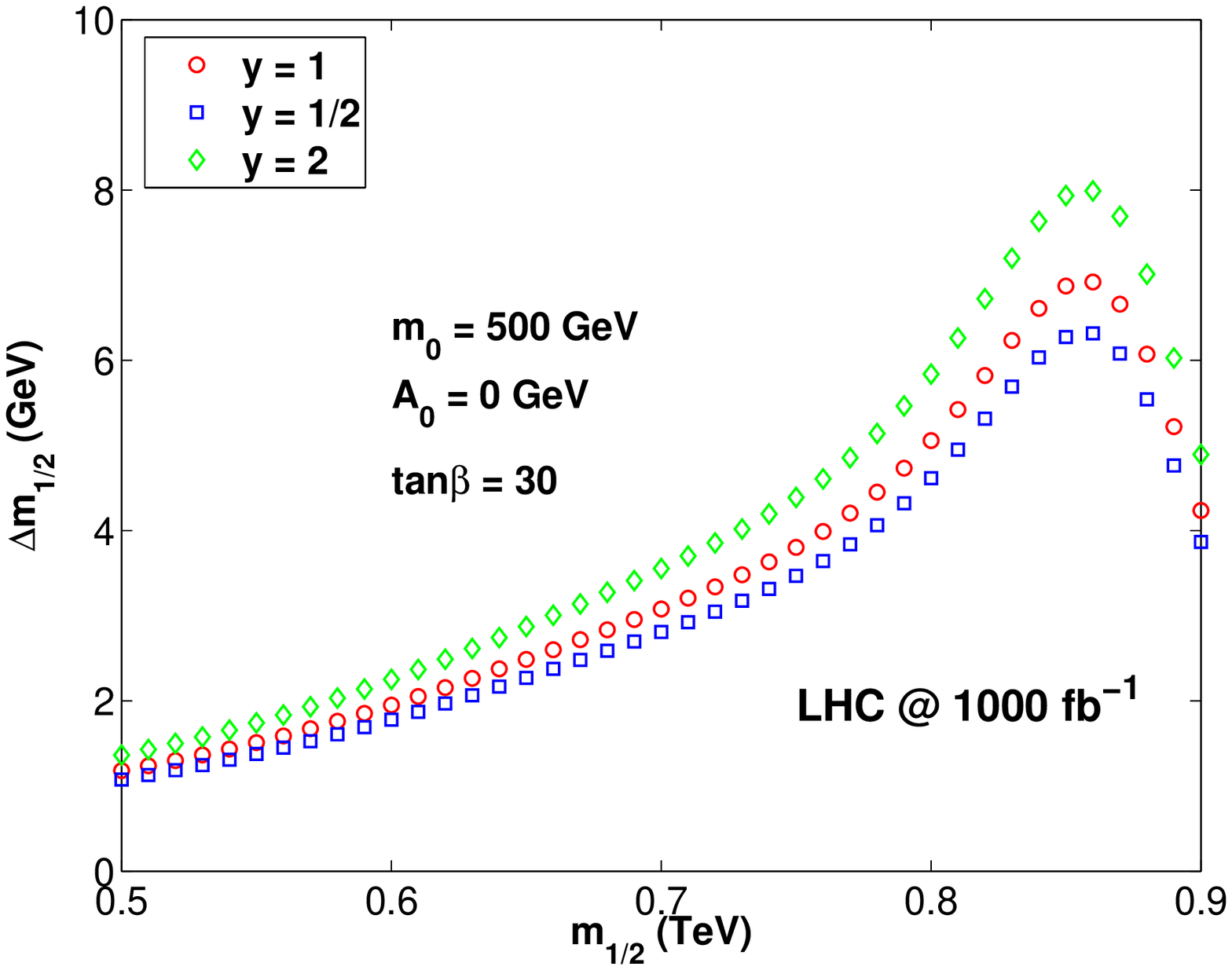}
\caption{An analysis showing the resolutions in $m_0$ and $m_{1/2}$
that can be reached with 1000 fb$^{-1}$ of integrated luminosity
under the REWSB constraints.
The two left panels give the number of SUSY events vs
$m_0$ (top left panel) and vs $m_{1/2}$ (lower left panel) for 1000
fb$^{-1}$ of integrated luminosity. The right panels give the
resolutions in $m_0$ (top right panel) and in $m_{1/2}$ (lower right
panel) using the left panels.
}
\label{fig:resolution}
\end{figure*}

We discuss now the issue of how well  we can resolve the points in
the parameter space $x_{\alpha}$ ($\alpha =1,..,p$) for a given
luminosity.
Consider Eq.(\ref{pulls1}) and set  $\delta N=\sqrt N$, and parameterize
the standard  model uncertainty  by $\delta N^{SM}=\sqrt y \delta N$.
Next we set the criterion for the resolution of two
adjacent points in the SUGRA parameter space separated by $\Delta x_{\alpha} $ so that
the separation in the signature space satisfies
\begin{equation}
\frac{\Delta N}{\sqrt{2N+yN}}= 5.
\end{equation}
Since $N=\sigma_{\rm susy}(x_{\alpha}){\cal{L}}_{\rm LHC}$, where $\sigma_{\rm susy}$ is the cross section
for the production of sparticles, and
${\cal{L}}_{\rm LHC}$
is the LHC integrated luminosity,
  the resolution achievable
   in the vicinity of SUGRA parameter point $x_{\alpha}$  at that luminosity
   is given by
\begin{equation}
\Delta x_{\alpha} = \frac{5}{2}{(2+y)}^{1/2}
{\cal{L}}_{\rm LHC}^{-1/2}
\left(\frac{\partial \sigma^{1/2}_{\rm susy}(x)}{\partial
x_{\alpha}}\right)^{-1}. \label{resolve}
\end{equation}
In Fig.(\ref{fig:resolution}) we give an illustration of the above
 when $m_0$ varies between
500 GeV and 2000 GeV while $m_{1/2}=500$ GeV, $A_0=0, \tan\beta=30$, and
$\mu>0$.
From Fig.(\ref{fig:resolution}) one finds that the resolution in $m_0$
strongly depends on the point in the parameter space
and on the luminosity.
Quite interestingly a resolution as small as  a few
GeV can be achieved for $m_0$ in the  range 500-1000  GeV with 1000
fb$^{-1}$ of integrated luminosity. A similar  analysis varying  $m_{1/2}$
in the range 500$\sim$900 GeV for the case when $m_{0}=500$ GeV, $A_0=0,
\tan\beta=30$ and $\mu>0$, shows that  a resolution in
$m_{1/2}$ as low as 1 GeV can be achieved with 1000 fb$^{-1}$ of
integrated luminosity.

\chapter{Higgs Production at Colliders}
\label{ch:higgs}

In this chapter we investigate the Higgs cross sections at the Tevatron and at the LHC. 
 The lightness of $A$ (and also of $H$ and $H^{\pm}$)
 in the Higgs Patterns implies that the Higgs  production cross sections can be large
 (for some of the previous analyses where light Higgses appear
 see \cite{Kane:2003iq,Carena:2006dg,Ellis:2007ss}).

\begin{figure}[t]
\begin{center}
\includegraphics[width=7cm,height=6cm]{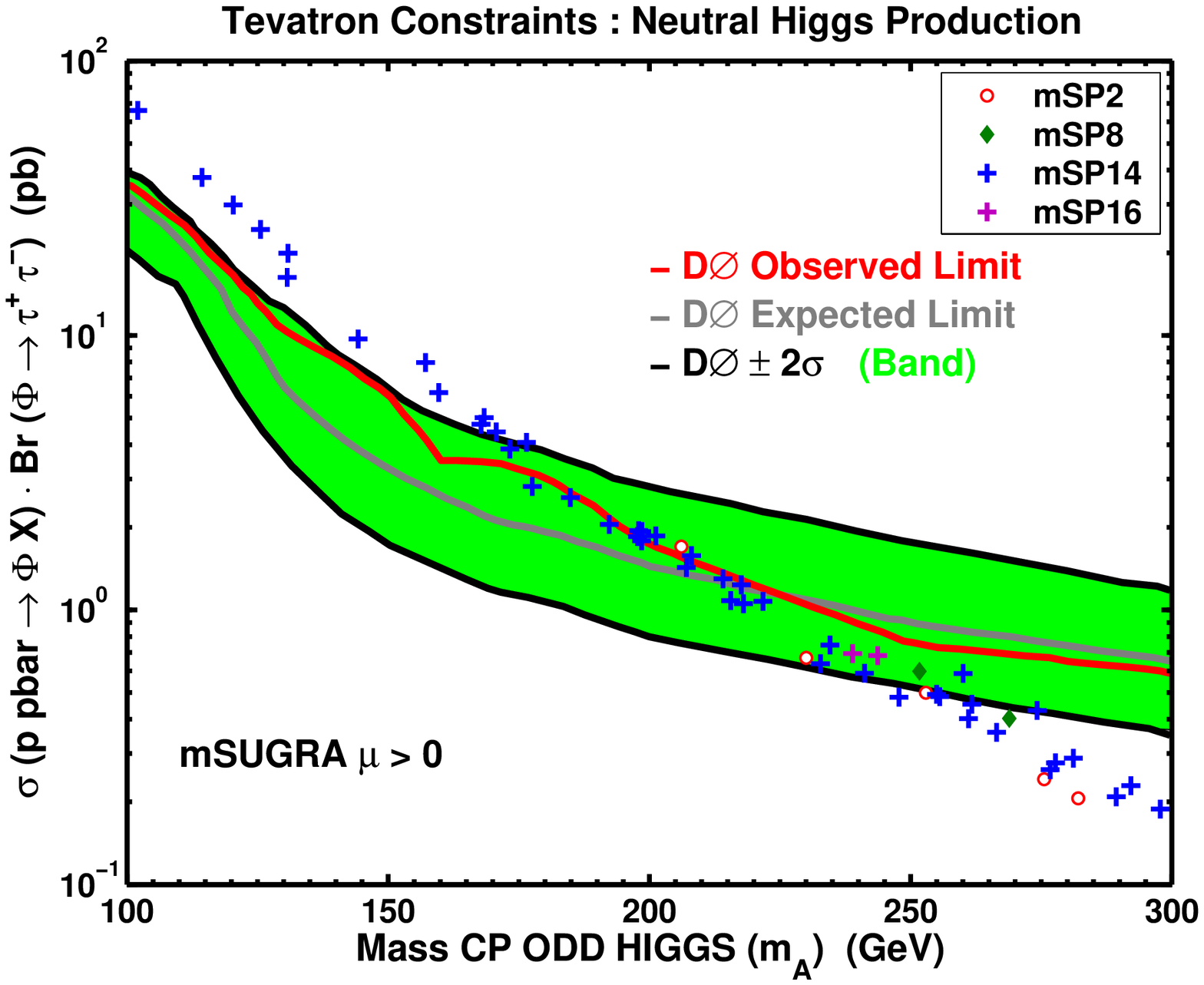}
\includegraphics[width=7cm,height=6cm]{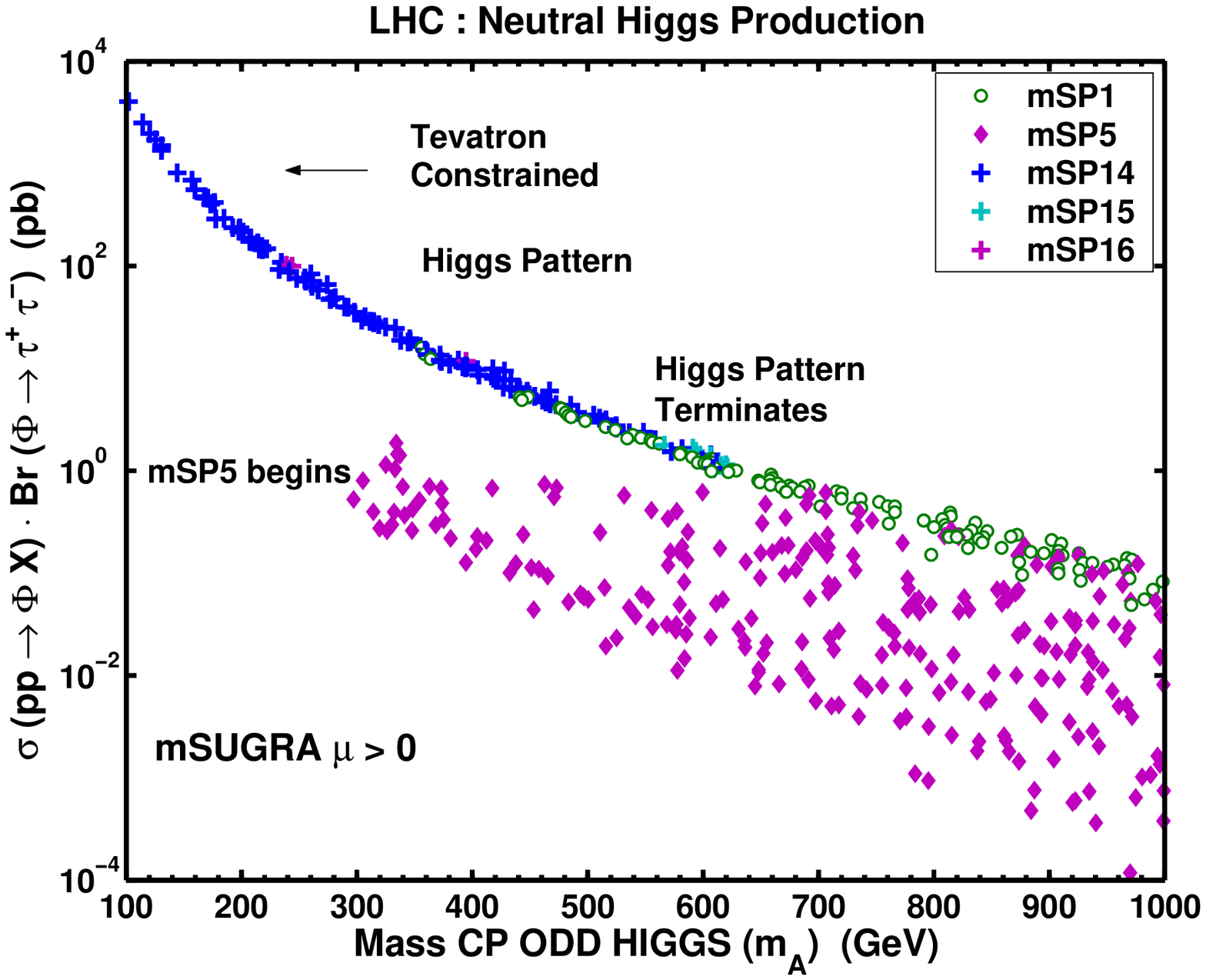}
\caption{ Left panel: Predictions for $
[\sigma(p\bar p \to \Phi){\rm BR}(\Phi \to 2\tau)]$
 in mSUGRA as a function of the CP odd Higgs mass $m_A$ for
the HPs at the Tevatron with CM energy of $\sqrt s =1.96$ TeV. The
limits from D\O\ are indicated \cite{Abazov:2006ih}.  Right panel:
Predictions for
 $[\sigma(p p \to \Phi){\rm BR}(\Phi \to 2\tau)]$
in mSUGRA as a function of $m_A$ at
the LHC  with CM energy of $\sqrt s =14$ TeV for the HPs, the
chargino pattern mSP1 and the stau pattern mSP5. The HPs are seen
to give the largest cross sections.}
 \label{higgstevlhc}
  \end{center}
\end{figure}

 Quite interestingly the recent Tevatron data is beginning to constrain
 the Higgs Patterns (HPs). This is exhibited  in the left panel of
  Fig.(\ref{higgstevlhc})   where the
 leading order
(LO) cross section for the sum of neutral Higgs processes
$\sigma_{\Phi \tau\tau} (p\bar p)=
[\sigma(p\bar p \to \Phi){\rm BR}(\Phi \to 2\tau)]$
(where sum over the neutral $\Phi$ fields is implied)
 vs the CP odd Higgs  mass is plotted for CM energy of  $\sqrt
s=1.96$ TeV  at the Tevatron. One finds  that the predictions of
$\sigma_{\Phi \tau\tau} (p\bar p)$ from the HPs are the largest and lie in a
narrow band  followed by those from the Chargino  Pattern mSP2.
The
recent data from the Tevatron is also shown\cite{Abazov:2006ih}. A
comparison of  the theory prediction with data shows that the HPs are
being constrained by experiment. Exhibited in  the right panel of
Fig.(\ref{higgstevlhc})  is
 $\sigma_{\Phi \tau\tau} (p p)=  [\sigma(p p \to \Phi){\rm BR}(\Phi \to 2\tau)]$
 arising from the HPs  (and also from other patterns which
make a comparable contribution) vs the CP odd Higgs  mass  with the
analysis done at  CM energy of  $\sqrt s=14$ TeV  at the LHC.  Again it is seen that
the predictions of  $\sigma_{\Phi \tau\tau} (p  p)$  arising
from the HPs are the largest and lie in a very narrow band
and the next largest predictions for
 $\sigma_{\Phi \tau\tau} (p  p)$ are typically from the Chargino Patterns (CPs).
The larger cross
sections for the HPs enhance the prospects of their detection.

\begin{figure}[t]
\begin{center}
\includegraphics[width=7cm,height=6cm]{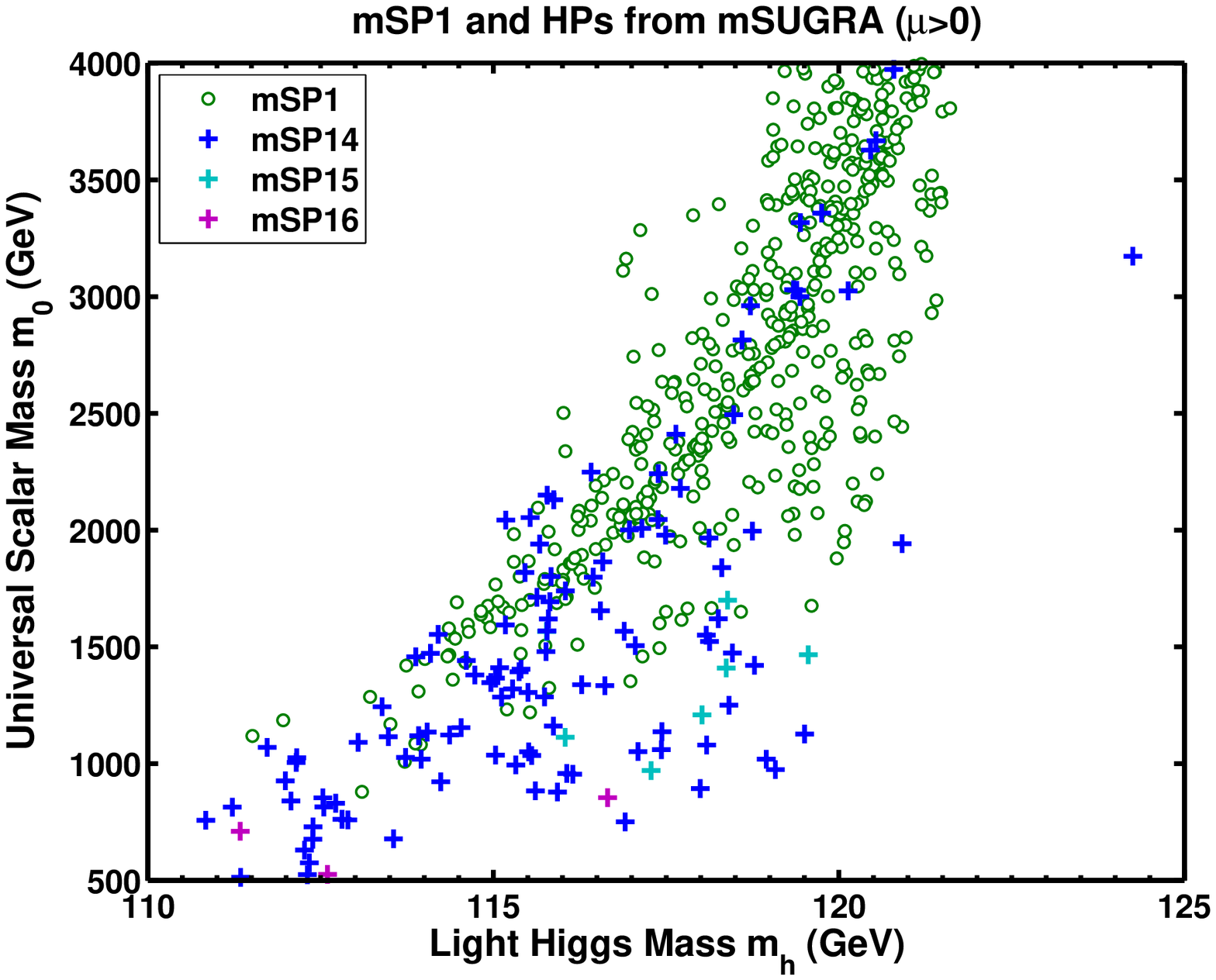}
\includegraphics[width=7cm,height=6cm]{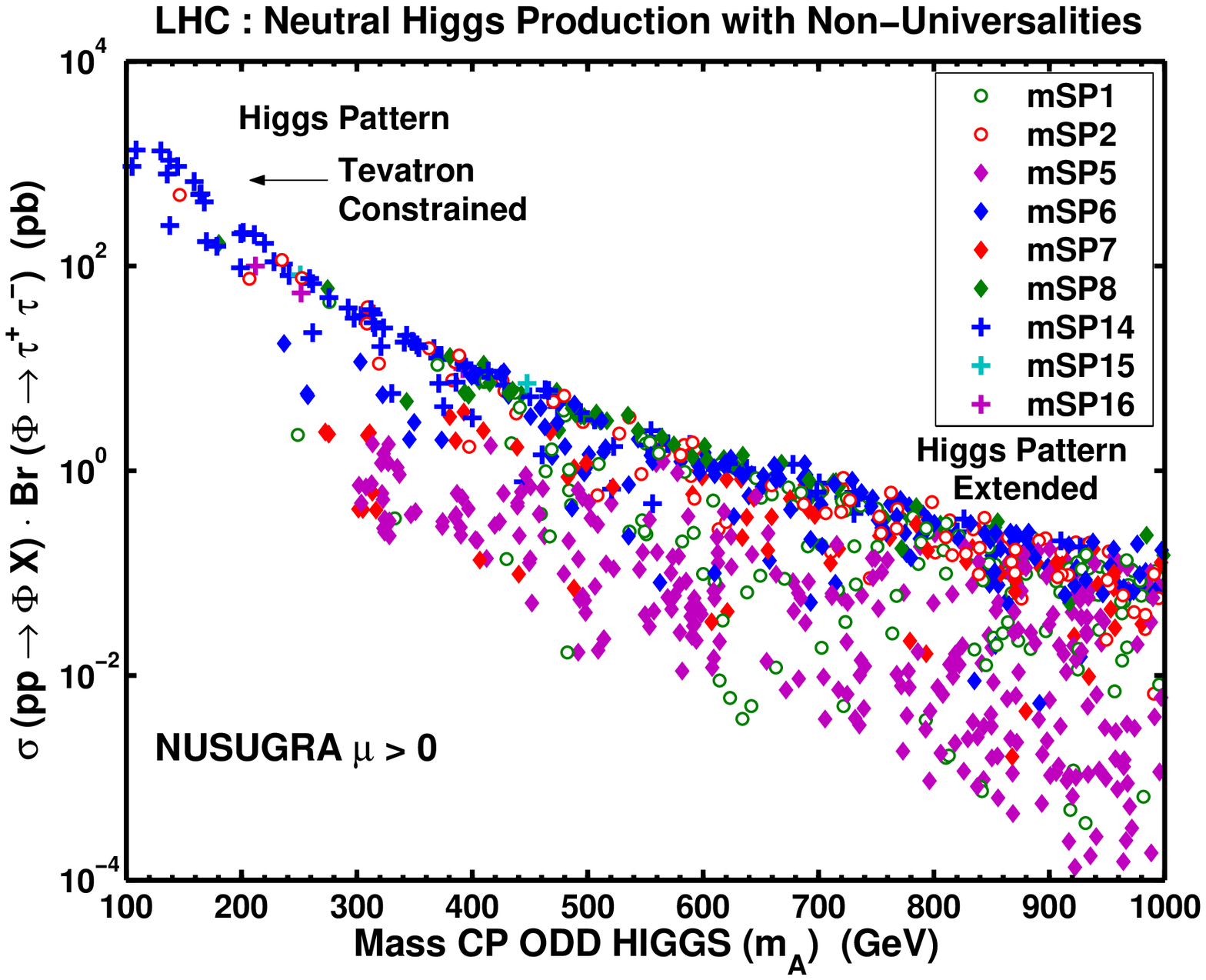}
\caption{ Left panel: mSP1 and HPs are plotted in
the $m_0$-$m_h$ plane in mSUGRA $\mu > 0$. Right panel: Predictions
for
$  [\sigma(p p \to \Phi){\rm BR}(\Phi \to 2\tau)]$
 in NUSUGRA (NUH,NUG,NU3) as a function of CP odd Higgs
mass at the LHC showing that the HPs extend beyond 600 ${\rm GeV}$
with non-universalities (to be compared with the analysis of
Fig.(\ref{higgstevlhc}) under the same naturalness assumptions).}
 \label{mandNUcross}
  \end{center}
\end{figure}

Since the largest Higgs production cross sections at the LHC
arise from the Higgs Patterns and the Chargino Patterns we exhibit the
mass of the light Higgs as a function of $m_0$ for these two patterns in
the left panel of Fig.(\ref{mandNUcross}). We note that many  of the Chargino Pattern
points in this figure appear to have large $m_0$ indicating that they originate from the
Hyperbolic Branch/Focus Point (HB/FP) region\cite{hb/fp}.

We discuss now briefly the Higgs to $b \bar b$ decay at the
Tevatron. From the parameter space of mSUGRA that enters in Fig.(1)
we can compute the quantity $[(p \bar p \to \Phi) {\rm
BR}(\Phi \to b \bar b)]$.
Experimentally, however, this quantity is
difficult to measure because there is a large background to the
production from $q\bar q, gg \to b \bar b$. For this reason one focuses on
the production $[(p \bar p \to \Phi b) {\rm BR}(\Phi \to b \bar
b)]$\cite{bh}.
For the parameter space of Fig.(1)   one gets
$[(p \bar p \to \Phi b)  {\rm
BR}(\Phi \to b \bar b)] \lesssim 1$~pb at ($\tan \beta = 55$,$M_A =
\rm 200~GeV$).
The preliminary CDF data \cite{CDFprelim} puts limits at
200~GeV, in the range (5-20)~pb over a $2 \sigma$ band at the tail
of the data set.
These limits are larger, and thus  less stringent,
than what one gets from  $\Phi\to \tau^+\tau^-$.
For the LHC, we find $[(p  p \to
\Phi b)  {\rm BR}(\Phi \to b \bar b)]\sim 200$~pb for the same
model point.  A more detailed fit requires a full treatment which is
outside the scope of the present analysis.

The neutral Higgs production cross section for the NUSUGRA case is given in the right panel of
  Fig.(\ref{mandNUcross}).  The analysis shows that the Higgs Patterns produce  the largest
  cross sections followed by the Chargino  Patterns as in mSUGRA case.
One feature which is now different is that the Higgs Patterns survive
significantly beyond the CP odd Higgs  mass of 600 GeV within our assumed naturalness
assumptions. Thus nonuniversalities tend to extend the CP odd Higgs beyond what one
has in the mSUGRA case.
		
\chapter{$B_s\to\mu^+\mu^-$ Constraints}
\label{ch:bsmm}

In this chapter, we investigate the $B_s\to\mu^+\mu^-$ constraints within 
the context of the sparticle pattern analysis. The process 
$B_s\to \mu^+\mu^-$ is dominated by the neutral Higgs exchange \cite{gaur}. 
The decay $B_{d'}^0\to \ell^+\ell^-$ ($d'$=$d$,$s$) is governed by 
the effective Hamiltonian \cite{Ibrahim:2002fx}
\begin{equation}
H_{\rm eff}=-\frac{G_Fe^2}{4\sqrt 2 \pi^2} V_{tb}V_{td'}^* 
(C_S O_S + C_P O_P +C_S' O_S'+ C_P' O_P'+ C_{10} O_{10})
\end{equation}
where 
\begin{eqnarray}
O_S&=& m_b  (\bar d'_{\alpha}P_Rb_{\alpha})(\bar \ell \ell),\\
O_P&=&m_b (\bar d'_{\alpha}P_Rb_{\alpha})(\bar \ell\gamma_5 \ell),\\
O_S'&=&m_{d'}  (\bar d'_{\alpha}P_Lb_{\alpha})(\bar \ell \ell),\\
O_P'&=& m_{d'}  (\bar d'_{\alpha}P_Lb_{\alpha})(\bar \ell \gamma_5 \ell),\\
O_{10}&=& (\bar d'_{\alpha}\gamma^{\mu}P_Lb_{\alpha})
(\bar \ell\gamma_{\mu}\gamma_5 \ell)
\end{eqnarray}
The branching ratio $B(B^0_{d'}\to\ell^+\ell^-)$
 is then given by
\begin{eqnarray}
B(B^0_{d'}\to\ell^+\ell^-) =
 \frac{G_F^2\alpha^2M_{B_{d'}}^5\tau_{B_{d'}}}{16\pi^3}
 |V_{tb}V_{td'}^*|^2 
  \sqrt{1-\frac{4 m_{\ell}^2}{M^2_{B_{d'}}}} \nonumber\\
  \times \left[\left(1-\frac{4 m_{\ell}^2}{M^2_{B_{d'}}}\right) |f_S|^2
  +|f_P+2m_{\ell} f_A|^2\right]
\end{eqnarray} 
where $f_i$ (i=S,P) and $f_A$ are defined as  follows
\begin{eqnarray}
 f_i  &=& -\frac{i}{2}f_{B_{d'}}(\frac{C_im_b-C_i'm_{d'}}{m_{d'}+m_b}), \\
 f_A &=& -\frac{if_{B_{d'}}}{2M_{B_{d'}}^2} C_{10}.
\end{eqnarray}
Specifically $C_S$ and $C_P$ have the form 
\begin{equation}
C_S=-\frac{m_{\ell}}{\sqrt 2 m_W^2 \cos^3\beta} 
\sum_{j=1}^{3}\sum_{s=1}^{2} m_{\chi_s^+}\frac{R_{j1}^2}{M_{H_j}^2} \psi_s ,
\end{equation}
\begin{equation}
C_P=\frac{m_{\ell}\tan^2\beta}{\sqrt 2 m_W^2 \cos\beta} 
\sum_{j=1}^{3}\sum_{s=1}^{2} m_{\chi_s^+}\frac{R_{j3}^2}{M_{H_j}^2} \psi_s .
\end{equation}
When $\tan\beta$ becomes large, one finds that the branching ratio is propotional to 
$\tan^6\beta$.

It is thus reasonable to expect that the Higgs patterns (HPs) will be constrained more
severely than other patterns by the $B_s\to \mu^+\mu^-$
experiment,  since HP points usually arise from the high
$\tan\beta$ region 
(however, as we noticed already, the nonuniversalities in the Higgs sector (NUH)
can also give rise to HPs for moderate  values of $\tan \beta$).

\begin{figure}[htb]
  \begin{center}
\includegraphics[width=7.5cm,height=6.5cm]{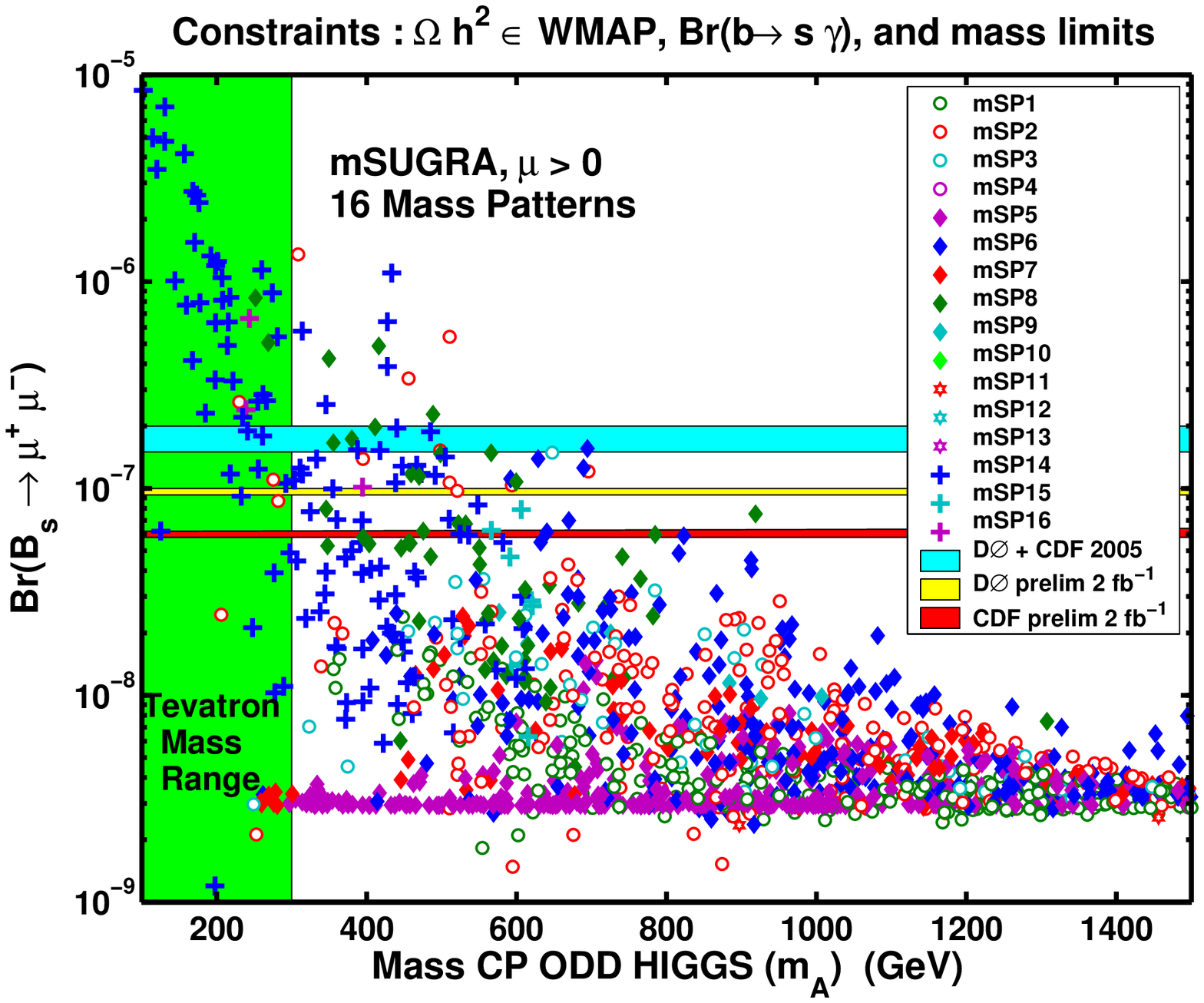}
\includegraphics[width=7.5cm,height=6.5cm]{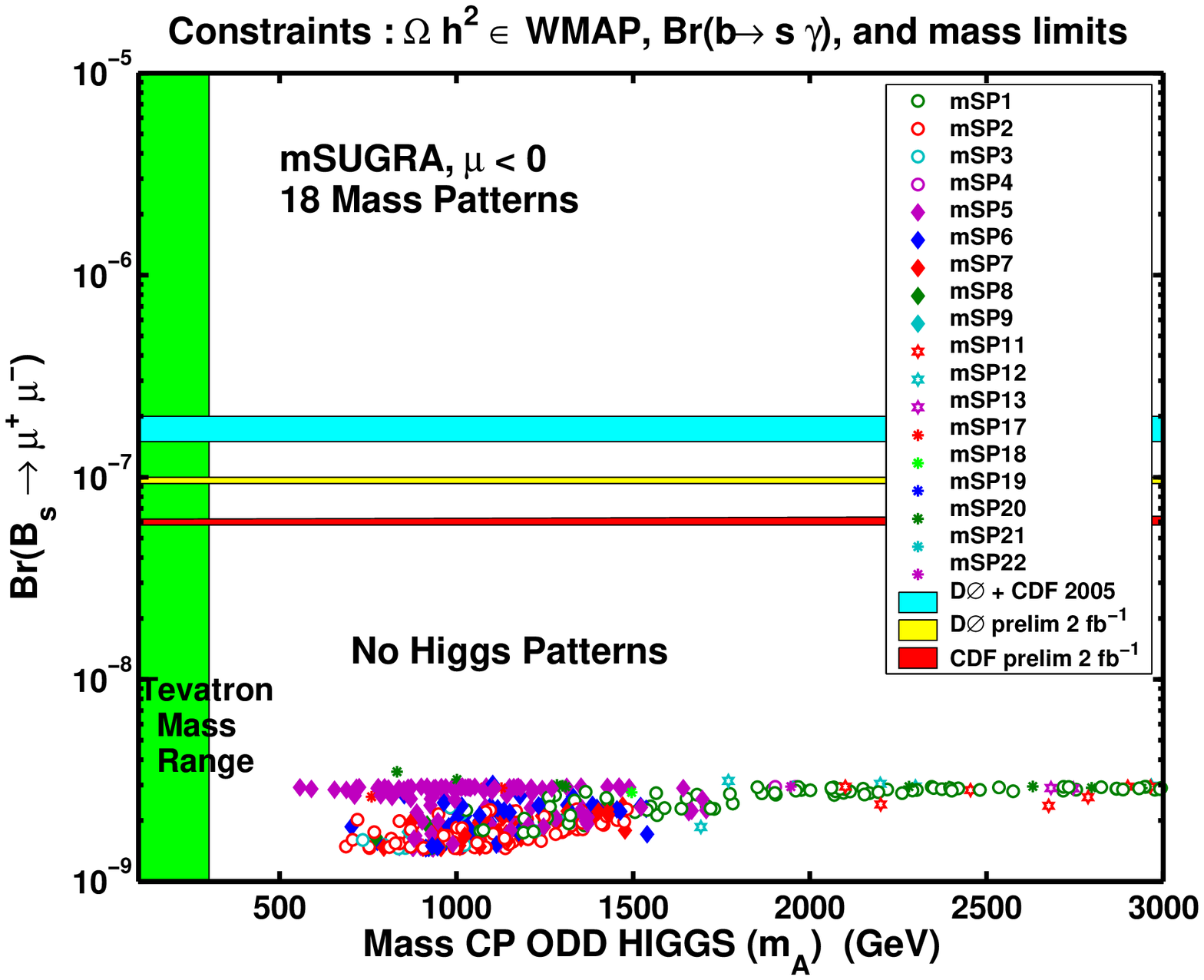}
\includegraphics[width=7.5cm,height=6.5cm]{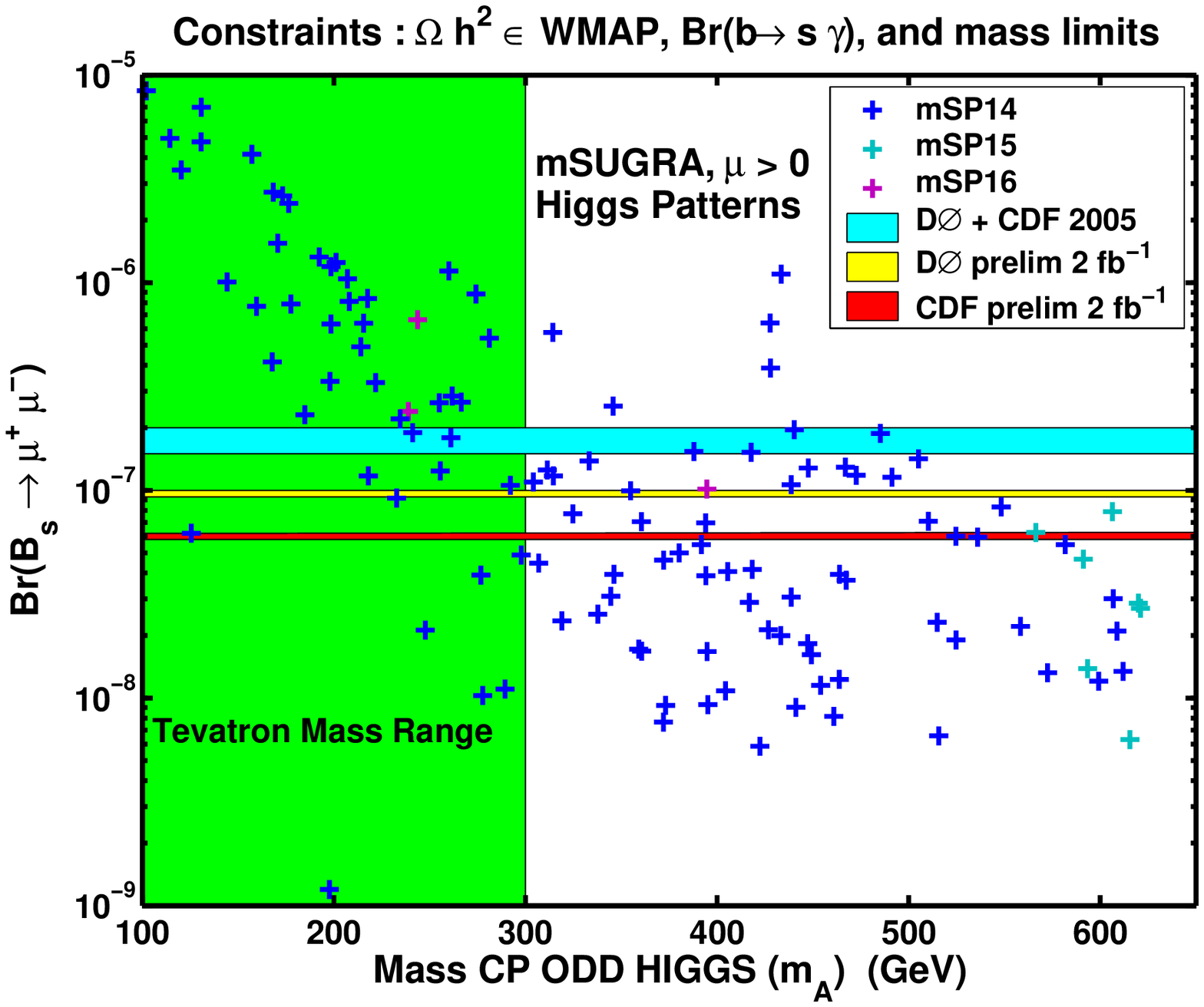}
\includegraphics[width=7.5cm,height=6.5cm]{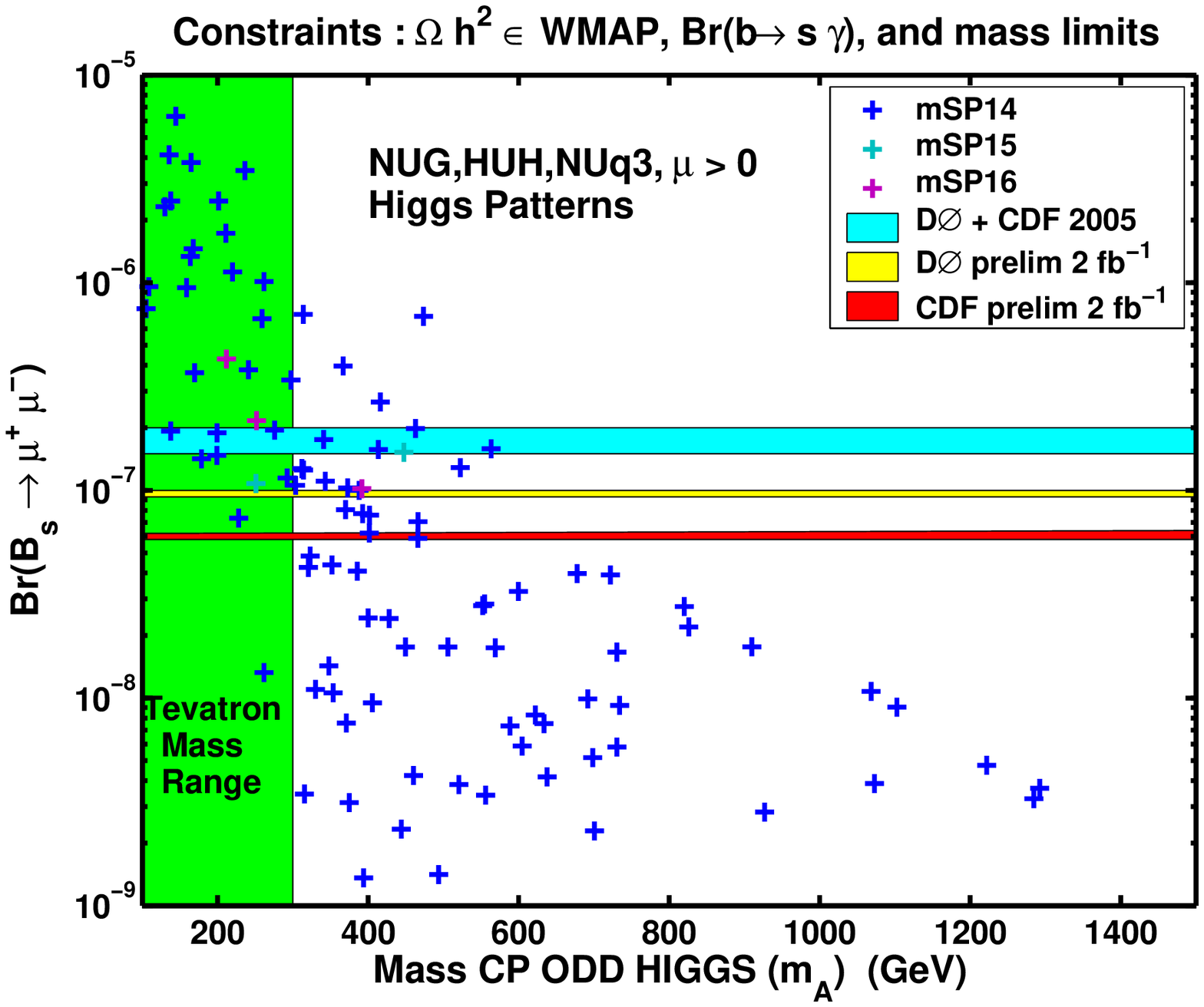}
\caption{
Predictions for the branching ratio $B_s\to \mu^+\mu^-$ in various
patterns in the SUGRA landscape. Upper left  panel: predictions are for the patterns
for $\mu>0$ in mSUGRA; upper right panel: predictions are for the patterns for $\mu<0$ in mSUGRA;
lower left panel: predictions for the Higgs Patterns alone for $\mu>0$ in mSUGRA;
lower right panel: predictions
for NUSUGRA models NUH, NUq3, and NUG for $\mu>0$.
The experimental limits are: top band
2005 \cite{Bernhard:2005yn,Abulencia:2005pw}, and the bottom two horizontal lines
are preliminary limits from the CDF and D\O ~ data \cite{bsmumu07}.  For convenience  we draw the limits
extending past the  observable mass of the CP odd Higgs at the Tevatron.}
\label{bsmumu}
  \end{center}
\end{figure}

In Fig.(\ref{bsmumu}) we carry out a detailed analysis 
where the branching ratio ${\mathcal Br}(B_s\to
\mu^+\mu^-)$ is plotted against the CP odd Higgs mass $m_A$.  The
upper left (right)  hand panel gives the analysis for the case of mSUGRA for
$\mu>0$ ($\mu<0$)
 for the Higgs Patterns as well as for several other
patterns, and the experimental constraints are also shown.
One finds that the constraints are very effective for $\mu>0$ (but not for $\mu<0$)
constraining  a part of the parameter space of the HPs
and also some models within the Chargino and the Stau Patterns are
constrained (see upper left and lower left panels of Fig.(\ref{bsmumu})).

From the analysis of Fig.(\ref{bsmumu}), it is observed that the strict imposition
of the constraint ${\mathcal Br } (B_s \to \mu^{+} \mu^{-}) < 1.5
\times 10^{-7}$ still allows for large  $\tan \beta$
in the mSUGRA model.  Thus all of the HP model points given
in Fig.(\ref{bsmumu}) that satisfy this constraint  for the mSUGRA $\mu
> 0$ case correspond to $\tan \beta$ in the range of 50 - 55. A similar
limit on $\tan \beta$ is also observed for the nonuniversal models.
We remark, however, that the HPs are not restricted to large $\tan
\beta$ in particular for the case of the NUH model.

\chapter{Direct Detection of Dark Matter}
\label{ch:directdark}

\section{mSUGRA}
We discuss now the direct detection of dark matter within the framework of the mSUGRA models. 
In direct detection experiments one measures the cross section 
of the WIMP scattering off the heavy nuclei such as germanium. 
The neutralino interacts with quarks in the target nuclei through a Higgs boson exchange, 
or a squark exchange. 
\begin{figure}[h]
\vspace{1cm}
\begin{center}
\begin{fmffile}{dddm}
		\begin{fmfgraph*}(80,80)
			\fmfstraight
			\fmfleft{i1,i2}
			\fmfright{o1,o2}
			\fmf{plain}{i2,v2,o2}
			\fmf{plain}{i1,v1,o1}
			\fmf{dashes,label=$h/H$}{v1,v2}
			\fmflabel{$q$}{i1}
			\fmflabel{$q$}{o1}
			\fmflabel{$\tilde{\chi}_1^0$}{i2}
			\fmflabel{$\tilde{\chi}_1^0$}{o2}
		\end{fmfgraph*}
		\hspace{2cm}
				\begin{fmfgraph*}(80,80)
			\fmfstraight
			\fmfleft{i1,i2}
			\fmfright{o1,o2}
			\fmf{plain}{i1,v1,i2}
			\fmf{plain}{o1,v2,o2}
			\fmf{dashes,label=$\tilde{q}$}{v1,v2}
			\fmflabel{$q$}{i1}
			\fmflabel{$q$}{o1}
			\fmflabel{$\tilde{\chi}_1^0$}{i2}
			\fmflabel{$\tilde{\chi}_1^0$}{o2}
		\end{fmfgraph*}
		\hspace{2cm}
				\begin{fmfgraph*}(80,80)
			\fmfstraight
			\fmfleft{i1,i2}
			\fmfright{o1,o2}
			\fmf{plain,tension=5}{i1,v1}
			\fmf{plain,tension=1}{v1,o2}
			\fmf{plain,tension=1}{i2,v2}
			\fmf{plain,tension=5}{v2,o1}
			\fmf{dashes,label=$\tilde{q}$}{v1,v2}
			\fmflabel{$q$}{i2}
			\fmflabel{$q$}{o2}
			\fmflabel{$\tilde{\chi}_1^0$}{i1}
			\fmflabel{$\tilde{\chi}_1^0$}{o1}
		\end{fmfgraph*}
	\end{fmffile}
	\end{center}
	\end{figure}
The neutralino-nucleus scattering cross-section typically is dominated by the scalar part of 
the neutralino-quark interaction and thus it is the quantity $\sigma_{\chi p}(scalar)$ 
that is of interest to us. 
The basic interaction governing the $\chi-p$ scattering is 
the effective four-fermi interaction given by (see e.g.\ \cite{Chattopadhyay:1998wb})
\begin{eqnarray}
{\cal L}_{\rm eff}=&\bar{\chi}\gamma_{\mu} \gamma_5 \chi 
\bar{q}\gamma^{\mu} (A P_L +B P_R) q
+ C\bar{\chi}\chi  m_q \bar{q} q  
+D  \bar{\chi}\gamma_5\chi  m_q \bar{q}\gamma_5 q \nonumber\\
&+E\bar{\chi}i\gamma_5\chi  m_q \bar{q} q
+F\bar{\chi}\chi  m_q \bar{q}i\gamma_5 q.
\end{eqnarray} 
The $\chi -p$  cross-section arising from scalar interactions 
$C\bar{\chi}\chi m_q\bar{q}q$ is given by 
\begin{equation}
 \sigma_{\chi p}({\rm scalar})=\frac{4\mu_r^2}{\pi}
 \left[\sum_{i=u,d,s}f_i^pC_i
 +\frac{2}{27}\left(1-\sum_{i=u,d,s}f_i^p\right)
 \sum_{a=c,b,t}C_a\right]^2.
\end{equation}
 Here $\mu_r$ is the reduced mass, $f_i^p$ (i=u,d,s quarks)
  are defined by
\begin{equation}
 m_pf_i^p=<p|m_{qi}\bar q_iq_i|p>,
\end{equation}
and C is given by
\begin{equation}
C=C_{h^0}+C_{H^0}+C_{\tilde{f}},
\end{equation}
where $C_{h^0},C_{H^0}$ are the contributions from 
  the s-channel $h^0$ and $H^0$ exchanges and 
$C_{\tilde{f}}$ is the contribution from the t-channel sfermion 
exchange. They are given by \cite{Chattopadhyay:1998wb} 
\begin{eqnarray}
C_{h^0}(u,d)&=& -(+)\frac{g^2}{4 M_W M^2_{h^0}}
\frac{\cos\alpha(sin\alpha)}{\sin\beta(cos\beta)} Re\sigma,\\
C_{H^0}(u,d)&=&\frac{g^2}{4 M_W M^2_{H^0}}
\frac{\sin\alpha(cos\alpha)}{\sin\beta(cos\beta)} Re \rho,\\
C_{\tilde{f}}(u,d)&=& -\frac{1}{4m_q}\frac{1}
{M^2_{\tilde{q1}}-M^2_{\chi}} Re[C_{qL}C^{*}_{qR}]
-\frac{1}{4m_q}\frac{1}{M^2_{\tilde{q2}}-M^2_{\chi}} Re[C^{'}_{qL}C^{'*}_{qR}].
\end{eqnarray}
Here (u,d) refer to the quark flavor, $\alpha$ is the Higgs mixing angle,  
 and $C_{qL}, C_{qL}'$ etc. are as defined in Ref.\cite{Chattopadhyay:1998wb}, 
 and  $\sigma$ and $\rho$ are defined by
\begin{equation}
\sigma= 
 X_{40}^*(X_{20}^*-\tan\theta_W X_{10}^*)\cos\alpha
+X_{30}^*(X_{20}^*-\tan\theta_W X_{10}^*)\sin\alpha,
\end{equation}
\begin{equation}
\rho=
- X_{40}^*(X_{20}^*-\tan\theta_W X_{10}^*)\sin\alpha
+X_{30}^*(X_{20}^*-\tan\theta_W X_{10}^*)\cos\alpha,
\end{equation}
 where $X_{n0}$ are the components of the LSP  
\begin{equation}
\chi=X^*_{10}\tilde B+X^*_{20}\tilde W_3+X^*_{30}\tilde H_1+X^*_{40}\tilde H_2.
\end{equation}
The coefficients $f_i^p$ are associated with some amount of uncertainties 
\cite{Corsetti:2000yq,Carena:2006dg}
\begin{eqnarray}
f_u^p &=& 0.020 \pm 0.004 ,\\
f_d^p &=& 0.026 \pm 0.005, \\
f_s^p &=& 0.118 \pm 0.062. 
\end{eqnarray}

\begin{figure}[htbp]
  \begin{center}
\includegraphics[width=7cm,height=6cm]{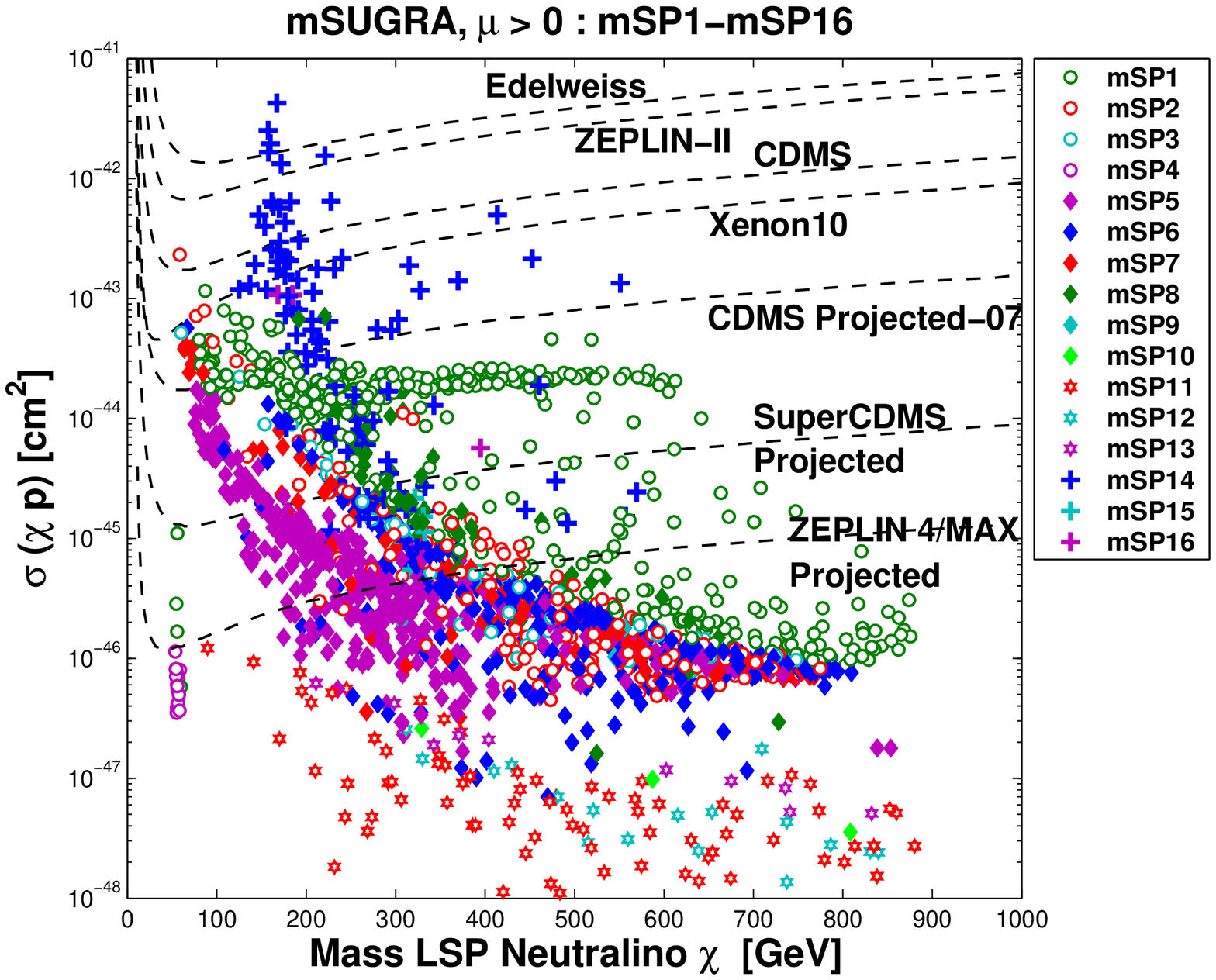}
\includegraphics[width=7cm,height=6cm]{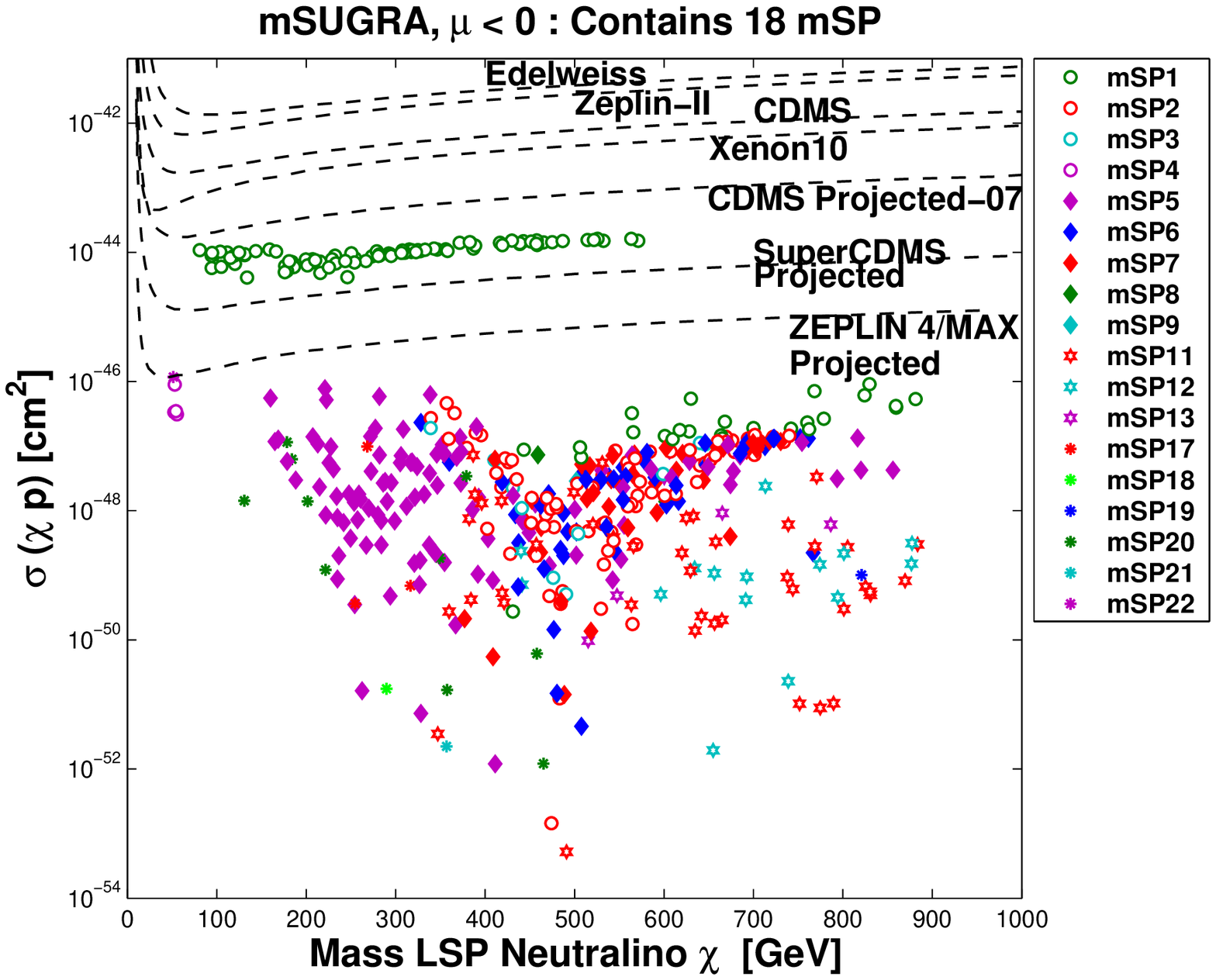}
\includegraphics[width=7cm,height=6cm]{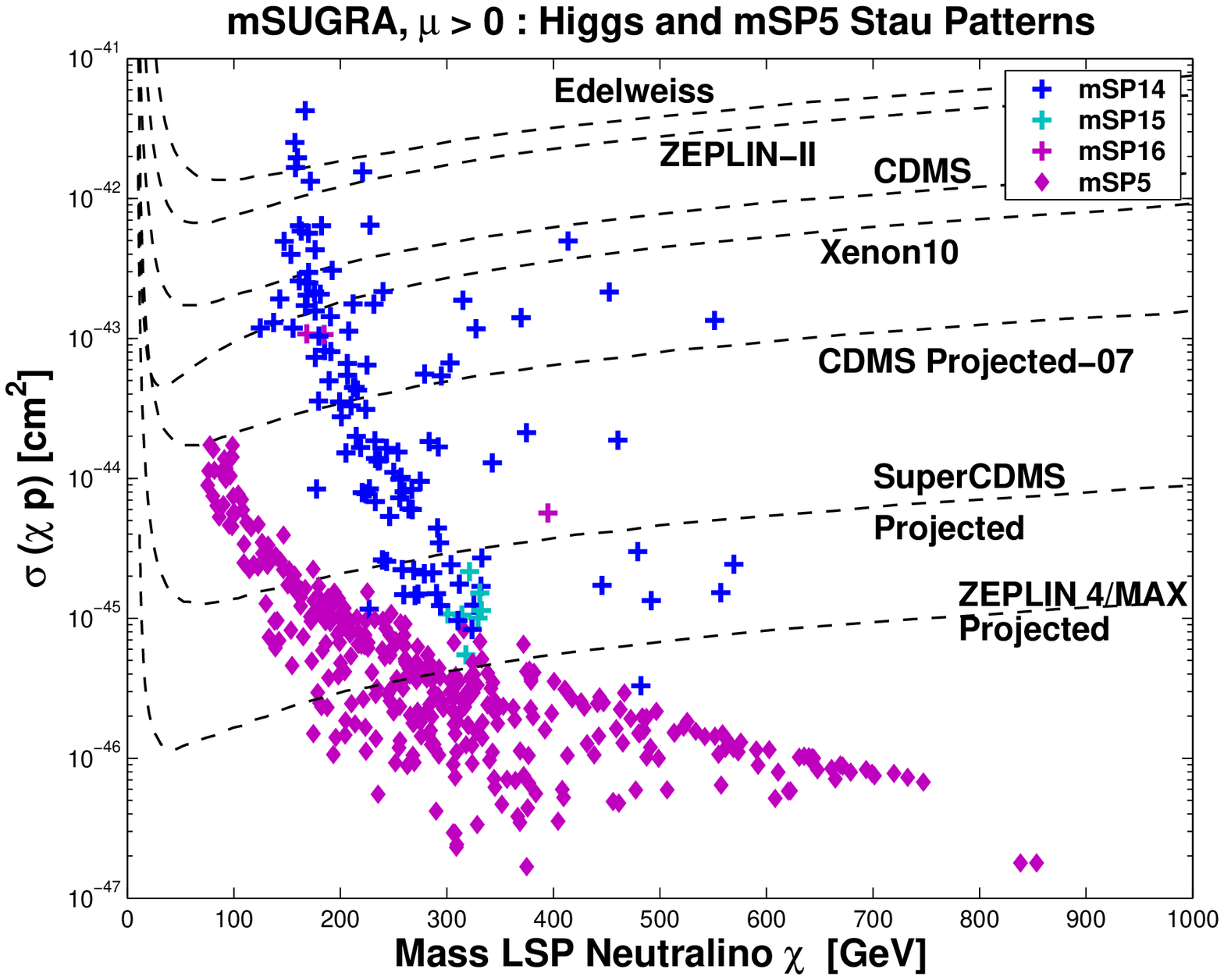}
\includegraphics[width=7cm,height=6cm]{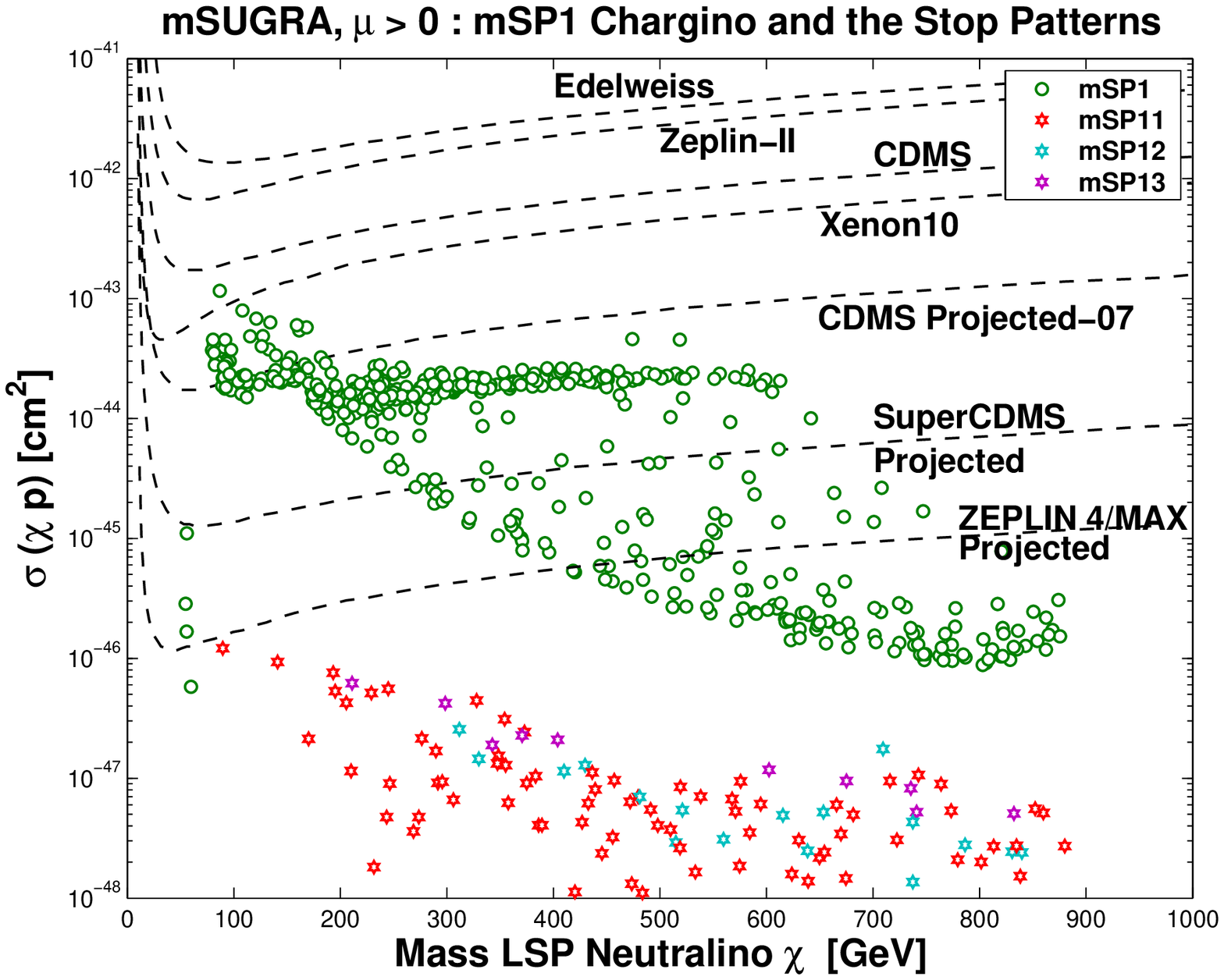}
\caption{  Analysis of $\sigma(\chi p)$ for mSUGRA: upper left panel:
$\mu>0$ case including  all patterns; upper right panel:     $\mu<0$
allowing  all patterns; lower left hand  panel: A comparison of  $\sigma(\chi p)$ for HPs and a
stau NLSP case which is of type mSP5 for
$\mu>0$;
 lower right panel:  a comparison of  $\sigma(\chi p)$ for the Chargino Pattern
mSP1 vs  the Stop Patterns mSP11-mSP13. The analysis shows  a
Wall  consisting of a clustering of points in the Chargino
Patterns mSP1-mSP4
with a $\sigma(\chi p)$ in the range $10^{-44\pm .5}$ cm$^2$
enhancing the prospects for the observation of dark matter by
SuperCDMS \cite{Schnee:2005pj},  ZEPLIN-MAX\cite{Atac:2005wv}
or LUX\cite{lux} in this region.}
\label{dcross}
  \end{center}
\end{figure}

In the absence of CP phases, in Fig.(\ref{dcross}) \cite{Feldman:2007fq} we give an
analysis of the scalar neutralino-proton cross  section
$\sigma({\tilde\chi_1^0 p})$ as a function of the LSP mass 
(for a sample of  Post-WMAP3 analysis of dark matter see \cite{modern,modern1}, 
and for more recent analysis see 
\cite{Feldman:2007fq,Allanach:2008iq,Barger:2008qd,Altunkaynak:2008ry}). 
The upper  left panel of Fig.(\ref{dcross}) 
gives the scalar $\sigma({\tilde\chi_1^0 p})$ for
the mSUGRA parameter space for $\mu>0$.
We  note  that the Higgs  patterns  typically give the largest dark matter cross
sections (see  the upper left and lower left panels of Fig.(\ref{dcross}))
and are the first ones to be constrained by experiment.
The second largest cross sections arise from the Chargino Patterns
which shows an embankment, or Wall, with a copious number of points with
cross sections in the range $10^{-44\pm .5}$cm$^2$ (see the upper left panel and
lower right panel), followed by Stau Patterns (lower left panel), with the
Stop Patterns producing the smallest cross  sections (upper left and lower right panels).
The upper right panel of Fig.(\ref{dcross}) gives the scalar  cross section $\sigma({\tilde\chi_1^0 p})$
for $\mu<0$ and here one finds  that the largest cross  sections arise  from the CPs
which also have a Chargino Wall with cross sections in the range $10^{-44\pm .5}$cm$^2$
(upper right panel).
The  analysis shows that altogether
the scalar  cross sections lie in an
interesting region and would be accessible to dark matter  experiments
currently underway and improved experiments  in the
future \cite{Bernabei:1996vj,Sanglard:2005we,Akerib:2005kh,Alner:2007ja,Angle:2007uj,lux}.
Indeed  the analysis of Fig.(\ref{dcross}) shows that some of the parameter
space of the Higgs Patterns is beginning to be constrained by the CDMS
and the Xenon10 data \cite{Angle:2007uj}.

\section{Nonuniversalities of Soft Breaking}

As already discussed in previous chapters, it is useful to consider 
other soft breaking scenarios beyond mSUGRA, since the nature of 
physics at the Planck scale is largely unknown. 
One such possibility is to consider nonuniveralities in the K$\ddot{\rm a}$hler potential,
which can give rise to nonuniversal soft breaking consistent with
flavor changing neutral current constraints.  We consider three
possibilities  which are nonuniversalities in    (i)   the Higgs
sector (NUH), (ii) the third  generation squark sector (NU3), and
(iii)  the gaugino sector (NUG) (for a sample of previous work on
dark matter analyses with nonuniversalities see \cite{Nath:1997qm}). 
We parametrize these nonuniversalities as in Eq.~(\ref{nonuni}). 
In each case we carry out a Monte Carlo scan of  $1 \times 10^6$  models. 
The above covers a very wide array of models.

\begin{figure}[htbp]
  \begin{center}
\includegraphics[width=7cm,height=6cm]{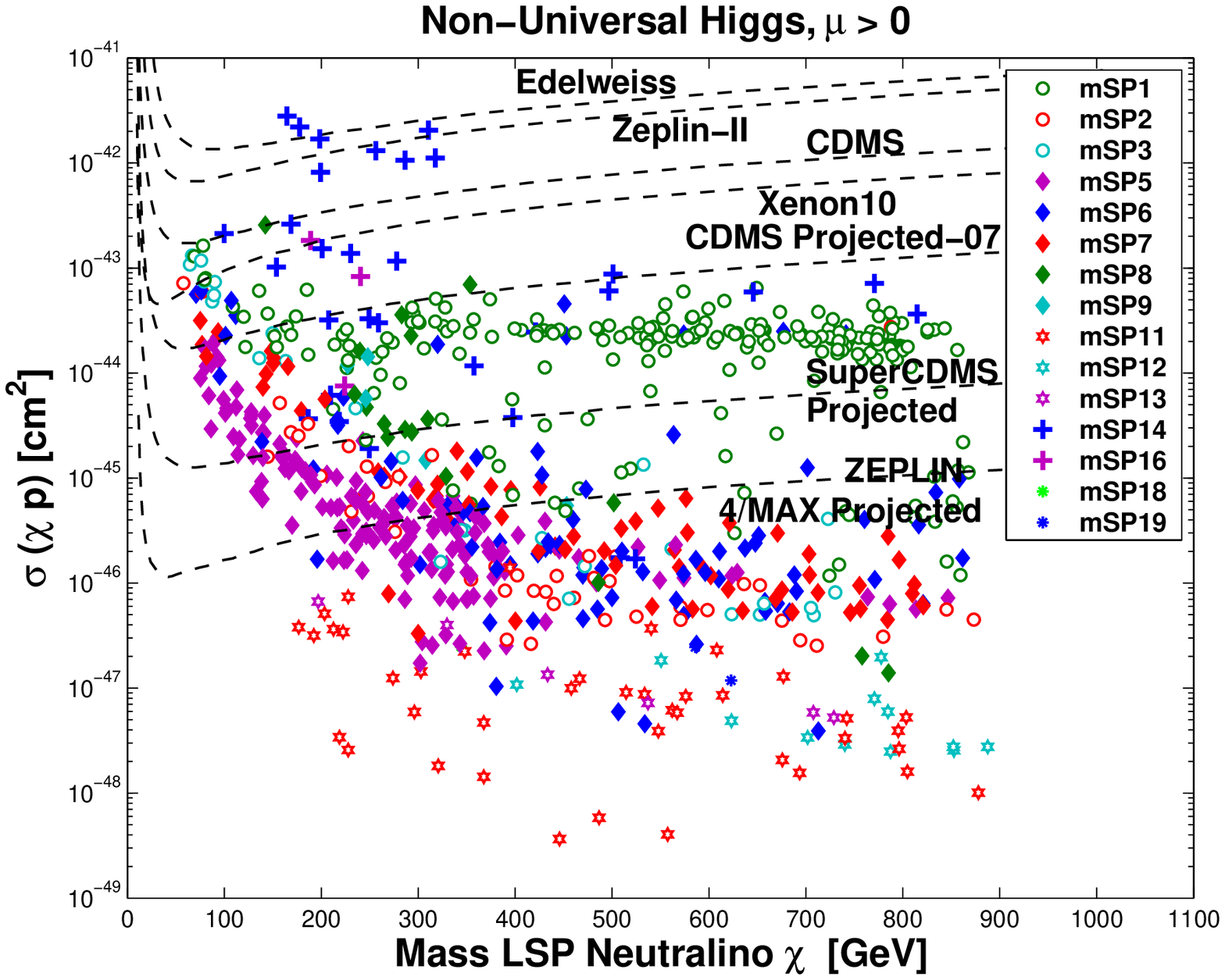}
\includegraphics[width=7cm,height=6cm]{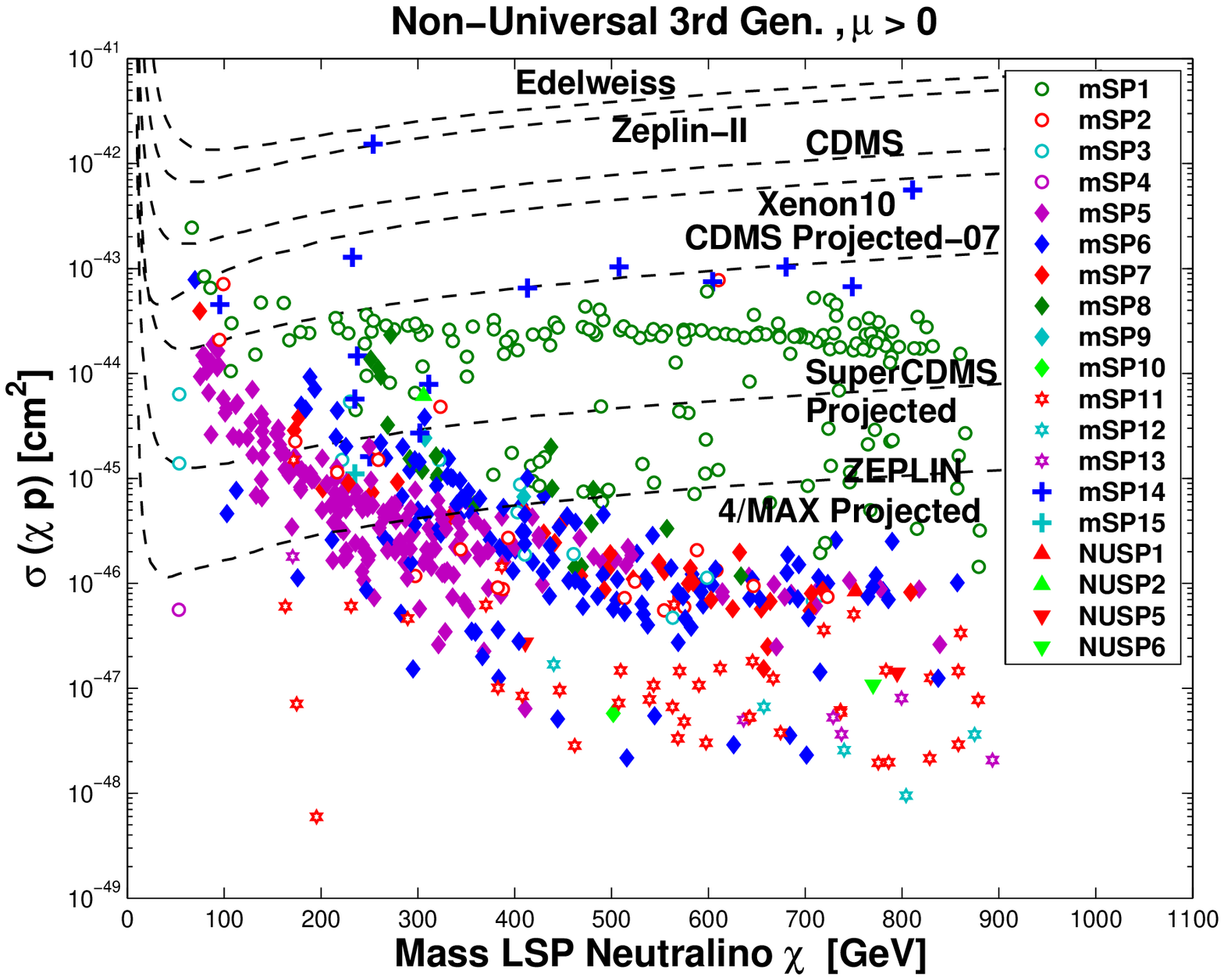}
\includegraphics[width=7cm,height=6cm]{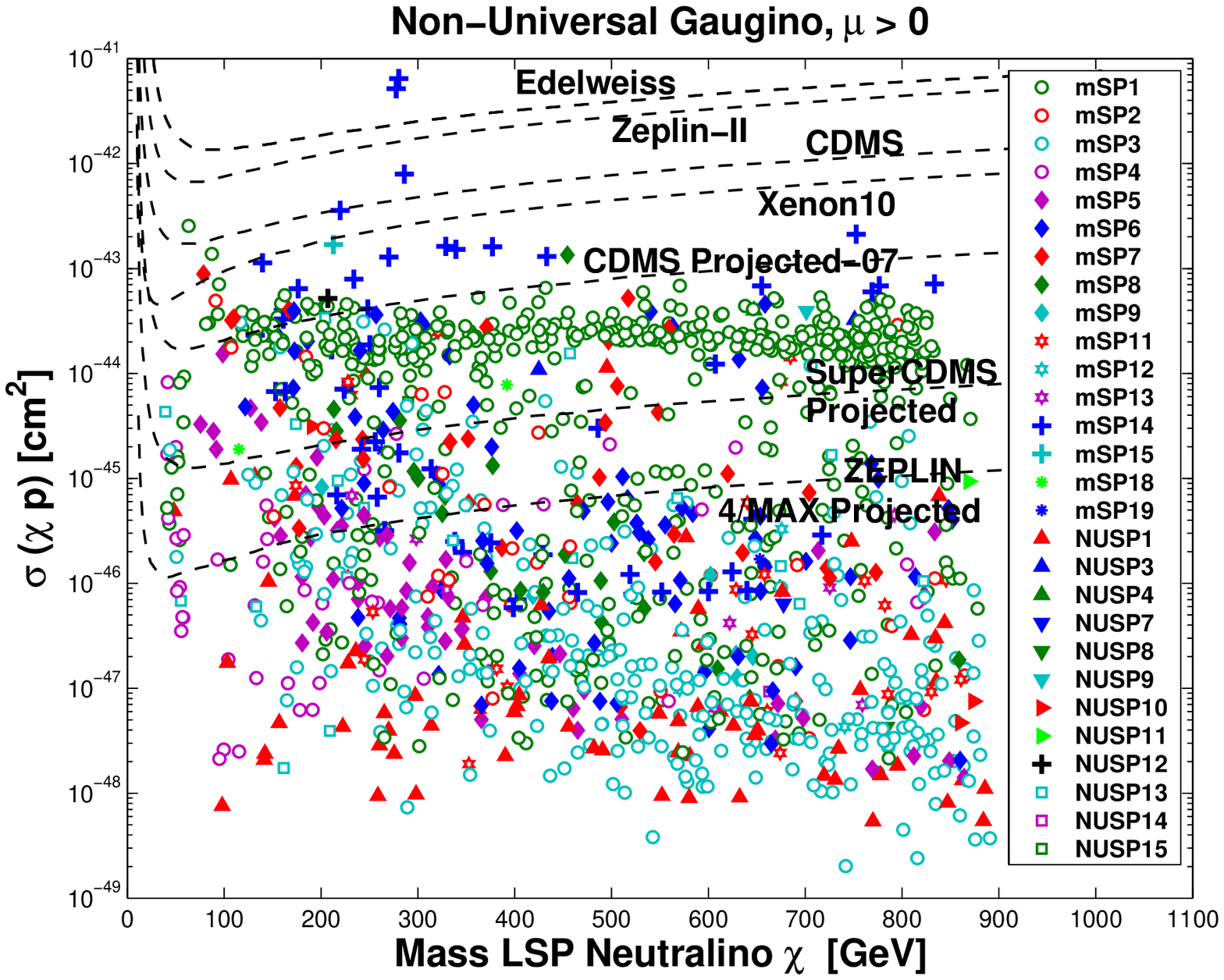}
\caption{  Analysis of the scalar cross section $\sigma(\chi p)$ for
NUSUGRA models:  NUH (upper left panel), NU3 (upper right  panel),
NUG  (lower panel).  As in Fig.(\ref{dcross})
the  Wall consisting of a clustering of points in the Chargino Patterns mSP1-mSP4 
persists  up to an LSP mass of about 900 GeV with a $\sigma(\chi p)$ in the range 
$10^{-44\pm .5}$ cm$^2$ enhancing the prospects for the observation of dark matter 
by SuperCDMS and ZEPLIN-MAX in this region.}
\label{dnonuni}
  \end{center}
\end{figure}

The analysis of the direct detection of  dark matter in NUSUGRA
 are presented in Fig.(\ref{dnonuni}). As in the mSUGRA case one finds
that the largest dark matter cross sections still arise from
the Higgs Patterns followed  by the Chargino Patterns
within the three types of nonuniversality models considered:
 NUH (upper left panel   of Fig.(\ref{dnonuni})),
 NU3 (upper right panel   of  Fig.(\ref{dnonuni})), NUG (lower panel
of Fig.(\ref{dnonuni})). Again the analysis within  NUSUGRA
shows the phenomenon of the Chargino Wall, i.e., the existence of a
copious number of Chargino  Patterns (specifically mSP1) in all cases 
with cross sections in the range $10^{-44\pm .5}$cm$^2$.
Most of the parameter  points  along the Chargino
Wall lie on the Hyperbolic Branch/Focus Point (HB/FP) region\cite{hb/fp}
where the Higgsino components of the LSP are substantial 
(for a review see \cite{Lahanas:2003bh}). Thus this
Chargino Wall presents an encouraging region of the parameter space
where the dark matter may become observable in improved experiments.


It is seen that Higgs Patterns (HPs) arising in a
wide range of models: in mSUGRA, and in NUSUGRA models 
are typically seen to lead to large Higgs production cross
sections at the Tevatron and at the LHC. It is also seen
that the HPs lead typically to the largest neutralino-proton cross
sections and would either be the first to be observed or the first
to be constrained by dark matter experiment.
The analysis presented here shows the existence of 
a Chargino Wall consisting of a copious number of parameter points 
in the Chargino Patterns where the NLSP is a chargino which give 
a $\sigma(\tilde\chi_1^0 p)$  at the level of $10^{-44\pm .5}$cm$^2$
in all models considered for the LSP mass extending up to 900 GeV in many cases.
These results  heighten the possibility for the observation of dark matter in improved dark
matter experiments such as  SuperCDMS\cite{Schnee:2005pj}, 
ZEPLIN-MAX\cite{Atac:2005wv}, and LUX\cite{lux} which are
expected to reach a sensitivity of $10^{-45}$ cm$^2$ or more.
Finally, we note that several of the patterns are well separated in the 
$\sigma(\tilde\chi_1^0 p)$- LSP mass plots,  providing important 
signatures along with the  signatures from colliders 
for mapping out the sparticle  parameter space.

\chapter{Conclusions}
\label{ch:conclusion}

The minimal supersymmetric Standard Model has 32 sparticle masses. Since the soft breaking 
sector MSSM is arbitrary, one is led to a landscape of as many as $10^{25}$ or more 
possibilities for the sparticle mass hierarchies. The number of  possibilities is
drastically reduced in well motivated models such as supergravity models, 
and one expects similar reductions to occur also in gauge and  anomaly mediated models, 
and in string and brane models.  We have analyzed the  mass hierarchies for the
first four lightest sparticle (aside from the lightest Higgs boson) for supergravity models. 
Specifically,  we analyzed the mass hierarchies for the mSUGRA model and for 
supergravity models with \non in the soft breaking in the Higgs sector, \non in the soft breaking in
the third generation sector, and \non in the soft breaking in the gaugino sector.
It is found that in each case only a small number of mass  hierarchies or patterns survive the rigorous
constraints of radiative breaking of the electroweak symmetry, relic density constraints on cold dark matter
from the WMAP data, and other experimental constraints from colliders.  These mass hierarchies can be
conveniently put into different classes labeled by the sparticle which is next heavier after the LSP. For
the SUGRA models we find six different classes: chargino patterns, stau patterns, stop patterns, Higgs
patterns, neutralino patterns, and gluino patterns.  

We discussed the techniques for the analysis of the  
signatures and the technical details on simulations of sparticle events. We also discuss 
the backgrounds to the SUSY phenomena arising from the Standard Model processes. 
Additionally we discussed the  identification of patterns based on 40 event identification criteria. 
It is found that these  criteria allow one to discriminate among most of the patterns.
An analysis of how one may lift degeneracies in the signature space, and how accurately
one can determine the soft parameters using the LHC luminosities is also given.
In addition, we also investigate the Higgs production at the Tevatron and at the LHC, and 
the direct detection of dark matter within the context of the sparticle pattern analysis. 

It is hoped that the analyses of the type discussed here would help not only in the search
for supersymmetry but also allow one to use the signatures to extrapolate back to the underlying 
supersymmetric model using the experimental data when such data from the LHC comes in.  
In the above our analysis was focused on supergravity unified models. However, the techniques 
discussed here have a much wider applicability to other models, including models 
based on  gauge and anomaly mediated breaking, as well as  string and brane based models.
	
\addcontentsline{toc}{chapter}{Appendix}

\newpage
\thispagestyle{plain}
\begin{flushleft}
\textbf{\Huge Appendix}
\end{flushleft}

\vspace{1cm}

\textbf{\large Dilepton Invariant Mass}

Here we give some further details of the analysis of the kinematic 
signatures discussed in chapter (\ref{ch:kin}). 
Specifically we study the kinematics of the dilepton invariant mass of 
SUSY chain decays.
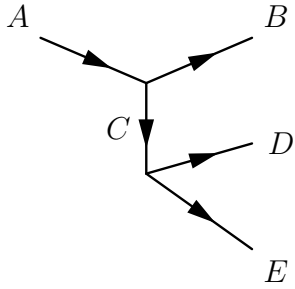
\begin{figure}[htb]
\vspace{1cm}
\begin{center}
 	\begin{fmffile}{ABCDE}
		\begin{fmfgraph*}(80,80)
			\fmfstraight
			\fmfleft{g1,g2,i3}
			\fmfright{o1,o2,o3}
			\fmf{fermion}{i3,v2,o3}
			\fmf{fermion}{v1,o1}
			\fmf{fermion}{v1,o2}
			\fmf{fermion,label=$C$}{v2,v1}			
			\fmf{phantom}{g1,v1,g2}
			\fmflabel{$A$}{i3}
			\fmflabel{$B$}{o3}
			\fmflabel{$D$}{o2}
			\fmflabel{$E$}{o1}
		\end{fmfgraph*}
	\end{fmffile}
	\vspace{0.5cm}
	\caption{Particle $A$ decays into particle $B$ and $C$, and particle $C$ continues to decay 
	into particle $D$ and $E$.}
	\label{fig:decay}
	\end{center}
\end{figure}
The process we consider here has two successive decays, 
$A\rightarrow B+C$ followed by $C\rightarrow D+E$ 
as shown in Fig.~(\ref{fig:decay}). 
Let particles $B$ and $D$ be the Standard Model particles and 
the $A$, $C$ and $E$ be the SUSY particles. 
We will make the approximation that 
all SM particles are massless (except for the top quark). 

Let us consider the decay process $A\rightarrow B+C$ 
in the rest frame of $A$. We will adopt the following notation 
$A\equiv m_A$, $B\equiv m_B$, and $C\equiv m_C$. 
Using energy-momentum conservation, one can obtain the following relations
\begin{eqnarray}
P&=&\frac{\sqrt{A^2-(B+C)^2}\sqrt{A^2-(B-C)^2}}{2A},\label{momentum1}\\
E_B&=&\frac{A^2+B^2-C^2}{2A},\label{momentum2}\\
E_C&=&\frac{A^2-B^2+C^2}{2A},\label{momentum3}
\end{eqnarray}
where $P=|\vec{P}_B|=|\vec{P}_C|$. We can do the same calculation for the 
second decay process $C\rightarrow D+E$ in the rest frame of $C$ by the following 
substitutions $A\to C$, $B\to D$, and $C\to E$. 
However the calculation for these two successive decay process 
is carried out in two different inertial frames. 
Thus we introduce the following notation $(P_B^{\mu})_A$ to indicate that the four 
vector momentum of particle $B$ is defined in the rest frame of particle $A$. 
So we rewrite the Eqs.~(\ref{momentum1}-\ref{momentum3}) 
for $A\rightarrow B+C$ and $C\rightarrow D+E$, 
and take the approximation that $B=D=0$.
\begin{eqnarray}
(P_{B,C})_A=\frac{A^2-C^2}{2A},~~~
(E_B)_A=\frac{A^2-C^2}{2A},~~~
(E_C)_A=\frac{A^2+C^2}{2A};\label{frameA}\\
(P_{D,E})_C=\frac{C^2-E^2}{2C},~~~
(E_D)_C=\frac{C^2-E^2}{2C},~~~
(E_E)_C=\frac{C^2+E^2}{2C}.
\end{eqnarray}

\begin{figure}[htb]
\begin{center}
 	\begin{fmffile}{theta}
		\begin{fmfgraph*}(160,80)
			\fmfstraight
			\fmfleft{i1}
			\fmfright{o1,o2,o3}
			\fmf{fermion,tension=3,label=$B$}{v1,i1}
			\fmf{fermion,tension=3,label=$C$}{v1,v2}
			\fmfblob{.05w}{v1}
			\fmfblob{.05w}{v2}
			\fmf{fermion,label=$E$}{v2,o1}
			\fmf{dashes}{v2,o2}
			\fmf{fermion,label=$D$}{v2,o3}	
			\fmfv{label=$A$,label.angle=90,label.dist=10}{v1}
			\fmfv{label=$\theta$,label.angle=22,label.dist=20}{v2}
			\fmflabel{$\hat{Z}$}{o2}
		\end{fmfgraph*}
	\end{fmffile}
	\caption{Angle $\theta$ is the angle between momentum $(\vec{P}_D)_C$ and $(\vec{P}_C)_A$. 
	The $z$ direction, or the longitudinal direction in the Eqs.~(\ref{LT1}-\ref{LT3}) 
	is along the direction of $(\vec{P}_C)_A$.}
	\label{fig:theta}
	\end{center}
\end{figure}
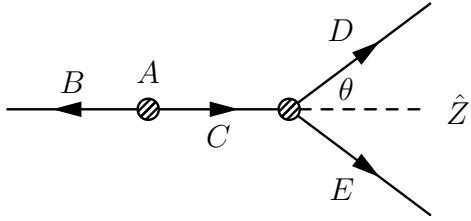

However, in order to reconstruct the invariant mass 
for the Standard Model particles $B$ and $D$, 
we have to obtain the energy-momentum four-vector 
for both particles in the one frame. 
Thus one performs some Lorentz transformations 
to convert the energy-momentum vectors to the same frame, 
for instance, transforming $(P_{D})_C$ to $(P_{D})_A$. 
To do this, one has to know $(P_{C}^{\mu})_A$ 
which has been done in Eq.~(\ref{frameA})
\begin{equation}
(E_C)_A=\frac{A^2+C^2}{2A}, ~~~(P_C)_A=\frac{A^2-C^2}{2A}.\label{CA}\\
\end{equation}
And the Lorentz transformations are as follows
\begin{eqnarray}
(P_D^T)_A&=&\sin\theta(P_D)_C,\label{LT1}\\
(P_D^L)_A&=&(\gamma_C)_A\left[\cos\theta(P_D)_C+(\beta_C)_A(E_D)_C\right],\label{LT2}\\
(E_D)_A	  &=&(\gamma_C)_A\left[(E_D)_C+\cos\theta(P_D)_C(\beta_C)_A\right],\label{LT3}
\end{eqnarray}
where the angle $\theta$ is the angle between momentum 
$(\vec{P}_D)_C$ and $(\vec{P}_C)_A$ as shown in Fig.~(\ref{fig:theta}); 
$\sin\theta(P_D)_C$ and $\cos\theta(P_D)_C$ are the transverse and longitudinal components 
of the momentum of particle $D$ in the rest frame of particle $C$; 
and the Lorentz transformation variables $\gamma$ and $\beta$ are defined as 
\begin{equation}
(\gamma_C)_A=\frac{(E_C)_A}{M_C}, ~~~(\beta_C)_A=\frac{(P_C)_A}{(E_C)_A}.\label{LTCA}
\end{equation}

In the rest frame of particle $A$, the Lorentz invariant quantity, the invariant mass of the 
Standard Model particles $B$ and $D$ now can be calculated easily so that 
\begin{equation}
M_{BD}=\sqrt{(E_B+E_D)^2-(P_D^T)^2-(P_D^L-P_B^L)^2}\label{mbd}
\end{equation}
where we have dropped the subscript $A$ in Eq.~(\ref{mbd}) and will do so in 
the subsequent analysis. The Eq.~(\ref{mbd}) can be simplified 
 \begin{equation}
M_{BD}=A\sqrt{1-\frac{C^2}{A^2}}\sqrt{1-\frac{E^2}{C^2}}\sqrt{\frac{1+\cos\theta}{2}}\label{mbd2}
\end{equation}
which has a maximum value when $\cos\theta=1$ so that 
\begin{equation}
M_{BD}^{max}=A\sqrt{1-\frac{C^2}{A^2}}\sqrt{1-\frac{E^2}{C^2}}\label{mbd3},
\end{equation}
and it has a vanishing minimum when $\cos\theta=-1$. The results discussed here 
are utilized in the analysis given in chapter (\ref{ch:kin}). 


\clearpage
\textbf{\large Benchmarks}

\begin{table}[htbp]
    \begin{center}
   {Chargino  Patterns (CPs)}
       \scriptsize{
\begin{tabular}{|c|c|c|c|c|c|c|c|c|c}                                                                  \hline  \hline \hline
 {\bf  \rm SUGRA}    &   $m_0$   &   $m_{1/2}$   &   $A_{0}$ &   $\tan {\beta}$  &   $\mu$   &   NUH &   NU3 &   NUG \\
{\bf  \rm Pattern}   &   (GeV)   &   (GeV)   &   (GeV)       &   ($v_u/v_d$)        & (sign)   & $(\delta_{H_u},\delta_{H_d})$ & $(\delta_{q3},\delta_{tbR})$ & $(\delta_{M_2},\delta_{M_3})$   \\
\hline  \hline
 ${\bf mSP1 }$  &   2001    &   411 &   0   &   30.0    &   +   &   (0,0)   &   (0,0)   &   (0,0)   \\ \hline
 ${\bf mSP1 }$  &   2366    &   338 &   -159    &   9.8 &   -   &   (0,0)   &   (0,0)   &   (0,0)   \\ \hline
 ${\bf mSP1 }$  &   1872    &   327 &   -1893   &   14.9    &   +   &   (0.107,0.643)   &   (0,0)   &   (0,0)   \\ \hline
 ${\bf mSP1 }$  &   1041    &   703 &   1022    &   11.6    &   +   &   (0,0)   &   (-0.524,-0.198) &   (0,0)   \\ \hline
 ${\bf mSP1 }$  &   1361    &   109 &   1058    &   14.4    &   +   &   (0,0)   &   (0,0)   &   (0.929,0.850)   \\ \hline \hline
 ${\bf mSP2 }$  &   1125    &   614 &   2000    &   50.0    &   +   &   (0,0)   &   (0,0)   &   (0,0)   \\ \hline
 ${\bf mSP2 }$  &   2365    &   1395    &   3663    &   42.2    &   -   &   (0,0)   &   (0,0)   &   (0,0)   \\ \hline
 ${\bf mSP2 }$  &   1365    &   595 &   3012    &   35.1    &   +   &   (0.116,-0.338)  &   (0,0)   &   (0,0)   \\ \hline
 ${\bf mSP2 }$  &   1166    &   507 &   -954    &   59.6    &   +   &   (0,0)   &   (0.325,0.458) &   (0,0)   \\ \hline
 ${\bf mSP2 }$  &   1414    &   221 &   -551    &   54.3    &   +   &   (0,0)   &   (0,0)   &   (0.156,0.968)   \\ \hline \hline
 ${\bf mSP3 }$  &   741 &   551 &   0   &   50.0    &   +   &   (0,0)   &   (0,0)   &   (0,0)   \\ \hline
 ${\bf mSP3 }$  &   1585    &   1470    &   3133    &   39.1    &   -   &   (0,0)   &   (0,0)   &   (0,0)   \\ \hline
 ${\bf mSP3 }$  &   694 &   674 &   -1564   &   27.0    &   +   &   (0.922,-0.293)  &   (0,0)   &   (0,0)   \\ \hline
 ${\bf mSP3}$   &   570 &   559 &   1042    &   41.3    &   +   &   (0,0)   &   (-0.482,-0.202) &   (0,0)   \\ \hline
 ${\bf mSP3 }$  &   392 &   312 &   320 &   41.3    &   +   &   (0,0)   &   (0,0)   &   (-0.404,0.908)  \\ \hline \hline
 ${\bf mSP4 }$  &   1674    &   137 &   1985    &   18.6    &   +   &   (0,0)   &   (0,0)   &   (0,0)   \\\hline
 ${\bf mSP4 }$  &   1824    &   127 &   -1828   &   6.4 &   -   &   (0,0)   &   (0,0)   &   (0,0)   \\ \hline
 ${\bf mSP4 }$  &   1021    &   132 &   -638    &   6.6 &   +   &   (0,0)   &   (-0.020,0.963)  &   (0,0)   \\ \hline
 ${\bf mSP4 }$  &   2181    &   127 &   -3859   &   3.9 &   +   &   (0,0)   &   (0,0)   &   (0.836,-0.248)  \\ \hline \hline
${\bf NUSP1}$   &   2738    &   1689    &   -4243   &   42.4    & + & (0,0)   &   (-0.828,-0.899) &   (0,0)   \\ \hline
${\bf NUSP1}$   &   540 &   1190    &   2516    &   13.9    &   +   & (0,0)   & (0,0) & (-0.408,-0.660) \\ \hline
${\bf NUSP2}$   &   845 &   726 &   -75 & 48.4    &   +   &   (0,0)   & (-0.694,-0.400) &   (0,0) \\ \hline
${\bf NUSP3}$  &   396 &   1018    &   -179    & 18.3    &   +   &   (0,0)   &   (0,0) &   (0.250,-0.452)  \\ \hline
${\bf NUSP4}$   &   400 &   1558    & 2511    &   5.9 &   +   & (0,0)   &   (0,0)   &   (-0.401,-0.607)
\\ \hline\hline \hline
\end{tabular}
}
 \caption{ Benchmarks for the class CP where the chargino $\cha$ is the NLSP
in mSUGRA and  in NUSUGRA  models. Benchmarks are computed with ${m_b}^{\overline{\rm MS}}(m_b) = 4.23$ {\rm GeV},
${\alpha_s}^{\overline{\rm MS}}(M_Z)=.1172$, and $m_t({\rm pole}) =
170.9$ ${\rm GeV}$. 
} \label{b1}
    \end{center}
 \end{table}



\begin{table}[htbp]
\begin{center}
  {Stau  Patterns (SUPs)}
   \scriptsize{
\begin{tabular}{|c|c|c|c|c|c|c|c|c|}                                                                  \hline  \hline \hline
{\bf  \rm SUGRA}   &   $m_0$   &   $m_{1/2}$   &   $A_{0}$ &   $\tan {\beta}$  &   $\mu$   &   NUH &   NU3 &   NUG \\
{\bf  \rm Pattern}  &   (GeV)   &   (GeV)   &   (GeV)   &   ($v_u/v_d$) & (sign)  &   $(\delta_{H_u},\delta_{H_d})$   & $(\delta_{q3},\delta_{tbR})$ & $(\delta_{M_2},\delta_{M_3})$   \\
\hline  \hline
${\bf mSP5}$    & 111 & 531 &   0   &   5.0 &   +   &   (0,0)   &   (0,0)   & (0,0)   \\ \hline
${\bf mSP5}$    &   162 &   569 &   1012    & 15.8    &   - & (0,0)   &   (0,0)   &   (0,0)   \\ \hline
${\bf mSP5}$    & 191 & 545 &   -722    &   17.2    &   +   & (-0.340,-0.332) & (0,0)   & (0,0)   \\ \hline
${\bf mSP5}$    & 114 &   440 &   -50 &   15.2    &   +   &   (0,0)   & (-0.204,-0.846) &   (0,0)   \\ \hline
${\bf mSP5}$    &   75  & 348 &   301 &   12.0    &   +   & (0,0)   &   (0,0)   & (0.234,-0.059)  \\ \hline\hline
${\bf mSP6}$ & 245 &   370 & 945 &   31.0    &   +   &   (0,0)   &   (0,0)   &   (0,0)   \\ \hline
${\bf mSP6}$    &   1452    &   1651    &   2821    &   38.5 &   -   &   (0,0)   &   (0,0)   &   (0,0)   \\ \hline
${\bf mSP6}$ & 356 &   545 &   927 &   31.7    &   +   &   (0.667,0.055)   & (0,0) &   (0,0)   \\ \hline
${\bf mSP6}$    &   442 &   463 & 1150    & 41.0    &   +   &   (0,0)   &   (-0.187,-0.546) & (0,0)   \\ \hline
${\bf mSP6}$    &   308 &   307 &   965 &   35.6 &   +   &   (0,0) & (0,0)   &   (-0.383,0.405)  \\ \hline\hline
${\bf mSP7}$    & 75  & 201 &   230 &   14.0    &   +   & (0,0)   &   (0,0)   & (0,0)   \\ \hline
${\bf mSP7}$    &   781 & 1423    &   983 &   36.8 &   -   & (0,0)   &   (0,0)   & (0,0)   \\\hline
${\bf mSP7}$ &   428 &   671 &   484 &   43.8 &   +   &   (-0.392,-0.808) & (0,0)   &   (0,0) \\ \hline
${\bf mSP7}$    &   226 &   426 & 944 &   27.1    &   +   &   (0,0)   & (0.176,-0.430)  &   (0,0)   \\ \hline
${\bf mSP7}$    &   143 & 425 &   266 &   23.4    &   +   & (0,0)   &   (0,0)   & (0.718,0.100)   \\ \hline\hline
${\bf mSP8}$ & 1880    &   877 &   4075    &   54.8    &   +   &   (0,0)   & (0,0) &   (0,0) \\ \hline
${\bf mSP8}$    &   994 &   1073    &   3761    &   38.1 &   -   &   (0,0)   &   (0,0)   &   (0,0)   \\ \hline
${\bf mSP8}$ & 602 &   684 &   805 &   49.6    &   +   &   (0.490,0.326)   & (0,0) &   (0,0)   \\ \hline
${\bf mSP8}$    &   470 &   624 & -88 &   55.4 &   +   &   (0,0)   &   (-0.531,-0.075) &   (0,0)  \\ \hline
${\bf mSP8}$    &   525 &   450 &   642 &   56.4    &   + &   (0,0)   &   (0,0)   &   (0.623,0.246)   \\ \hline\hline
${\bf mSP9}$    &   667 &   1154    &   -125    &   51.0    &   +   & (0,0)   &   (0,0)   &   (0,0)   \\ \hline
${\bf mSP9}$    &   560 & 1156    &   -1092   &   39.5    &   -   &   (0,0)   &   (0,0)   & (0,0)   \\ \hline
${\bf mSP9}$    &   362 &   602 &   268 &   37.0 & +   &   (0.969,-0.232)  &   (0,0)   &   (0,0)   \\ \hline
${\bf mSP9}$    &   496 &   731 &   679 &   49.3    &   +   &   (0,0)   & (-0.241,-0.452) &   (0,0)   \\ \hline
${\bf mSP9}$    &   485 & 478 &   -128    &   52.8    &   +   &   (0,0)   &   (0,0)   & (0.971,0.653)   \\ \hline\hline
${\bf mSP10}$   &   336 &   772 & -3074   &   10.8    &   +   &   (0,0)   &   (0,0)   &   (0,0)   \\ \hline
${\bf mSP10}$   &   738 &   1150    &   -4893   &   15.5    & +   &   (0,0)   &   (0.802,0.343)   &   (0,0)   \\ \hline\hline
${\bf mSP17}$   &   908 &   754 &   5123    &   25.4    &   -   & (0,0)   &   (0,0)   &   (0,0)   \\ \hline\hline
${\bf mSP18}$   & 344 &   686 &   -2718   &   13.8    &   -   &   (0,0)   &   (0,0) & (0,0)   \\ \hline
${\bf mSP18}$   &   322 &   806 &   -3069   & 9.3 &   +   &   (0.526,-0.707)  &   (0,0)   &   (0,0)   \\ \hline
${\bf mSP18}$   &   60  &   290 &   -339    &   5.2 &   +   & (0,0)   & (0,0)   &   (0.967,-0.074)  \\ \hline\hline
${\bf mSP19}$   &   1530 &   1875    &   13081   &   16.3    &   -   & (0,0)   &   (0,0)   & (0,0)   \\ \hline
${\bf mSP19}$   &   1828 &   1326    &   -5102   & 32.3    &   +   &   (0.592,-0.213)  & (0,0)   &   (0,0)   \\ \hline
${\bf mSP19}$   &   782 &   637 & 2688    &   37.9    &   +   & (0,0)   &   (0,0)   & (0.451,-0.551)  \\ \hline\hline
${\bf NUSP5}$ & 649 &   955 & -1984   &   33.5    &   +   &   (0,0)   & (-0.763,0.701)  & (0,0)   \\ \hline
${\bf NUSP6}$ &   1360    &   1736 &   -2871   & 46.1    &   +   &   (0,0)   &   (-0.466,0.694)  &   (0,0)   \\\hline
${\bf NUSP7}$ &   1481    &   1531    &   -3169   &   42.2    & +   &   (0,0)   &   (0,0)   &   (0.117,-0.463)  \\ \hline
${\bf NUSP8}$ &   670 &   1788    &   371 &   57.9    &   +   &   (0,0)   & (0,0)   &   (-0.223,0.931)  \\ \hline
${\bf NUSP9}$ &   46  &   1938 & -48  &   13.0    &   +   &   (0,0)   &   (0,0)   & (-0.412,-0.650) \\ \hline
                                                                    \hline \hline
\end{tabular}
} \caption{ Benchmarks for the class SUP where the stau $\sta$ is
the NLSP  in mSUGRA and in NUSUGRA. } \label{b2}
    \end{center}
 \end{table}


\begin{table}[htbp]
\begin{center}
  {Stop Patterns (SOPs)}
       \scriptsize{
\begin{tabular}{|c|c|c|c|c|c|c|c|c|}\hline  \hline \hline
 {\bf  \rm SUGRA}   &   $m_0$   &   $m_{1/2}$   &   $A_{0}$ &   $\tan {\beta}$  &   $\mu$   &   NUH &   NU3 &   NUG \\
{\bf  \rm Pattern}   &   (GeV)   &   (GeV)   &   (GeV)   &   ($v_u/v_d$) &
(sign)  &   $(\delta_{H_u},\delta_{H_d})$   &
$(\delta_{q3},\delta_{tbR})$ & $(\delta_{M_2},\delta_{M_3})$   \\
\hline  \hline
 ${\bf mSP11 }$ &   871 &   1031    &   -4355   &   10.0    &   +   &   (0,0)   &   (0,0)   &   (0,0)   \\ \hline
 ${\bf mSP11 }$ &   1653    &   909 &   7574    &   5.9 &   -   &   (0,0)   &   (0,0)   &   (0,0)   \\ \hline
 ${\bf mSP11 }$ &   1391    &   1089    &   8192    &   14.9    &   +   &   (0.470,0.632)   &   (0,0)   &   (0,0)   \\ \hline
 ${\bf mSP11 }$ &   2204    &   933 &   -1144   &   35.6    &   +   &   (0,0)   &   (0.642,-0.400)  &   (0,0)   \\ \hline
 ${\bf mSP11 }$ &   1406    &   1471    &   -2078   &   8.3 &   +   &   (0,0)   &   (0,0)   &   (-0.130,-0.690) \\ \hline \hline
 ${\bf mSP12 }$ &   1371    &   1671    &   -6855   &   10.0    &   +   &   (0,0)   &   (0,0)   &   (0,0)   \\ \hline
 ${\bf mSP12 }$ &   1054    &   1372    &   -5754   &   13.7    &   -   &   (0,0)   &   (0,0)   &   (0,0)   \\ \hline
 ${\bf mSP12 }$ &   915 &   927 &   -3993   &   20.7    &   +   &   (0.078,0.833)   &   (0,0)   &   (0,0)   \\ \hline
 ${\bf mSP12 }$ &   826 &   1016    &   -3926   &   12.8    &   +   &   (0,0)   &   (-0.630,-0.490) &   (0,0)   \\ \hline
 ${\bf mSP12 }$ &   1706    &   1287    &   -4436   &   29.7    &   +   &   (0,0)   &   (0,0)   &   (0.416,-0.260)  \\ \hline \hline
 ${\bf mSP13 }$ &   524 &   800 &   -3315   &   15.0    &   +   &   (0,0)   &   (0,0)   &   (0,0)   \\ \hline
 ${\bf mSP13 }$ &   765 &   1192    &   -4924   &   12.0    &   -   &   (0,0)   &   (0,0)   &   (0,0)   \\ \hline
 ${\bf mSP13 }$ &   1055    &   1601    &   -6365   &   13.6    &   +   &   (0.277,-0.820)  &   (0,0)   &   (0,0)   \\ \hline
 ${\bf mSP13}$  &   1073    &   1664    &   -6528   &   11.6    &   +   &   (0,0)   &   (0.728,0.060)   &   (0,0)   \\ \hline
 ${\bf mSP13 }$ &    540 &   774 &   -2432   &   5.3 &   +   &   (0,0)   &   (0,0)   &   (0.705,-0.201)  \\ \hline \hline
 ${\bf mSP20 }$ &   1754    &   840 &   7385    &   13.3    &   -   &   (0,0)   &   (0,0)   &   (0,0)   \\\hline
 ${\bf mSP21 }$ &   792 &   845 &   6404    &   12.6    &   -   &   (0,0)   &   (0,0)   &   (0,0)   \\ \hline
${\bf NUSP10}$  &   718 &   467 &   1657    &   19.0    &   +   & (0,0)   &   (0,0)   &   (0.023,-0.810)  \\ \hline
                                                                    \hline \hline
\end{tabular}
} \caption{Benchmarks for the class SOP where the stop  $\ta$
is the NLSP  in mSUGRA and in NUSUGRA models.  } \label{b3}
    \end{center}
 \end{table}


\begin{table}[h]
    \begin{center}
    Higgs Patterns (HPs)
    \scriptsize{
\begin{tabular}{|c|c|c|c|c|c|c|c|c|}
                                                                    \hline  \hline \hline
 {\bf  \rm SUGRA}   &   $m_0$   &   $m_{1/2}$   &   $A_{0}$ &   $\tan {\beta}$  &   $\mu$   &   NUH &   NU3 &   NUG \\
{\bf  \rm Pattern}   &   (GeV)   &   (GeV)   &   (GeV)   &   ($v_u/v_d$) &
(sign)  &   $(\delta_{H_u},\delta_{H_d})$   &
$(\delta_{q3},\delta_{tbR})$ & $(\delta_{M_2},\delta_{M_3})$   \\
\hline  \hline
 ${\bf mSP14 }$ & 1040 & 560 & 450 & 53.5 & + & (0,0)          & (0,0) & (0,0) \\ \hline
 ${\bf mSP14 }$ & 760 & 515 & 2250 & 31.0 & + & (0.255,-0.500) & (0,0) & (0,0) \\ \hline
 ${\bf mSP14 }$ & 740 & 620 & 840   & 53.1 & + & (0,0)     & (-0.530,-0.249) & (0,0)\\ \hline
 ${\bf mSP14 }$ & 1205 & 331 & -710 & 55.0 & + & (0,0) &(0,0) & (0.380,0.250)\\ \hline\hline
 ${\bf mSP15 }$ & 1110 & 760 & 1097 & 51.6 & + & (0,0)          & (0,0) & (0,0) \\ \hline
 ${\bf mSP15 }$ & 1395 & 554 &-175 & 59.2  & + &  (0,0)& (-0.040,0.918) & (0,0)\\ \hline
 ${\bf mSP15 }$ & 905  & 500 & 1460 & 54.8 & + & (0,0) & (0,0)& (-0.350,-0.260)\\ \hline\hline
${\bf mSP16 }$ & 520 & 455 &620 &55.5 & + & (0,0) & (0,0)& (0,0)\\ \hline
 ${\bf mSP16 }$ &   282 &   464 &   67  &   43.2    &   + &   (0.912,-0.529)  &   (0,0)   &   (0,0)   \\ \hline
${\bf NUSP12}$  &   2413    &   454 &   -2490   &   48.0    & + & (0,0)   &   (0,0)&   (-0.285,-0.848) \\ \hline
                                                                    \hline \hline
                                                                     \end{tabular}
} \caption{ Benchmarks for  the class HP where the Higgs boson
$(A,H)$ is the next nearest  heavy particle after the LSP
 in mSUGRA and in NUSUGRA. The LSP and $(A,H)$ sometimes are seen to switch.} \label{b5}
    \end{center}
 \end{table}


\begin{table}[h]
\begin{center}
  {Gluino  Patterns (GPs)}
     \scriptsize{
\begin{tabular}{|c|c|c|c|c|c|c|c|c|}\hline \hline\hline
   {\bf  \rm SUGRA}     &   $m_0$   &   $m_{1/2}$   &   $A_{0}$ &   $\tan {\beta}$  &   $\mu$   &   NUH &   NU3 &   NUG \\
   {\bf  \rm Pattern}    &   (GeV)   &   (GeV)   &   (GeV)   &   ($v_u/v_d$)   &   (sign)  &   $(\delta_{H_u},\delta_{H_d})$   &   $(\delta_{q3},\delta_{tbR})$   &   $(\delta_{M_2},\delta_{M_3})$   \\ \hline  \hline
 {\bf NUSP13}    &  2006    &   1081 &   -2027    &   21.1   &   +   &   (0,0)   &   (0,0)   &   (0.207,-0.844)  \\ \hline
  {\bf NUSP14}    &   3969    &   1449    &   -6806   &   29.3   &   +   &   (0,0)   &   (0,0)   &   (0.611,-0.834)   \\ \hline
 {\bf NUSP15}    &   1387    &   695 &   2781    &   50.5   &   +   &   (0,0)   &   (0,0)   &   (0.136,-0.827)  \\ \hline
                                                                        \hline \hline
\end{tabular}
} \caption{Benchmarks for the class GP where the  gluino $~\g  $ is
the NLSP.  Such a pattern was only seen to appear in NUSUGRA models with non universal
gaugino masses. An analysis of light gluinos in the MSSM can be seen in \cite{Profumo:2004wk}.
} \label{b4}
    \end{center}
 \end{table}

\singlespacing		
\addcontentsline{toc}{chapter}{Bibliography}
\bibliography{ref}

\end{document}